%% file: main.tex
\documentclass[runningheads]{llncs}
\usepackage{graphicx} 
\usepackage{fullpage}
\usepackage{url}
\usepackage{xcolor}
\usepackage{longtable}
\usepackage[round]{natbib}
\usepackage{hyperref}
\usepackage{amssymb}
\usepackage{booktabs}
\usepackage{tabularx}
\usepackage{algorithm} 
\usepackage{algpseudocode} 
\usepackage{lscape}
\usepackage[a4paper,margin=1in]{geometry}
\usepackage{xr}
\title{Reliability and predictability of phenotype information from functional connectivity in large imaging datasets}

\author{
Jessica Dafflon\inst{1, 2} \and 
Dustin Moraczewski \inst{1} \and
Eric Earl \inst{1} \and
Dylan M. Nielson \inst{2} \and
Gabriel Loewinger \inst{2} \and
Patrick McClure \inst{3} \and
Adam G. Thomas\inst{1} \and
Francisco Pereira\inst{2}
}

\authorrunning{Dafflon et al.}

\institute{
Data Science \& Sharing Team, National Institute of Mental Health, Bethesda, MD, USA \and 
Machine Learning Team, National Institute of Mental Health, Bethesda, MD, USA \and
Naval Postgraduate School, Monterey, CA, USA
}

\begin{document}

\maketitle
\label{sec:title}
\begin{abstract}

One of the central objectives of contemporary
neuroimaging research is to create predictive models that can disentangle the
connection between patterns of functional connectivity across the entire brain
and various behavioral traits. Previous studies have shown that models trained
to predict behavioral features from the individual's functional connectivity
have modest to poor performance. 
In this study, we trained models that predict observable individual traits (phenotypes) and their corresponding singular
value decomposition (SVD) representations -- herein referred to as \emph{latent
phenotypes} from resting state functional connectivity. For this task, we predicted phenotypes in two large neuroimaging datasets: the Human Connectome Project (HCP) and the Philadelphia
Neurodevelopmental Cohort (PNC). We illustrate the importance of regressing out confounds, which could significantly influence phenotype prediction. Our findings reveal that both phenotypes and their corresponding latent phenotypes yield similar predictive performance. Interestingly, only the first five latent phenotypes were reliably identified, and using just these reliable phenotypes for predicting phenotypes yielded a similar performance to using all latent phenotypes. This suggests that the predictable information is present in the first latent phenotypes, allowing the remainder to be filtered out without any harm in performance. This study sheds light on the intricate relationship
between functional connectivity and the predictability and reliability of
phenotypic information, with potential implications for enhancing predictive
modeling in the realm of neuroimaging research.

\end{abstract}

\vspace{.5cm}
\section{Introduction}

One of the central objectives of contemporary neuroscience is to elucidate a
connection between brain structure/function and behavior. This connection
could provide clinically relevant metrics, and be used for prognosis and
treatment decisions. In order to identify such connections, many neuroimaging studies have focused
on predicting observable characteristics or behavior -- broadly, {\em
phenotypes} -- from brain measurements. Despite the popularity of this
question, most studies report a small correlation of $r=0.1-0.4$ between the
phenotypes and their predictions from brain measurements; moreover, the
prediction performance for similar or related phenotype measures may vary substantially across studies,
\citep{Greene-Clinical, Sui-2020Neuroimaging-based, Poldrack-2020Establishment,
Pervaiz-2020Optimising, Genon-2022Linking}.

The situation described above can be attributed to several factors, for instance intrinsic variations of scanning equipment, noise at various stages of acquisition, or natural and pathological differences between study participants, among others. Moreover, the effect of these factors can be exacerbated by variations in sample size across studies. 
In general, the use of machine learning methods in typical clinical neuroimaging datasets
is severely constrained by their size (tens to low hundreds of subjects), in
comparison with their dimensionality (tens of thousands of features). Determining how much data is needed to train models
to predict brain-behavior associations is still an open question. Recent research by
\cite{Marek-2022Reproducible} suggested that at least a thousand subjects are
needed to accurately measure univariate brain-behavior associations, and that small
studies fail to replicate and produce inflated effect sizes. This controversial
study prompted an avalanche of responses showing that the brain-behavior
relationships could be predicted from smaller datasets by using multivariate models and hold-out samples
 \citep{Spisak-2023Multivariate} and thoughtfully
choosing the brain data and behavioral targets of prediction
\citep{Cecchetti-2022Reproducible, Gratton-2022Brain-behavior,
Rosenberg-2022How}. 

In addition, recent studies \citep{Nikolaidis-2022Suboptimal, Gell-2023The} argue that increasing sample size might improve the prediction performance, but not be sufficient for consistent results. They argue that the reliability of the prediction targets will limit predictive performance, and that the reliability of clinical and cognitive phenotypes should be better assessed. In particular, \cite{Nikolaidis-2022Suboptimal} showcased that, while inconsistency in results is aggravated by small sample sizes, the use of more reliable measures can allow for more robust estimates. The authors' suggested solution to improve reliability is to aggregate across multiple raters and/or measurements.

Traditionally, brain-behavior models have primarily focused on individual behaviors or symptoms examined in isolation.  The
increasing recognition of the integrated nature of the brain has led to a
methodological shift, however, and researchers have begun to consider the complex interplay of variables across various demographic, clinical, and behavioral phenotypes \citep{Holmes-2018The}. Following this line of thought,
\cite{Chen-2023Relationship, Chen-2022Shared} averaged the scores of selected
behavioral phenotypes into three categories (Cognition, Personality, and Mental
Health). Their main idea was that the combination would reflect not only a single phenotypic measure and its noise, but a hopefully more robust combination of many measures. In addition, because scores across different behavioral tasks may be  correlated, many studies process them using dimensionality reduction techniques like factor analysis \citep{Ooi-2022Comparison, Schottner-2023Exploring} and principal
component analysis \citep{Chopra-2022Reliable}. The goal of dimensionality reduction is to find a reduced set of (potentially orthogonal) latent variables that contain most of the information contained in the original phenotypes. \cite{Ooi-2022Comparison} showed that latent phenotypes derived from factor analysis were more predictable than the original phenotypes in the HCP and ABCD datasets. Therefore, in this study, we attempted to replicate this finding in a different dataset using a different dimensionality reduction algorithm: singular value decomposition (SVD). We chose to use SVD because it produces the optimal low-rank approximation to the data, in addition to having other desirable properties that will be discussed later in this paper. In addition, \cite{Ooi-2022Comparison} only used the top three components. Therefore, an additional question that was left unanswered, and we address in this paper, is what is the impact of using all latent phenotypes and if the latter might be more reliable than the former, and hence lead to better predictive results. This is particularly
relevant as prior research has identified a limited set of latent phenotypes that
connect brain imaging data with a wide array of phenotypes, including
cognition, mental health, demographic characteristics, and clinical phenotypes
\citep{Chen-2022Shared, Smith-2015A, Goyal-2022The, Miller-2016Multimodal}.

An additional motivation for this work is analyzing the robustness of findings in the literature when using a different dimensionality reduction method and applying this dimensionality reduction to more than one dataset. There is an inherent flexibility in the process of designing and executing scientific experiments and, in particular, analytical pipelines, and  this can be yet another source of variability in the performance of brain-behavior
prediction models. It is now widely acknowledged that a single research
question can be approached through a diverse array of analytical pipelines,
often resulting in different outcomes \citep{Dafflon-2022A,
Botvinik-Nezer-2020Variability, Carp-2012On}. In young research areas, such as functional neuroimaging, in which
many of the ground truths are yet to be discovered, there are limited foundations to favor one option over its alternatives. As a result,
researchers must consider various factors when designing imaging analyses,
including data preprocessing, feature selection, functional connectivity
computation, and prediction model of choice. A study by \cite{Pervaiz-2020Optimising}
explored the impact of various choices on model performance, provided
insights into the influence of these choices, and offered recommendations about which parcellations to use, metrics for estimating functional connectivity and how those choices depend on the target of interest. It's worth noting, however, that while this study employed a
substantial sample size, it primarily focused on a limited set of cognitive
measures (fluid intelligence and neuroticism score), age, and sex.
Consequently, whether the findings and suggestions generalize to other
phenotypes remains uncertain. Many studies individually focused on one dataset
or one phenotype prediction, and the absence of a gold
standard complicates comparisons between studies that adopt different
analytical approaches.

In this study, we examined the questions above in the context of predicting rich
phenotypic information from functional connectivity data, in the
Human Connectome Project (HCP) and Philadelphia Neurodevelopmental Cohort (PNC)
datasets. First, we analyzed the importance of regressing out confounding\footnote[1]{While not confounds in the strict definition of confounding from the causal inference literature \citep{hernan_causal_2023}, we refer to inflation of phenotype prediction by age and sex as confounding in this paper to reflect the common usage in the machine learning field.} covariates
such as age or sex at birth, since predictive performance for variables correlated with these may be inflated by the known predictability of age and sex from functional connectivity.

We then attempted to replicate previous findings that latent phenotypes are more predictable than original phenotypes \citep{Ooi-2022Comparison}.
Specifically, we compared the performance of models predicting latent phenotypes from functional connectivity to the performance of models predicting the original phenotypes from functional connectivity.

Since the reliability of the predicted variable limits the maximal observable predictive performance, we also evaluated the reliability of the latent phenotypes across different participant samples, and the relationship between reliability and predictability from functional connectivity.
Finally, we tested whether the phenotype information predictable from functional connectivity is primarily contained in reliable latent phenotypes.  To this effect, we reconstructed phenotypes from the subset of reliable latent variables, and evaluated the performance of models trained to predict these reconstructed phenotypes from functional connectivity. Overall, our study
provides insights into the relationship between prediction accuracy and
reliability and extends our understanding of the extent to which latent, and reconstructed phenotypes can be predicted from resting state fMRI.

\section{Methods}
\label{sec:methods}

\subsection{Datasets}

In the analyses described in this paper, we used resting state fMRI data and
phenotype information from two datasets: the Human Connectome project
\citep{Van-Essen:2012aa} and Philadelphia Neurodevelopmental Cohort
\citep{Satterthwaite-2016The}. We will focus on predicting phenotypes only from
functional connectivity (FC) measures derived from the fMRI data, as prediction
models derived from FC outperform other modalities in general
\citep{Ooi-2022Comparison}. We provide dataset-specific details in the
following subsections, and describe the common extraction of FC matrix after
that. 
The code used for the analysis, together with a list of all subjects used for training, validation, and hold-out, can be found at the following repo: \href{https://github.com/JessyD/bblocks-phenotypes}{https://github.com/JessyD/bblocks-phenotypes}.
Our work relies on the recently released
``dataset-phenotypes" tool
\footnote{https://github.com/nimh-dsst/dataset-phenotypes}, which outputs BIDS tabular
phenotypic data dictionaries and transforms tabular phenotypic data to BIDS TSVs for
common neuroimaging datasets.
 
\subsubsection{Human Connectome Project (HCP)} We used the behavioral and
imaging data from the HCP Young Adult 1200 Subject release \citep{Van-Essen:2012aa}(N=1071
; 485 males/586 females aged $28.8 \pm 3.7$ years). 
As described by \cite{Barch-2013Function}, the functional connectivity data was acquired with a 32-channel head coil on a 3T Simens Skyra with TR=720 s, TE=33.1 ms, flip
angle=52 deg, BW=2290 Hz/Px, in-plane FOV=208 × 180 mm, 72 slices, 2.0 mm isotropic
voxels, with a multi-band acceleration factor of 8. 

Subjects were chosen if their imaging preprocessing pipeline was completed without error. While for PNC, we removed frames that were considered as high-motion using their framewise displacement and DVARS (as described below), we did not use quality control metrics to select HCP subjects. Subjects were split into a training set (N=856; 385
males/471 females aged $28.88 \pm 3.70$ years), a validation set (N=108; 45
males/63 females aged $28.16 \pm 3.50$ years), and a separate hold-out set (N=107; 55 males/52 females aged $28.69 \pm 3.96$).
The hold-out set was not used for the results reported in this paper, and is
being kept for a future pre-registered analysis. 
Subjects were assigned to the training, validation, and hold-out sets in a two-stage procedure. First, all families with two or more siblings were included in the training set. This was done to prevent leakage of information between twins or other siblings, which could happen if they were split between training and the other two datasets. Second, all subjects were assigned at random to the three datasets, until the specified number of participants was obtained in each.

\paragraph{Imaging Data} We used the cleaned version of the HCP S1200 release,
and used the grayordinate resting-state functional MRI data processed with
ICA-FIX and MSM-All provided by \cite{Glasser-2013The-minimal}, a pipeline
which was developed and optimized for this dataset. 

\paragraph{Behavioral Phenotypes}
\label{sec:behavioral-hcp}
We used 83 phenotypes scores (Appendix~A\ref{tab:phenotypes-hcp} includes a
full list of all scores used), which span across behavioral domains of
cognition and personality and some additional variables that we included to be consistent with prior work. To facilitate comparisons, we included all
behavioral measures previously used by \cite{Ooi-2022Comparison}, \cite{
Kong-2019Spatial} and \cite{He-2020Deep}. All behavioral scores were individually
z-scored across participants, and outliers that were above and below 3 standard
deviations were treated as missing data as described below. We incorporated
behavioral phenotypes categorized as ``age-adjusted behaviors", which had
undergone prior age adjustments by the HCP team, alongside the raw, unadjusted
behaviors. The difference between ``age-adjusted behaviors" and ``unadjusted"
behavior is that, while the Unadjusted Scale Score evaluates a participant's
performance compared to the entire NIH Toolbox Normative Sample, the
Age-Adjusted Scale Score assesses a participant's performance in the context of
a specific age group within the Toolbox Norming Sample (e.g., 18-29 or 30-35)
\citep{Slotkin-2012NIH}. Both the outliers and the missing data were imputed
using the IterativeImputer function from  \verb+scikit-learn+
\citep{Pedregosa-2011Scikit-learn:}, which models each
feature as a function of other features, and uses the model to predict its missing values, iterating over features in a round-robin fashion. To make sure that
we keep the development and validation sets completely separate, we trained the
imputation function on the development set and applied it to the validation
set. 

\subsubsection{Philadelphia Neurodevelopmental Cohort (PNC)} 
As described by \cite{Satterthwaite-2016The}, the PNC dataset was acquired using a 3T Siemens TIM Trio, with a 32-channel head coil with TR=3000 ms, TE=32 ms, flip angle=90 deg, BW=2056, FOV=192 x 192 mm, voxel resolution of 3x3x3mm with 46 slices. 
For our analysis,
we used 1082 subjects. Similar to the HCP dataset, we split the
subjects into a development set (N=864; 404 males/460 females aged $15.670 \pm
3.376$ years), validation set (N=109; 48 males / 61 females aged $16.231 \pm 3.364$),
and an unused hold-out (N=109, 47 males/ 62 females aged $15.97 \pm 3.38$) dataset reserved for future pre-registered analyses. Subjects were assigned to the training, validation, and hold-out sets in the same two-stage process used in the HCP data.

\paragraph{Imaging Data} We preprocessed the PNC dataset using fmriprep version
21.0.2 \citep{Esteban-2019fMRIPrep:} (additional information about all software
used can be found in Appendix A.\ref{fmriprep-pnc-boilerpalette}). For motion
censoring, time points were considered individually. If a time point exceeded
0.2 mm frame-wise displacement or a derivative root mean square (DVARS) above
75, it was marked as a point to be censored. Intervals of less than five points
between censor points were also censored. The six estimated
head-motion parameters, their derivatives, the average signal within the
anatomically-derived white matter, and cerebrospinal fluid masks obtained from
fmriprep were used as nuisance variables and regressed from the fMRI signal
before we computed the FC. 

\paragraph{Behavioral Phenotypes}
\label{sec:behavioral-pnc}
Although the PNC dataset includes surveys, cognitive and clinical phenotypes, for
this analysis we did not use surveys or clinical phenotypes. We selected 39  summary scores from cognitive tasks from the PNC dataset \citep{Satterthwaite-2016The, Gur-2010A-cognitive, Roalf-2014Within-individual}. See Appendix A.~\ref{tab:phenotypes-pnc} for a full list of all scores used. Similar to the procedure used for the
HCP dataset,  missing
values and outliers were imputed using the IterativeImputer function from
\verb+scikit-learn+, and each measure was z-scored across participants.

\subsection{Generation of datasets for prediction experiments}
\label{sec:fc-matrix}

\begin{figure}[h!] \centering
\includegraphics[width=\linewidth]{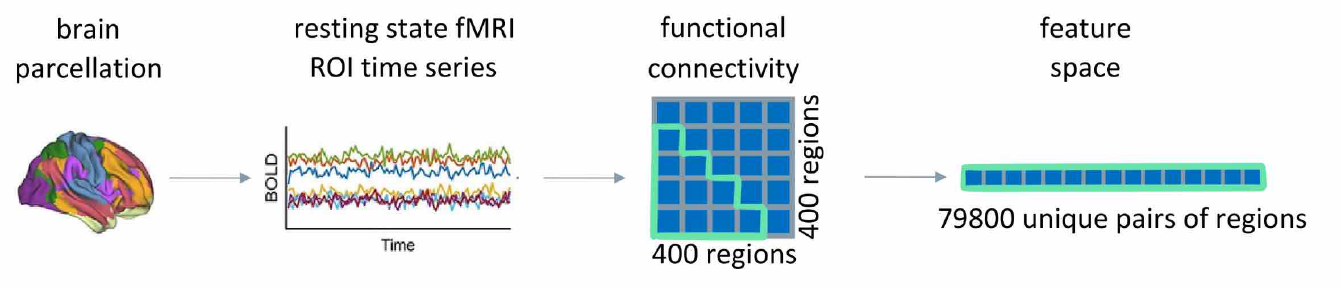}
\caption{Schematic overview of how the data matrix is constructed. First, the
brain activity is parcellated using an atlas of choice (here we used the Schaefer
400 parcellation). The functional connectivity matrix is created by computing
the Pearson correlation between the average resting-state fMRI time series in
each pair of regions. Because the FC matrix is symmetric between its upper and
lower triangular entries, we flatten only the lower triangular entries of the
functional connectivity matrix. }
\label{fig:image_feature extraction}
\end{figure}

We constructed functional connectivity matrices by computing the Pearson
correlations between the average time series for each pair of brain regions.
Regions were defined using the 17-network Schaefer 400 parcellation
\citep{Schaefer-2018Local-global}. If more than one resting state session was
available for a subject, we averaged the resulting functional connectivity
matrices across sessions and used the resulting one in our analyses. If any of the runs had more than 50\% of the runs flagged as time points to be censored, it was excluded from the computation of functional connectivity. Across
this paper, we will refer to the data matrix $X^{n \times d}$), where $n$ is
the number of subjects available and $d$ the number of pairwise connections
after vectorizing the lower diagonal (in our case 79800). These pairwise
connections will be the input features for the prediction model described in
Section~\ref{sec:prediction-model}. The process is illustrated in
Figure~\ref{fig:image_feature extraction}. 

\subsection{Dimensionality reduction of phenotype data}
\label{sec:svd-pheno}

\subsubsection{Linear dimensionality reduction}
The following paragraphs compare different types of linear dimensionality reduction. They introduce those methods and compare them to SVD, the method we used in further analysis. 

In our experiments, we structure phenotype data as a matrix with $n$ rows (participants) and $d$ columns (phenotype measurements).
Each row is therefore a $d$-dimensional observation for a participant. A linear dimensionality reduction method expresses that matrix as a product of a $n \times k$ latent variable matrix, and a $k \times d$ matrix that expresses observation vectors as a linear combination of latent variables (often called the mixing or loading matrix). The latent variable matrix can be viewed as a reduced dimensionality representation of the original matrix.
The underlying assumption behind this framing is that each observation does not vary independently in $d$ different ways. Instead, those $d$ measures are driven by one or more latent variables, and that relationship holds across participants. If measurements were the summary scores of different instruments, the latent variables could correspond to factors that drive variance across them; in that case, latent variables would be helpful to elicit relationships between instruments.
Aside from their role in psychology research, latent variables are potentially useful as prediction targets from imaging, as discussed earlier. This is because they are estimated from multiple phenotype measures they are associated with, and hence, are potentially less noisy. Factorizations can also be used for denoising, in that the product of the latent variable matrix and the mixing matrix can be viewed as an approximation that tries to keep the essential characteristics of the data. This is usually done by considering fewer than the maximum number of latent variables that can be estimated from the data matrix.

\subsubsection{Linear dimensionality reduction methods}

There are many linear dimensionality reduction methods. They differ primarily in their assumptions about the relationship between latent variables, and their connection to the measured variables. While it is beyond the scope of this paper to review all the methods available, we briefly do so for those that have been used for experiments similar to those we carry out.
 Factor Analysis (FA) finds latent variables (factors) and a mixing matrix (loadings) such that between factor correlation is minimized. FA is used in the development of psychometric questionnaires, where the goal is to relate answers to questions to hypothesized latent constructs driving behavior. FA usually includes a step called varimax rotation \citep{Kaiser-1958The}, where the goal is to further transform the factor/loading relationship so that each factor loads on as few phenotype measurements as possible, to facilitate interpretability. 

Independent Component Analysis (ICA) finds latent variables (components) and a mixing matrix such that the factors are statistically independent, not orthogonal, but uncorrelated. Neither FA nor ICA have an intrinsic criterion for choosing the number of dimensions $k$ analogous to the percentage of variance explained in PCA and SVD. 

Principal Component Analysis (PCA) finds successive orthogonal projection directions of the data that maximize variance after projection. Each latent variable is defined as the projection of all the data points into one dimension; the mixing matrix is derived as part of the process of finding the projection directions. The intuition for this approach is that important latent variables should drive a lot of variance in the observations. Similar to FA, a varimax rotation can also be applied to a PCA to simplify the interpretation of the components. 

Singular Value Decomposition (SVD), the technique we will use in this paper, finds the same projection directions as PCA if the inputs to the SVD are mean-centered. Both PCA and SVD share a useful characteristic that other methods do not have: the percentage of variance explained provides an order of importance of the latent variables. In addition, SVD has a number of practical advantages over PCA (e.g., it does not require computation of a covariance matrix), as well as useful theoretical properties. The one that is most important for this paper is that the data reconstruction through SVD, using $k$ latent variables, is the best rank $k$ approximation of the data, in the least squares sense. We provide an illustrated introduction to SVD, its mathematical properties, and its relationship with PCA in Section~\ref{sec:svd-maths} of the Appendix. Finally, there are methods such as reduced rank regression that implicitly perform an SVD/PCA of a dataset of target variables, as part of a multivariate, multiple regression model. Given that the results are similar to those obtained by computing an SVD of the target variables, predicting each latent variable independently, and reconstructing predictions of the targets, we do not consider these further.

\subsection{Controlling for age/sex}
\label{sec:confounds}

Before training the prediction models we regressed out age and sex at birth terms from each
phenotype measure in both datasets: $\textit{age}$, $\textit{sex}$, $\textit{sex}^2$, and the interactions  $\textit{age} \times \textit{sex}$ and $\textit{age}^2 \times
\textit{sex}$. Sex at birth was
coded as a 0/1 binary variable and age was z-scored. The phenotype measure values used in the prediction experiments
were the residual values after fitting this regression model. The purpose of regressing age and sex information out of the phenotype targets was to determine the degree to which their predictability was a combination of a) their being predictable from age/sex (e.g. a developmental phenotype measure) and b) age/sex being predictable from the imaging data. We deliberately did not regress age/sex information out of the imaging data. Our plan is to train neural networks that predict age/sex, in tandem with other phenotypes where age/sex have been regressed out. For that application, age/sex should not be regressed out of the imaging data. As these experiments are meant to produce baseline results using linear models, against which neural network results can be compared, we opted not to regress age/sex out of the imaging data in this case as well.

We also conducted a paired two-sided t-test to determine if there was a significant difference in performance before and after adjusting for age and sex. Before running the t-test, we calculated the correlations between the actual and predicted values for models both with and without the regression adjustments. We then applied the Fisher z-transformation to these correlations and calculated the average across multiple repetitions.

\subsection{Prediction experiments} 
\label{sec:prediction-model}

\paragraph{Prediction model} We used a Ridge regression model to find the
linear relationship between the imaging data and each of the selected
behavioral phenotypes, or the latent variables derived from them. We chose a model with $L_2$-regularisation, as
implemented by the \verb+scikit-learn+ library
\citep{Pedregosa-2011Scikit-learn:}, version 1.2.2, as it can handle ill-posed
problems that result from extremely correlated features. The reason for only using a linear prediction model in this work is that our primary goal is to obtain baseline results for future nonlinear models.

\paragraph{Experimental setup} The predictive results reported reflect the mean and
standard deviation -- which is an estimate of the standard error of the mean -- across 100 bootstrap samples, taking age, gender and family information into
account. In this procedure, we first split the training data into a part that
would be resampled (80\% of the total dataset) and a part that would be used to evaluate the performance of models with different regularization parameters, which was
kept fixed (10\% of the total dataset). The remaining 10\% was kept as a hold-out dataset and was not used in this study. We did not use nested cross-validation. For each resampled training set, we found the optimal setting of
the $l_2$-regularisation term ($\alpha$) using a grid search over the search
space [$10^2$, $10^7$], applying the model to the fixed part of the dataset used for evaluation. The model
with the best $l_2$-regularisation term was then applied to the validation set,
yielding one result of the 100 in the sample for predicting that particular
phenotype measure. We performed the SVD of the phenotypes within each resampled
training set. These were z-scored in a column-wise fashion, i.e., we computed
the mean and standard deviation across the training set to normalize that
specific phenotype, so that the features with large means or variance would not
dominate the results, and repeated this procedure for all phenotypes. We then
used the mean and standard deviation from the training dataset to z-score
the phenotype measures in the validation set. 

\paragraph{Model evaluation}
\label{sec:critical-diff}
The performance of each model was evaluated using the coefficient of
determination ($r^2$) and Pearson's correlation between the predicted measure
and the actual measure in the validation set. While Pearson's correlation assumes
values between -1 and 1, the $r^2$ can assume values between $(-\infty, 1]$
where 1 corresponds to the best possible score, as it is being computed on a
separate dataset. 

We used the Autorank library \citep{herbold2020autorank, demvsar2006statistical} to evaluate if there was any statistical difference in the performance of the prediction algorithms using all of the components or a subset thereof. The Autorank library first assesses the normality and heteroscedasticity of the data before selecting the most suitable group level and post-hoc test to determine differences between the groups. The family-wise significance level for these tests was set at $\alpha$ = 0.05. We provided inputs to the Autorank library by calculating the correlation between the predictions and the actual values, applying the Fisher-z transformation to these correlations, and then computing the average correlation across multiple bootstraps. 

\subsection{Comparing predictive performance on original phenotypes and latent phenotypes}
The latent phenotypes are composed of linear combinations of the original phenotypes, so directly comparing predictive performance between the two is not straightforward. Instead, we computed absolute difference between z-scored values of the prediction and true phenotypes (i.e., the error) for the component with the highest prediction, taking the average over all bootstraps. We also computed the same difference for the best performing latent phenotype. We then used a paired t-test to test the difference in error between the best performing latent phenotype and the best performing original phenotype.
 
\subsection{Latent phenotypes reliability analysis}
\label{sec:gale-algo}
We conducted analyses to assess the reliability of the latent phenotypes produced by the SVD separately in the HCP and PNC datasets. Specifically, we wanted to determine how similar each latent phenotype would be if the SVD was conducted in distinct sets of subjects with the same phenotypic measures. We accomplished this by combining the training and validation subsets and then randomly splitting these combined datasets into two halves (without replacement) 1000 times. 
Splits were applied at the family level so that all members of a family were in the same subsample in every splitting. Since the latent phenotypes produced by the SVD are ordered by percent variance explained, the order may differ between samples. To ensure that the order was consistent across all subsample, we first ran an SVD on the combined dataset, saving the latent phenotypes. 
Then the latent variables from each subsample were reordered to match the order of latent phenotypes in the combined dataset. Components were matched via the Gale–Shapley stable marriage algorithm \citep{gale-shapely} using the correlation between latent phenotype weights as the preferences. 
Once the latent phenotypes from each subsample within a split were aligned, we took the correlation between the latent phenotype's weights on the features (following the SVD notation, this would be $V_T$) as our measure of latent phenotype reliability. 
In all cases age and sex were regressed out of the phenotypes, values more than three standard deviations from the mean were censored, and missing values were imputed as described above prior to running the SVDs. These preprocessing steps were carried out separately in each subsample. Empirical 95\% confidence intervals were determined from the distribution of inter-subsample correlations across splits. Correlation coefficients were Fisher z-transformed prior to averaging across splits and then transformed back for plotting/analysis.

\section{Results}

\subsection{Predictability of phenotype measures from functional
connectivity}
\label{sec:prediction-pheno}

The goal of this experiment was to ascertain the degree to which each phenotype
measure is predictable from functional connectivity, in both HCP and PNC.

\begin{figure}
    \centering
    \includegraphics[width=\linewidth]{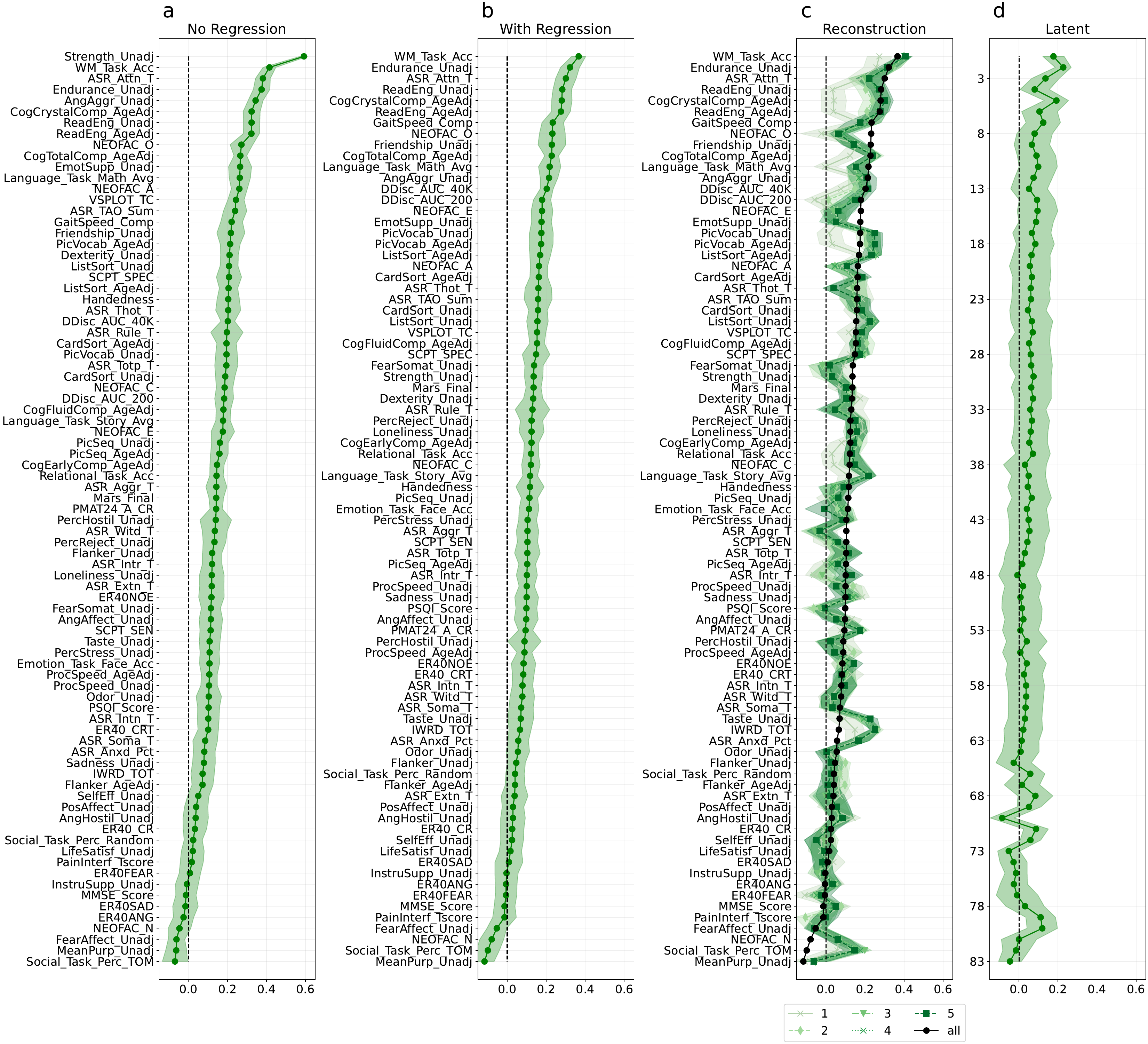}
\caption{Predictability of phenotype variables in the HCP validation set before regressing out confounds (a), after confound removal (b), reconstructed using only a subset of the latent phenotypes (c), and of the latent phenotypes (d). Predictability is quantified as the correlation between true phenotypes and prediction from functional connectivity data. Panel c shows the correlation of the predicted phenotypes with true values when the model was trained with a subset of latent phenotypes (1, 2, 3, 4, 5 and all, represented by the different shades of green).  The shaded regions represent the standard deviation across
resamplings. The phenotypes (y-axis) are ordered by the prediction performance, with the exception of (d), which is ordered by the variance explained by the latent phenotypes. (a) Before regressing out, the phenotypes with the best performance is ``Strength unadjusted"; after regressing age and sex, cognitive phenotypes have the best performance. In addition, using latent phenotypes (d) or untransformed phenotypes (b) does not impact the performance substantially.}
\label{fig:all_hcp}
\end{figure}

\begin{figure}
    \centering
    \includegraphics[width=\linewidth]{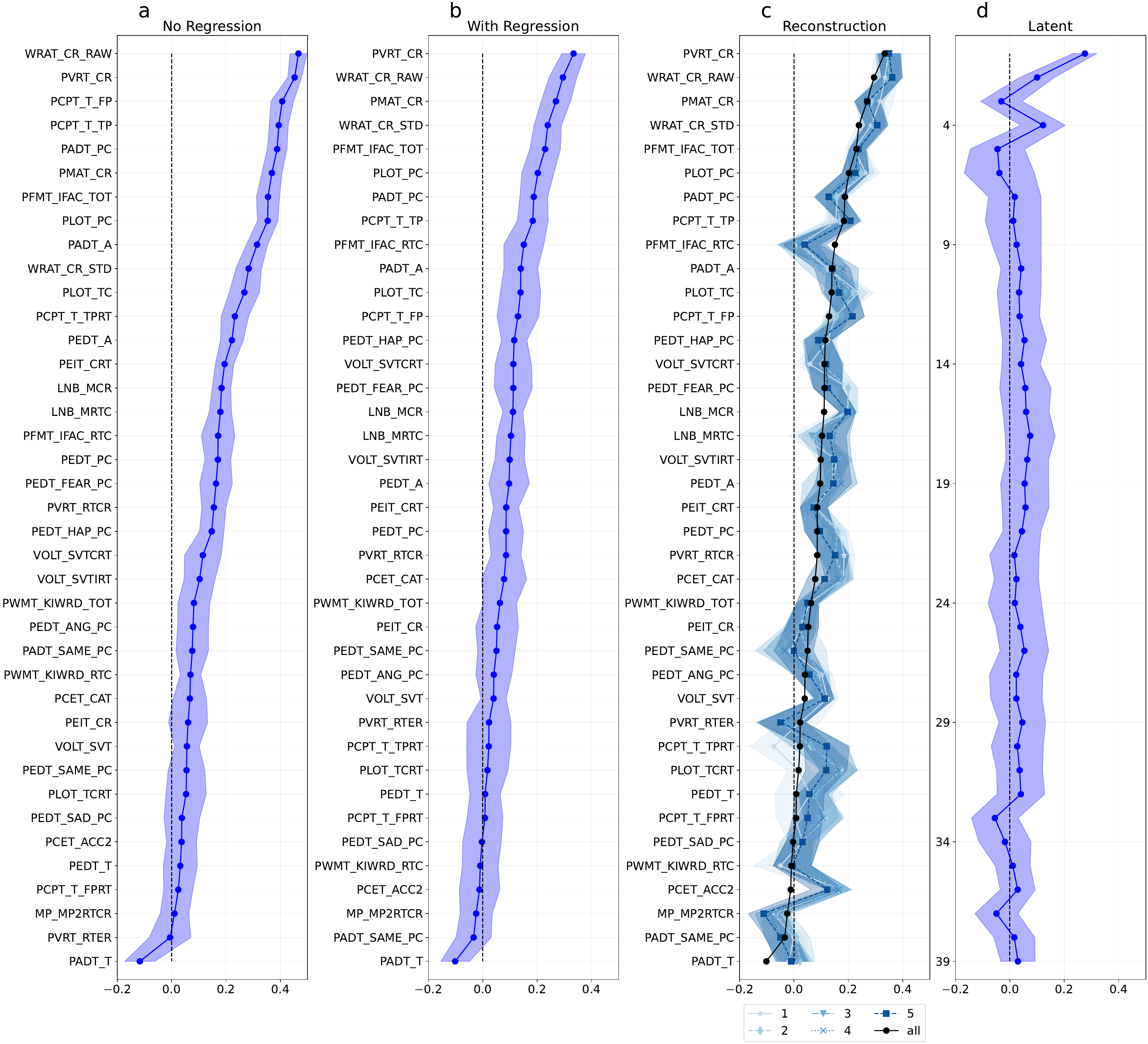}
\caption{Predictability of phenotype variables in the PNC validation set, before regressing out confounds (a), after confound removal (b), reconstructed using only a subset of the latent phenotypes (c), and of the latent phenotypes (d). Predictability is determined by the correlation between actual phenotypes and predictions from functional connectivity. The shaded regions represent the standard deviation across
resamplings. Similarly to Figure~\ref{fig:all_hcp}, the correlation is higher before regressing out the confounds (a) than in the remaining plots (b-d). As HCP and PNC have different prediction ranges, the x-axes here are different than in Figure~\ref{fig:all_hcp}}
\label{fig:all_pnc}
\end{figure}

As described in Section~\ref{sec:prediction-model}, we trained ridge regression
models to predict each phenotype measure across 100 bootstrap resamplings; we
then used them to generate predictions for participants in the validation set, which
was fixed rather than resampled. This resampling scheme allows us to estimate the variability in performance across potential training sets for the fixed validation set. We averaged prediction results across
resamplings, and used those results to estimate the standard deviation of the
mean estimate. Figure~\ref{fig:all_hcp}.a and Figure~\ref{fig:all_pnc}.a show the average correlation
between the phenotype predictions and their true values, with the
measures sorted by predictability.  We can see that for the HCP data the most predictable phenotype is
\emph{Strength Unadjusted}, which measures the Grip Strength (Figure~\ref{fig:all_hcp}a). Because of the relation between strength and sex at birth, we repeated the same analysis after regressing out age and sex and their interaction as confounds. Appendix Figure~\ref{fig:all_hcp_r2} shows the
corresponding results using $r^2$ as the metric of performance.

\subsection{Predictability of phenotype measures from functional
connectivity after regressing the confounds}
\label{sec:covariates}

\begin{figure}[htbp]
\includegraphics[width=\linewidth]{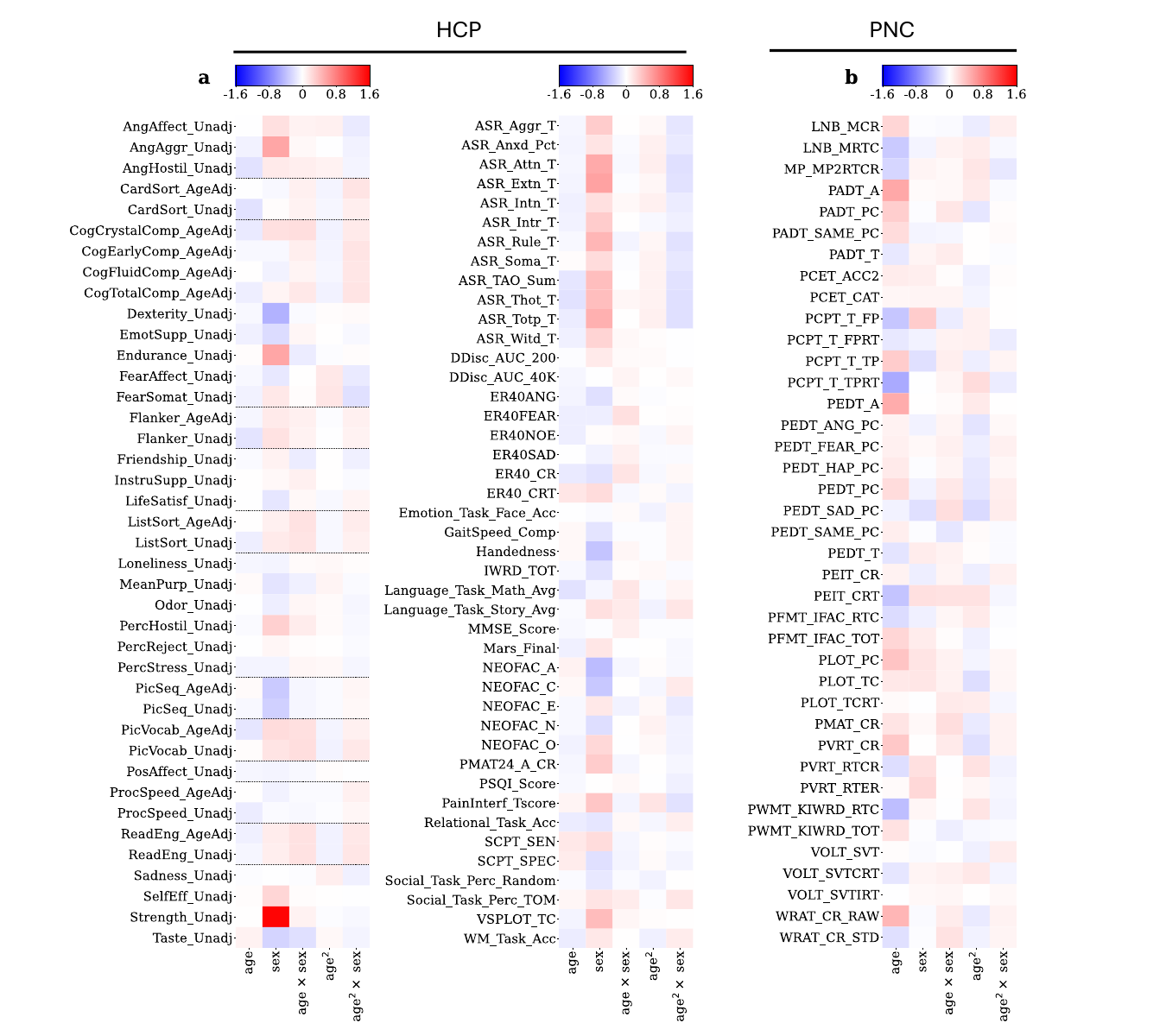}
\caption{Beta-weights of the linear regression used to regress out age, sex,
and their quadratic interactions from HCP (\textbf{a}) and PNC (\textbf{b}) over the repetions.
(\textbf{a}) Dotted lines include pairs of phenotypes included in the analysis
that were both Unadjusted and Age-adjusted (\emph{AgeAdj}).}
\label{fig:beta-coefficients}
\end{figure}

One of the issues in determining how predictable a phenotype measure is from
imaging is the presence of potential confounds, such as participant sex at birth,
age, and time of scan, which are to some extent predictable from imaging data.
Given this, a phenotype measure might be predictable from imaging data because
a) the confound can be predicted from imaging data, and b) the phenotype
measure can be predicted from the confound.  To try to minimize the impact of confounding
variables on our predictions, we regressed out age, sex, and age$\times$sex
interaction, as well as their squares, from each phenotypic measure.

Figure~\ref{fig:beta-coefficients} illustrates the regression beta coefficients
for each phenotype measure in HCP (Figure~\ref{fig:beta-coefficients}a) and PNC
(Figure~\ref{fig:beta-coefficients}b). The magnitude of the beta coefficient
indicates the strength of the relationship between the independent variable and
the dependent variable. A list of all the phenotypes used, the acronyms and a
short description can be found in Table~\ref{tab:phenotypes-hcp} (HCP) and
Table~\ref{tab:phenotypes-pnc} (PNC).

In the HCP analysis, we did not see a large dependence between the variables
and age, and the highest beta coefficient was observed for ``Strength
Unadjusted", where there was a high interaction with sex. This aligns with our expectations as HCP is a young adult cohort while PNC is a neurodevelopmental cohort.
Note that we included
behavioral phenotypes that had previously been adjusted for age by the HCP team
(referred to as ``age-adjusted behaviors") and the raw unadjusted behaviors.
Our initial aim was to compare the beta coefficients between the adjusted and
unadjusted variables provided by the HCP consortia. We observed that if we had trained our model without correcting for the
confounds, some of the predictive performance would be due to their presence.
For the PNC dataset (Figure~\ref{fig:beta-coefficients}b), the highest beta
coefficients are observed between age and Total Correct Response for All Test Trials (PEDT\_A), Median Response Time for All Test Trials (PADT\_T), and Total Raw Score (WRAT\_CR\_RAW).

We also compared predictive performance for phenotypes in which we regressed out age and sex confounds or not. Regressing out age and sex results in a significant decrease in mean prediction across all phenotypes (before regression: 0.172 $\pm$ 0.143 (mean $\pm$ standard deviation); after regression 0.096 $\pm$ 0.095; t-statistic:-6.37, value:1.75e-07, df:38)

\subsection{Predictability of SVD latent variables from functional
connectivity}
\label{sec:prediction-svd}

As described earlier, SVDs
were fit to each resampled training set and used to produce latent
phenotype scores for that training set and the fixed validation set. Contrary to our
expectations, we found that SVD latent phenotypes were not more predictable than
individual phenotype measures (correlation in Figure~\ref{fig:all_hcp}d and Figure~\ref{fig:all_pnc}d; $r^2$ in Appendix Figure~\ref{fig:all_hcp_r2}d and Figure~\ref{fig:all_pnc_r2}d). We evaluated this by performing a paired t-test to compare the absolute error between predicted and true z-scored values for the most predictable latent factor and the most predictable phenotypes. In both datasets the error was not significantly different between the best phenotype (HCP: phenotype mean error (std):0.890 (0.600)/ PNC: phenotype mean error (std): 0.891 (0.654)) and the best latent phenotype (HCP: latent phenotype mean error (std): 1.001 (0.645) / PNC: latent phenotype mean error (std): 0.950 (0.641); HCP: t-statistic=-1.78, p-value=0.078, df=107 / PNC: t-statistic=-1.138, p-value=0.258, df=108).

\begin{figure}[htbp]
\centering
\begin{minipage}{.45\textwidth}
    \includegraphics[width=\linewidth]{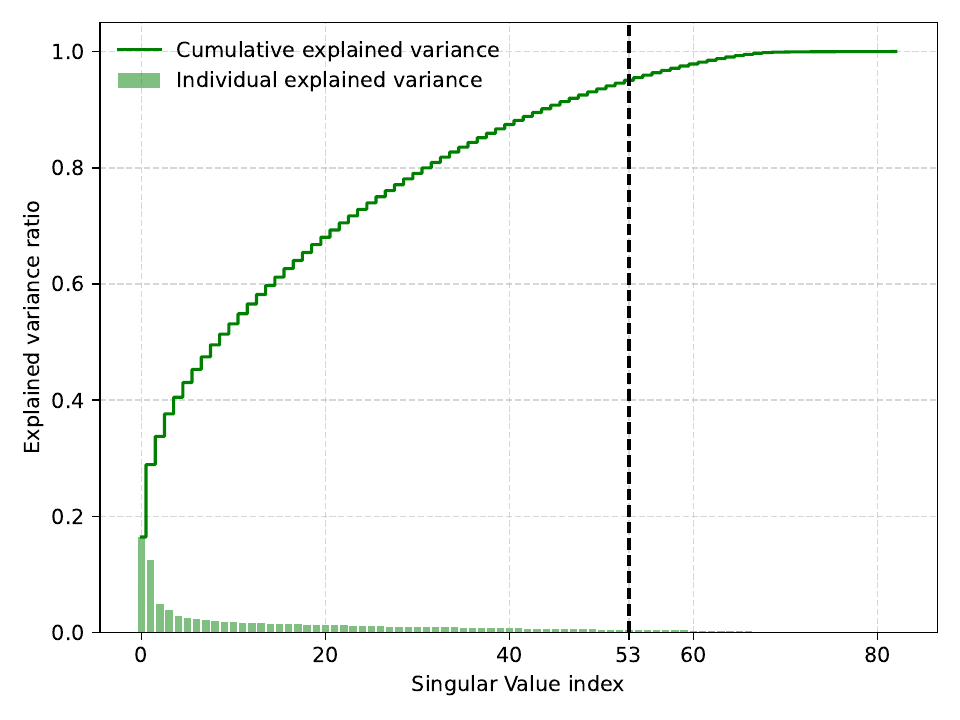}
\end{minipage}
\begin{minipage}{.45\textwidth}
    \includegraphics[width=\linewidth]{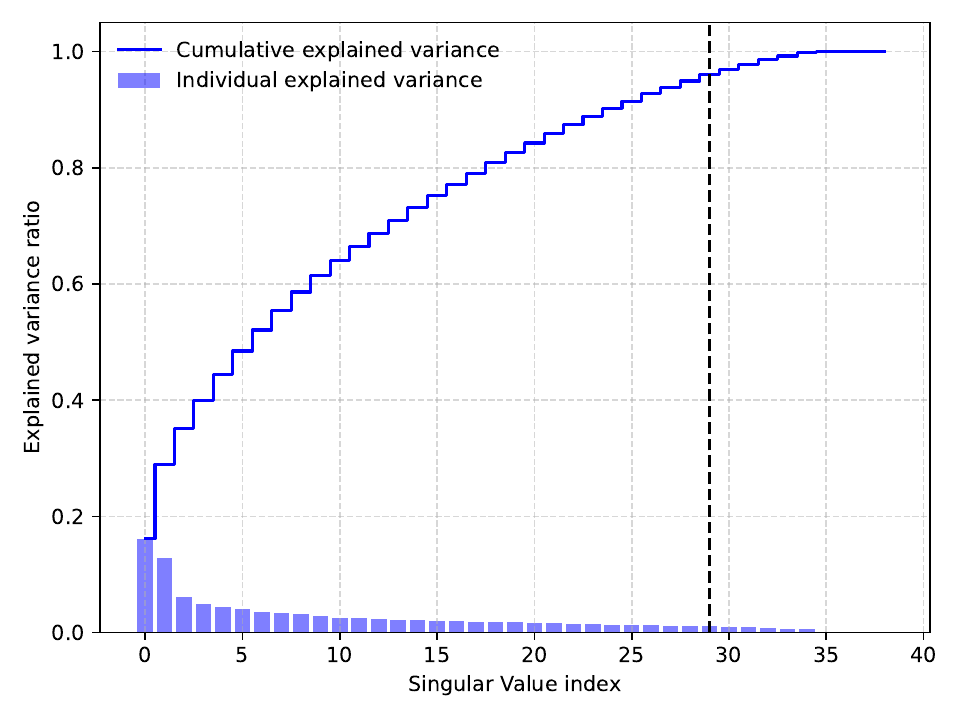}
\end{minipage}
\caption{Explained variance for each of the latent phenotype
variables obtained after applying an SVD to the behavioral variables from HCP
(green, left) and PNC (blue, right). In this experiment, the SVD was computed using the training and the validation data. The bar graphs show the variance explained by each latent variable, while the line plots show the cumulative variance explained. For both datasets, a large proportion of
the variance is explained by the first two latent phenotypes; however, most
latent phenotypes are needed to reach 95\% of the variance, as represented by the dashed line in the figure. Note that the x-axis for these two plots are different as HCP has 83 phenotypes and PNC 39.}
\label{fig:explained_variance}
\end{figure}
\subsection{Low-dimensional representations of phenotype variables and their reliability}
\label{sec:svd-reliability}
One of our goals was to understand the dependency structure between different
phenotype measures, as captured through the SVD of the dataset. This analysis is important to determine if the phenotypes can be represented using a smaller set of uncorrelated latent phenotypes. 

For both datasets, we can see that while the first two
latent phenotypes explain about 30\% of the variance, this decays rapidly for
successive latent phenotypes (Figure ~\ref{fig:explained_variance}). To attain a comprehensive 95\% explanation of
the variance, a substantial number of latent phenotypes are required. Considering that each dataset uses a diverse array of tests to cover various cognitive domains, we were not surprised to see that many of the components were required to explain 95\% of the variance.

\begin{figure}[hbtp]
\centering
    \includegraphics[width=\linewidth]{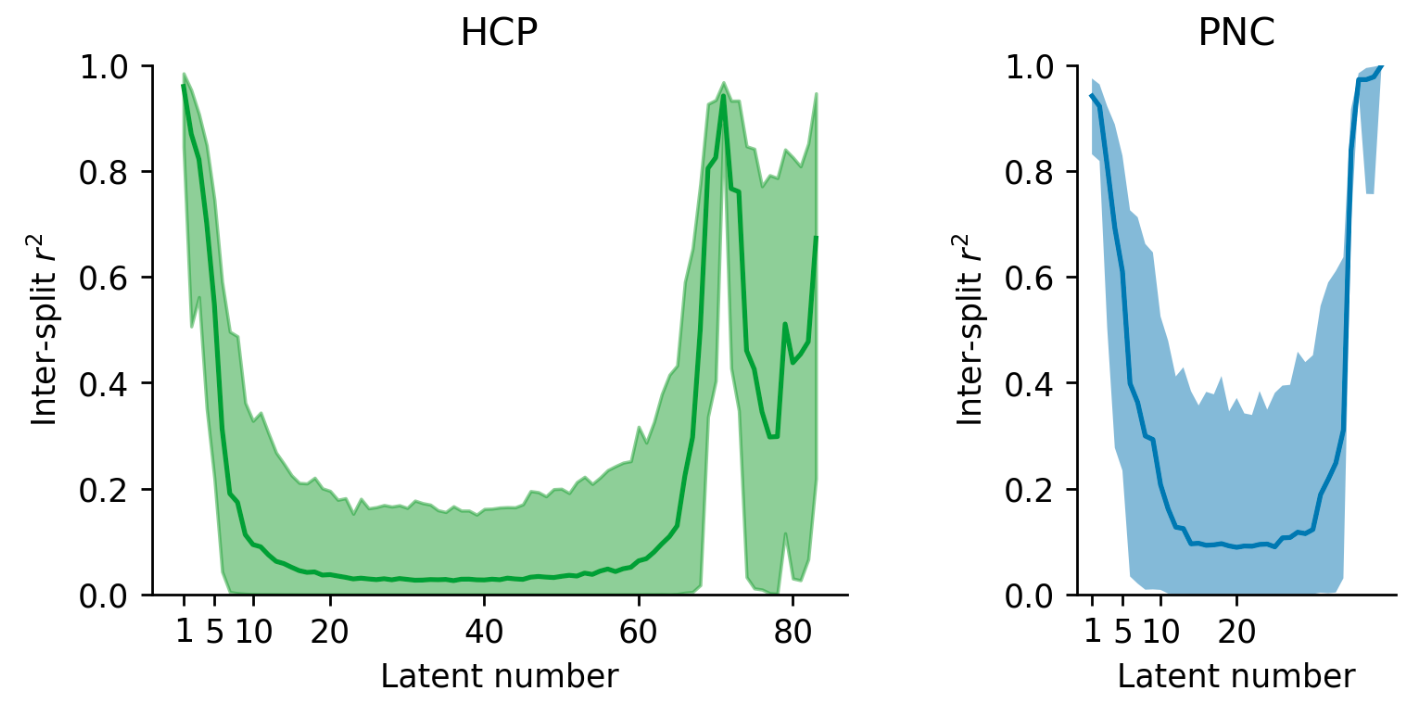}
\caption{Latent phenotypes reliability for both datasets HCP (in green; left) and PNC (in blue; right). The loadings of the first 5 SVD components were reliably identified within different splits using the Gale–Shapley stable marriage algorithm. The shaded area represents the 95\% confidence intervals.}
\label{fig:smudge_plot}
\end{figure}

We also conducted an experiment to examine the reliability of the
latent phenotypes, i.e., how many of them would replicate if the SVD was applied to
two independent samples drawn from the same population. To that effect, we
simulated this situation with an experiment where we repeatedly split the
phenotype dataset into two halves (i.e., sub-samples), and independently computed the SVD of each
split. To ensure that the order was consistent across all subsample, we first ran an SVD on the combined dataset, saving the latent phenotypes. We then reordered the latent phenotypes in each split to match the order of latent phenotypes in the original dataset based on the Gale–Shapley stable marriage algorithm \citep{gale-shapely} using the correlation between latent phenotype weights as the preferences. The matching procedure was repeated for 1000 random splits of the data. The first five
latent phenotypes in each dataset had an inter-split $r^2$ greater than 0.2 in 95\% or more or random splits (Figure~\ref{fig:smudge_plot}, Figure~\ref{fig:smudge_plot_all} in the appendix for correlation). Both datasets also had some reliable low-variance latent phenotypes (69-73 and 83 in HCP, 35-39 in PNC). Outside of these latent phenotypes, the inter-split correlation of the weights was low. We
postulate that this could be due to the loadings for remaining factors being
driven by sample idiosyncrasies or the remaining relationships between
phenotype measures being weak enough that noise can perturb their
estimation.
The predictability analysis reveals that the first three HCP latent phenotypes and the fifth HCP latent phenotype, along with the first PNC component, exhibit a correlation mean between true and predicted values that is above 0.15 (Figures~\ref{fig:all_hcp} and~\ref{fig:all_pnc}). That predictability decays quickly beyond them with an increase in the prediction performance for a few of the last components that do not explain much variance.

\subsection{Predictability of original phenotypes after training on phenotypes 
reconstructed from latent space}
\label{sec:prediction-recons}

The previous sections investigated the predictability of individual phenotypes
measures, and of latent phenotypes derived from them. We saw that
only the first five latent phenotypes in each dataset were reliable (the first 5 components explain 40\% of the variance in HCP and 45\% on PNC), but  explaining
95\% of the variance would require most of the latent phenotypes (53 on the HCP and 28 in PNC). If we assume that variance explained by unreliable latent phenotypes will not be predictable from functional connectivity data, then these results suggest the intriguing possibility that {\em most} of
the variance present across phenotype measures might not be predictable from
functional connectivity data (at least using linear prediction models in a moderately sized sample of about 1000 subjects for both HCP and PNC). We can test this by comparing the predictive performance of models trained using progressively larger proportions of the variance in the original phenotypes to the performance of models trained on the original phenotypes themselves. If a model trained to predict a reduced variance representation of the phenotypes does as well as models trained with the original phenotypes, then this will indicate that that reduced variance is the maximally predictable portion of the variance.  

To test this possibility, we took advantage of the fact that the SVD of a
dataset can be used to generate the best possible rank $k$ approximation of
that dataset, in the least squares sense (
Appendix, section~\ref{sec:svd-maths}). This corresponds to reconstructing each phenotype
measure by considering only those relationships to all other phenotype measures
that explain the most variance. Specifically, we used the first five
latent phenotypes, as those were reliability identified
(Figure~\ref{fig:smudge_plot}),  to produce rank-1 through rank-5 approximate
reconstructions of the phenotype measures in each resampling training set. We then evaluated models trained to predict the reconstructed phenotypes against the true, original phenotypes in the validation set. For both datasets, the reconstructed prediction of rank-1 to rank-5 was very similar to that obtained using the original phenotypes. 
(PNC: Figure~\ref{fig:all_pnc}c, HCP:  Figure~\ref{fig:all_hcp}c).

\begin{figure}[h]
\centering
\includegraphics[width=.8\linewidth]{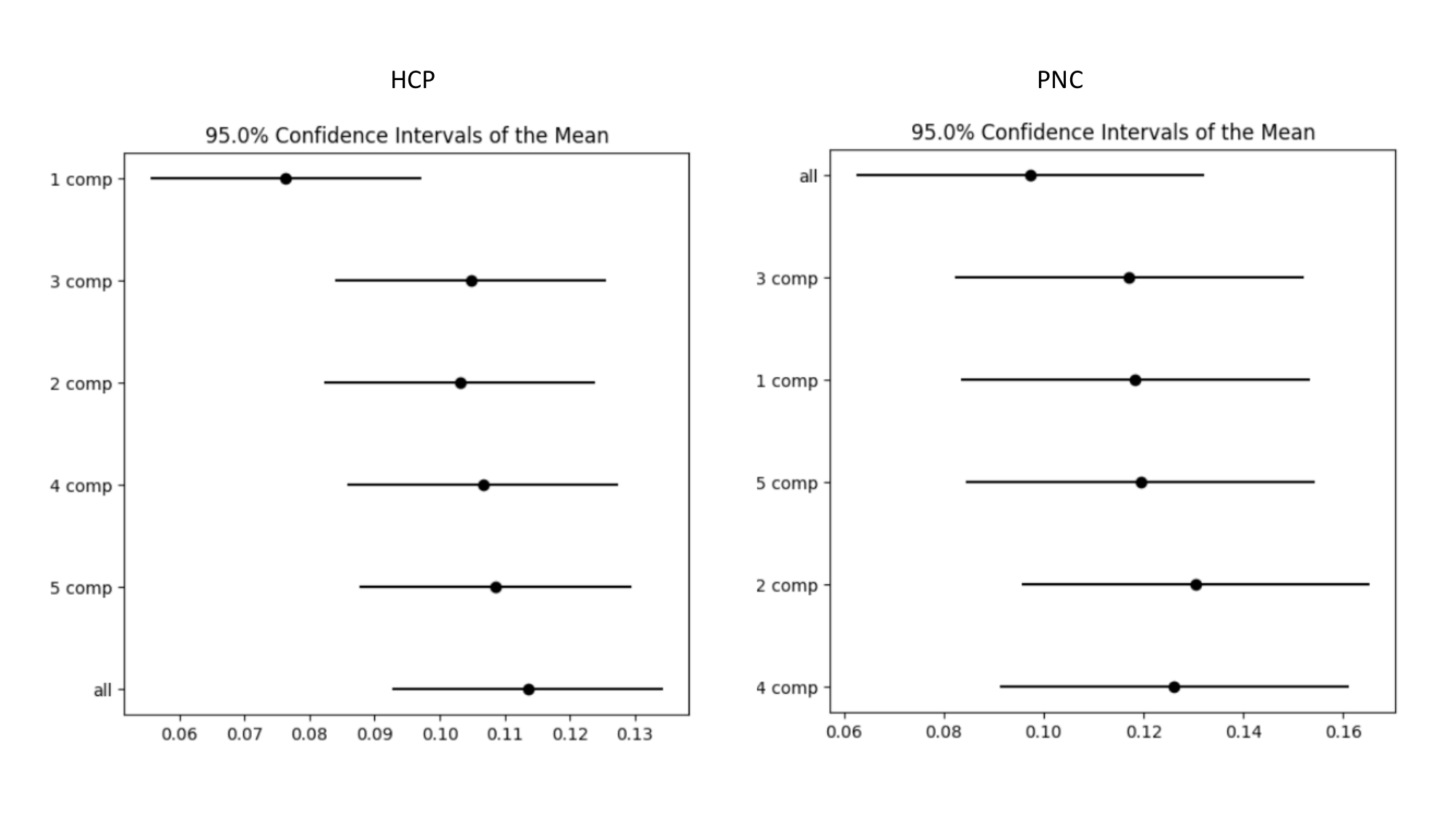}
\caption{Critical difference diagrams for HCP (left) and PNC (right).
For the HCP data, the post-hoc Tukey HSD showed that using just one
latent phenotype had a significantly worse performance than all other groups. There
is, however, no significant difference in using all or a subset of latent phenotypes for
the PNC dataset}
\label{fig:autoref}
\end{figure}

To perform a quantitative comparison
between the 5 reconstruction models and the baseline one, we performed a
critical difference analysis \citep{demvsar2006statistical}. This method is
used to statistically compare the prediction performance of several models across
many different prediction problems (phenotype measures, in this case), to
determine which subsets have significant differences in performance relative to
other subsets. In the case of HCP data, the post-hoc
Tukey HSD test revealed a significant decline in performance when employing a
single latent phenotype compared to all other groups (Figure~\ref{fig:autoref}a). Nevertheless, there was no
statistically significant difference in performance across all measures when
employing 2, 3, 4, 5, or all latent phenotypes. For the PNC dataset, there were no
significant differences in performance between utilizing all latent phenotypes and any of the latent phenotypes.

\section{Discussion}
\label{sec:discuss}

\paragraph{Phenotype measures are predictable in both HCP and PNC.}

In this study, we examined prediction of behavioral phenotypes from functional connectivity data. 
Several studies have trained different kinds of models to predict a variety of phenotype targets, but have mainly focused on one dataset \citep{Chen-2023Relationship, Bertolero-2020Deep} or have used PCA with varimax transformation for dimensionality reduction \citep{Ooi-2022Comparison}. 
As a first experiment, we tested phenotype prediction from functional connectivity in both the Human Connectome Project (HCP) and Philadelphia Neurodevelopmental Cohort (PNC) datasets.
We found that phenotypes can be predicted from resting state functional connectivity in both datasets (Figure~\ref{fig:all_hcp}b and Figure~\ref{fig:all_pnc}b). Similar to other studies, we observed correlations between predicted and true values in the range of 0.1-0.4 for the majority of phenotypes. For example, the top three phenotypes that could be predicted with the highest performance were Working memory score across all memory tasks (WM\_Task\_ACC), 2-min walk endurance test (Endurance\_Unadj), Attention problems (ASR\_ATTN\_T) in the HCP (Figure~\ref{fig:all_hcp}b) and Professional Verbal Reasoning Test Total Correct Responses for All Test Trials (PVRT\_CR) Wide Range Achievement Test's Wide Range Assessment Test 4 Total Raw Score (WRAT\_CR\_RAW) and Primary Mental Abilities total correct response (PMAT\_CR) in the PNC dataset (Figure~\ref{fig:all_pnc}b).

\paragraph{Regression of the confounds is important.}

Studies predicting phenotypes from imaging data do not always control for demographic confounds in their analyses \citep{Chyzhyk-2022How}. While this may make sense in some cases (as we discuss below), failure to control for these confounding variables can inflate estimates of how predictable some phenotypes are if those phenotypes are correlated with the demographic confounds. To evaluate the degree to which this could be an issue, we considered the two most common potential confounds: age and sex. We regressed age, sex, age squared and the interactions of age and age squared with sex from our
prediction targets -- phenotype measures -- and evaluated differences in predictive performance.
This showed that confounds can inflate the prediction results if not accounted for; this is particularly visible, for instance, in the relation between the phenotypic variable ``strength unadjusted" and sex in the HCP dataset ($r=0.6$ before; $r=0.1$ after adjustment; Figure~\ref{fig:all_hcp}a and Figure~\ref{fig:all_hcp}b).
It is also important to note that it is essential to handle confounding separately for the training and test data. Otherwise, attempting to control for these variables across the entire dataset can undermine the independence of the
training and testing data \citep{Chyzhyk-2022How}.

Finally, it is worth noting that it may be useful to keep the confounding
variables in either the prediction targets or the imaging data, depending on
the purpose of the prediction model. One example would be age, which is sometimes regressed out as a confounding variable, and sometimes added as an
additional predictor to explain part of the variance. For example, in the
context of predicting Alzheimer's disease, not regressing age out of imaging
might help prediction models that can consider combinations of age and other
features. It is still an open question, and left as an additional researcher's
degree of freedom, if the confounds should be removed from the image, targets,
or both. In this manuscript, we explored the effects of removing the confounds from the phenotypes but future work should consider regressing the confounds from the images. 

\paragraph{SVD latent variables are not more predictable than individual phenotype
measures.}

Many of the phenotype measures collected in large batteries are correlated to some extent. As discussed earlier, dimensionality reduction techniques can be applied to transform phenotype measures into a latent variable representation. These variables are usually uncorrelated or independent and, informally, explain different aspects of the phenotype measures. Beyond the scientific reasons for generating them, latent variables may also be cleaner than individual phenotype measures, if the individual measures are taken to be noisy measurements of an underlying construct. And, finally, reducing dimensionality also reduces the number of prediction targets to consider.

Several studies have constructed uncorrelated latent variable representations, with each variable explaining different aspects of the phenotypes \citep{Schottner-2023Exploring, Ooi-2022Comparison, Chen-2022Shared}. For example, \cite{Ooi-2022Comparison} computes a factor analysis of the behavioral scores and predicts the latent phenotypes instead of the raw phenotypes. They observed that latent phenotypes, in particular the first three, are more predictable than using individual measures and that functional connectivity outperforms other modalities to predict behavioral phenotypes. In this study, we also transformed the phenotype measures into a latent variable representation, and used those latent variables as targets for prediction experiments. Figure~\ref{fig:all_hcp}d and Figure~\ref{fig:all_pnc}d illustrates that even though the first components have the highest prediction performance, the obtained performance is not higher than that obtained for the untransformed phenotype measures (Figure~\ref{fig:all_hcp}b and Figure~\ref{fig:all_pnc}b). One important point is to
make a distinction between predictability and reliability, as a perfectly reliable phenotype might still not be predictable or, even worse, might not be related to the phenomena being studied.
\cite{Finn-2021Beyond} illustrates this point by making an analogy between functional connectivity (FC)
and barcodes. Barcodes are unique patterns, but they have no intrinsic
connection to any noteworthy traits of the items they label. Therefore barcodes
can have excellent accuracy for identification purposes while offering limited
value in predicting behavior. In simpler terms, FC might be one-of-a-kind yet
fail to provide meaningful insights (high reliability but low validity).

\paragraph{The majority of SVD latent variables are needed to explain 95\% of phenotype variance, but only the first few can be reliably estimated.}
Phenotype prediction is a challenging task, with
performance often being negligible, in terms of $r^2$, if not undistinguishable from chance level.
\cite{Ooi-2022Comparison} showed that latent variables derived from phenotypes
using factor analysis could be more predictable than using the untransformed
phenotypes in the HCP and ABCD datasets. We wanted to assess if this finding
would replicate with a different dimensionality reduction algorithm, and generalize to other datasets. 

The present study employed SVD because it is the most efficient approach to producing low-rank data
approximations. The questionnaires used to produce the phenotype measures in
each study are chosen to span a range of domains of cognition. However, it is
quite likely that each phenotype measure draws from more than one latent
construct, leading to a complex correlation structure between them. This
correlation structure -- identified with SVD, PCA, or FA -- allows robust
identification of latent variables, and has been used to guarantee the
performance of transfer learning algorithms \citep{He-2022Meta-matching}. The
resulting latent variables may also be more interpretable because they are
uncorrelated. 

For both datasets, the majority of the latent variables were required to account for 95\% of the variance; at the same time, the first variables explain more variance than most others (Figure~\ref{fig:explained_variance}). Because of this behavior, we hypothesised that the predictability of each latent variable would be correlated with how much variance it explains. As discussed above and visible from Figure~\ref{fig:all_hcp}d, the observed phenomenon is more complicated than that. Over the first few latent variables, there was no clear relationship between the amount of variance explained and predictability. This is particularly visible for the HCP data, where the first latent variable, which explains 16\% of the variance, was not the most predictable one. At the same time, predictability decays rapidly after the first few latent variables (Figure~\ref{fig:all_hcp}d).

This latter finding prompted us to study the reproducibility of the loadings associated with each latent variable. To do this, we used the Gale–Shapley stable marriage algorithm \citep{gale-shapely}, where we used the correlation to match latent phenotypes across different splits. After further investigation, we observed that, for both datasets, only the loadings for the first five components were reliably identified on 1000 repetitions (Figure~\ref{fig:smudge_plot}), and reliability quickly decays for the remaining factors. We think that due to the mathematical properties of the SVD (that requires all components to be orthonormal), we see an increase of reliability in the low-variance phenotypes again. The increase in reliability for the first components might be related to their increased predictability compared to the other latent phenotypes. Further highlighting this complex relation between predictability and reliability, we also noticed that, in the HCP dataset, some of the variables explaining the least variance are most predictable. This experiment is
important as it showcases that, beyond five latent phenotypes, there would be no
guarantee of finding the same latent phenotypes in different samples. If we cannot
even ensure reliable splits within the same dataset, our ability to uncover
relationships in out-of-sample associations becomes increasingly uncertain. Our
results raise an interesting question: why do latent features beyond the fifth
one become unreliable, and why do they explain so little variance? One simple
guess would be that starting from the fifth component, what is predominantly
being captured is either noise or variance that occurs on very few subjects, as
opposed to meaningful, structured information inherent in the data.

\paragraph{A low-dimensional representation of phenotype variables has similar predictability to the original phenotypes.}

The unreliability from the fifth latent phenotype onward
prompted us to question how much information could be predicted using only the reliable latent variables. Figure~\ref{fig:all_hcp}c and Figure~\ref{fig:all_pnc}c demonstrate
that only a few latent variables are necessary to adequately reconstruct the training set phenotype measures, and achieve similar performance on the validation set as a model trained using the original phenotype measures. This finding suggests that
relations between phenotype measures that are present across many participants
may be captured primarily on the first few latent variables. Therefore, SVD could
serve as a denoising algorithm, providing means to work with fewer latent variables
in training but still yielding as good or better prediction performance as
using the original phenotype measures. 

The relationship between noise and reliability and predictive models is crucial but, while the importance of sample size for statistical power is widely
acknowledged, the consideration of measurement reliability when estimating the
required sample sizes remains an under-addressed aspect
\citep{Zuo-2019Harnessing}. Following this logic, \cite{Gell-2023The, Nikolaidis-2022Suboptimal} propose
that focusing on the reliability of phenotypic measurements during target
selection and choosing only those with high reliability might improve the
performance of brain-behavior models. However, due to the notable variability
in reliability, the proposed sample size requirements may not universally apply
\citep{Rosenberg-2022How, Chen-2023Relationship}. As a significant
consequence, when the reliability of a measurement diminishes, a larger number
of participants will be needed to accurately identify a correlation between two
measures \citep{Zuo-2019Harnessing}. Some studies have found that a larger
sample size helps obtain a higher prediction performance, but carefully
choosing reliable targets has a bigger impact on the model's performance
\citep{Gell-2023The, Nikolaidis-2022Suboptimal}. One remaining question is whether using low-rank SVD
reconstructions of the phenotype measures lessens the sample sizes needed to
obtain a given prediction performance. This could be tested by running
resampling experiments with smaller sample sizes, but this is beyond the scope
of the present paper.

In this paper, we did not assess the reliability of the features that are important for prediction. 
Many works that trained brain-behavior models use a variety of interpretation techniques -- e.g., feature weights or saliency maps -- to
analyze which brain regions were more relevant for prediction
\citep{Finn-2015Functional, Jiang-2020Gender, Chen-2022Shared}. 
While
reliability of the prediction targets has not been the focus of some of these studies \citep{Finn-2015Functional, Jiang-2020Gender, Chen-2022Shared}., the feature
weights of models frequently show moderate to low test-retest reliability as well.
Surprisingly, a linear transformation to these feature weights (i.e.,
Haufe-transformed weights \citep{Haufe-2014On}) demonstrated greater
reliability compared to untransformed weights \citep{Tian-2021Machine,
Chen-2023Relationship}.

\paragraph{Limitations of the current work}
To overcome the scarcity of data, a few studies have used transfer learning to
improve the performance of brain-behavior models \citep{Holderrieth-2022Transfer, He-2022Meta-matching,
Chopra-2022Reliable, Wulan-2024Translating} and shown that information from large datasets
with rich phenotypic data can be used to improve performance on smaller
datasets. However, there are still many implementation details that need to be
further investigated before being able to transfer phenotype
models from large to small datasets successfully.
Generalizability, the broad applicability of a model, is closely linked to the concept of transfer learning. The main difference is that while generalizability refers to the model's ability to perform well on new, unseen data, in transfer learning, a model predicting a phenotype is first trained on a large dataset and then the model is refined to improve its performance on specific smaller datasets, combining the general information learned from the larger dataset with information specific to the smaller dataset. A few studies have evaluated the
generalizability of models trained on one dataset to other datasets, for a variety of brain-behavioral phenotypes, and have
shown that fluid intelligence is one of the few targets for which models have shown moderate performance when tested on new datasets
\citep{Wu-2022Cross-cohort, tong2022transdiagnostic}. In this study, we did not explore transfer learning, as we explored the idea shown by previous research \citep{Nikolaidis-2022Suboptimal, Gell-2023The} that by using more reliable prediction targets (in our case obtained from latent components), we could improve the model prediction but as we have seen the using latent phenotypes did not improve the performance.
 
When pre-processing the phenotypes, we z-scored all phenotypes, set values that were above and below 3 standard deviations to zeros and treated those values as missing by the imputation algorithm. This choice of how to deal with outliers could have an impact on our analysis and was one of the possible researcher degrees of freedom in our analysis. In particular, by making this choice, we are eliminating the most severe scores. A possible alternative would have been to use winsorizing instead of treating the severe scores as missing. 
Another limitation of our preprocessing stems from the fact that we used different processing pipelines from HCP and PNC. While for the HCP we were using the already pre-processed data for HCP, we pre-processed the data using fMRIPrep. Because of this choice, we had additional information about motion parameters for PNC and used this information to censor time points that had excessive motion, but we did not censor time points with excessive motion in HCP.

Another limitation of the current work stems from our reliance on a ridge
regression model, as this can only capture a linear relationship between the
functional connectivity data used as input and the phenotype measures used as
prediction targets.  It is still debated if the brain-behavior relationship can
be better predicted using linear, and therefore more explainable models, or
benefit from the non-linear models. While \cite{Bertolero-2020Deep} showed that
deep neural networks can model simple and complex relationships between brain
and behavior, \cite{He-2020Deep} defended that deep neural networks and kernel
regression yield similar levels of accuracy when it comes to predicting
behavior and demographics through functional connectivity analysis. On the
other hand, \cite{Pervaiz-2020Optimising} and \cite{Schulz-2020Different}
showed that simple linear models perform very similarly to more complex ones.
The second limitation is the use of a linear method (SVD) to derive latent
variables from the phenotype measures. SVD captures covariance relationships, a
linear measure of association between those measures. It is possible that there
are also non-linear relationships between the phenotypes, which could be
identified with non-linear dimensionality reduction approaches such as
auto-encoders. Here, we chose to use both a linear model and a linear
dimensionality reduction approach to set a baseline for future analyses and
facilitate comparison with previous work. 

In conclusion, we showed the importance of controlling for age and sex confounds when creating a brain-behavior model. Failure to remove them could lead to artificially inflated prediction results. Our main
contribution is showing that, by reconstructing phenotype
variables using the first few latent SVD variables, we could obtain a very similar performance training a model on the original variables. This suggests that most of the information about phenotype that can be identified using a linear model trained on functional connectivity data is present only in the first
latent phenotypes.  We suggest that future research should further explore the usage of latent variables derived from phenotypes.

\paragraph{Acknowledgements} This research was supported by the National Institute of Mental Health Intramural Research Program: ZIC-MH002968 (Dylan Nielson, Gabriel Loewinger, Patrick McClure, Francisco Pereira) and ZIC-MH002960 (Jessica Dafflon, Eric Earl, Dustin Moraczewski, Adam Thomas). This research utilized the computational resources of the NIH HPC Biowulf cluster (http://hpc.nih.gov).

\paragraph{Authorship contribution statement} 
Each author contributed as follows:
\newline
JD: data curation, design the study, analyzed the data, contributed to the manuscript\\ 
DMN: design the study, analyzed the data, contributed to the manuscript \\ 
DM: design the study, data curation, contributed to the manuscript \\
EE: data curation, contributed to the manuscript \\
GL: methodology feedback, contributed to the manuscript \\
AGT: design the study, contributed to the manuscript \\ 
PM: design the study, contributed to the manuscript \\
FP: data curation, design the study, contributed to the manuscript

\paragraph{Data and Code availability} All code is available at
\href{https://github.com/JessyD/brain-phenotypes}{https://github.com/JessyD/brain-phenotypes}.
There the user can also find the lists of participants' IDs and behavior scores
utilized for both datasets. Both
\href{https://www.med.upenn.edu/bbl/philadelphianeurodevelopmentalcohort.html}{PNC}
and
\href{https://www.humanconnectome.org/study/hcp-young-adult/document/1200-subjects-data-release}{HCP}
are publicly available and can be obtained after acceptance of their respective
data agreement. 

\bibliographystyle{abbrvnat} \bibliography{main.bib}

\section{Appendix}

\subsection{Additional details on fmriprep pre-processing pipeline}
\label{fmriprep-pnc-boilerpalette}
\input{fmriprep_boiler.tex}

\subsection{Singular Value Decomposition (SVD)}
\label{sec:svd-maths}

\subsubsection{Definition}

Given a $n \times d$ data matrix X, the SVD of X
is \[ X = U S V' \] where $U$ is $n \times k$, $S$ is $k \times k$, and $V$ is
$d \times k$, for a given number of latent variables $k$. The matrices $U$ and
$V$ are called the left and right singular vector matrices (the columns are the
{\em singular vectors}), and $S$ is the singular value matrix (the diagonal
contains the {\em singular values}).

The matrices $U$, $S$, and $V$ have various useful properties:
\begin{itemize}
\item $U$ and $V$ are orthonormal: their columns have norm 1, and are orthogonal to each other.
\item As orthonormal matrices, $U$ and $V$ can be viewed as {\em bases} for the
spaces spanned by columns or rows of X, respectively, and where coordinates for
X would be $SV'$ or $US$.
\item S is a diagonal matrix, with different values in each entry of the diagonal.
\item $k$ can only be as large as $\mathtt{min}(n,d)$; if $X$ is of rank lower than that -- which can happen if
columns or rows are linearly dependent -- it can be smaller.
\end{itemize}

\subsubsection{Dimensionality reduction with SVD}

In the dimensionality terminology used in the paper, $US$ is the latent variable matrix, and $V'$ is the mixing matrix. The variables in the low-dimensional space are uncorrelated, which makes them suitable for various applications (e.g., visualization, and preprocessing for other methods that assume uncorrelated variables).

In addition, the matrix $V$ can be used {\em both} to convert $X$ (or other examples) in the $d$-dimensional space into the $k$-dimensional space, as

well as convert any examples in the low-dimensional space back into the
$d$-dimensional space by multiplying them by $V'$. 

\begin{figure} \includegraphics[width=\textwidth]{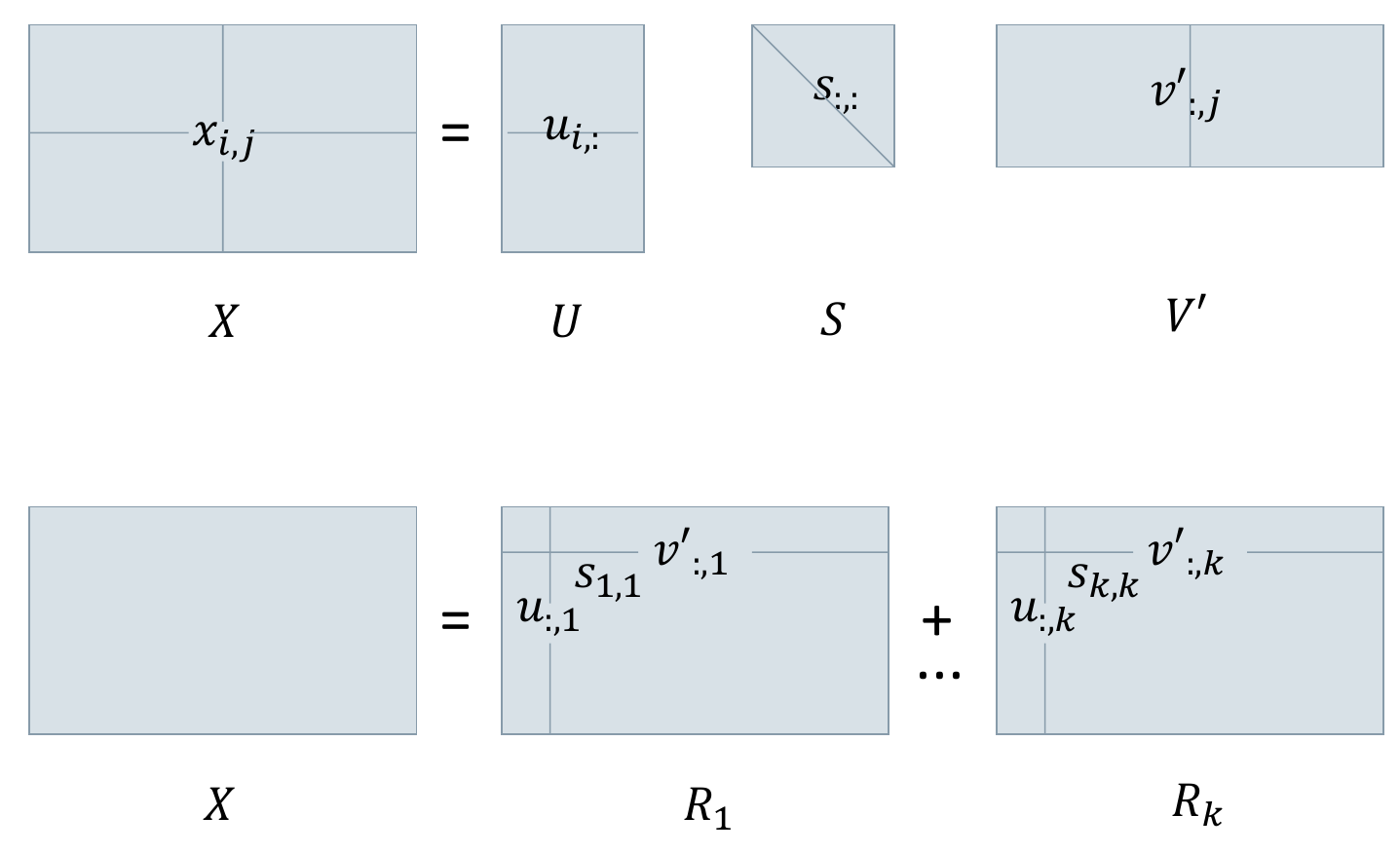}
\caption{{\bf Top:} The support vector decomposition of matrix $X$ as a product
of orthonormal matrix $U$, diagonal matrix $S$, and the transpose of
orthonormal matrix $V$. Each entry $x_{ij}$ in matrix $X$ is reconstructable as
a multiplication of the $i^{th}$ row of $U$ ($u_{i,:}$, scaled by the entries
in the diagonal of $S$, by the $j^{th}$ column of $V'$. {\bf Bottom:} The
matrix $X$ can be expressed as a sum of rank 1 matrices $R_1,\ldots,R_k$, where
$R_l$ is the outer product $u_{:,l} \otimes v'_{:,l}$ multiplied by the scalar
$s_{l,l}$. } \label{fig:matrices} \end{figure}

Figure~\ref{fig:matrices} (top) shows how each individual entry $x_{i,j}$ in
matrix $X$ is reconstructable as a multiplication of the $i^{th}$ row of $U$
($u_{i,:})$, scaled by the entries in the diagonal of $S$, and by the $j^{th}$
column of $V'$. 

An essential step to take into consideration before applying SVD is
normalization as (1) it ensures scale-invariance in the SVD results, preventing
larger-scale variables from dominating the analysis and distorting the
interpretation of relationships between variables; (2) without normalization,
it becomes challenging to understand the relative importance and contributions
of variables due to varying scales, making the interpretation less intuitive;
(3) it enhances the numerical stability of SVD computations, reducing the risk
of numerical errors that can arise when data scales vary widely.

\subsubsection{Relationship between SVD and the data matrix}

SVD provides the best approximation of the matrix X at any specified rank (i.e.,
one can take the full SVD and expand it to show that it reconstructs X as a sum
of $k$ rank 1 matrices, in other words, $R_1,\ldots,R_k$: \begin{eqnarray*} X &
= & R_1 + \ldots + R_k \\ & = & U_{:,1} S_{1,1} V_{:,1}' + \ldots + U_{:,k}
S_{k,k} V_{:,k}' \end{eqnarray*}

as shown in Figure~\ref{fig:matrices} (bottom)). Each of these matrices explains
some of the variance in $X$. If we sort the dimensions of the SVD from those
that explain the most to those that explain the least, adding $k$ matrices will
yield the best rank $k$ approximation of $X$, in terms of minimizing squared
error.

If we take the diagonal of the matrix S, ${\bf s}=\mathtt{diag}(S)$, then the
percentage of variance accounted for by dimension $i$ is
$\frac{s_i^2}{\sum_{j=1}^k s_j^2}$. Adding up the variance for dimensions $1-i$
will give you the total variance explained by the rank $i$ approximation. This
is useful in data compression (best approximation), denoising (many dimensions
with little variance are capturing noise), or matrix completion (if there are
missing entries, the SVD reconstruction can be a good multivariate imputation
method, subject to certain conditions). All of this results from the fact that,
if minimizing squared error to the original matrix $X$, the SVD with $k$
latent phenotypes is the best rank $k$ approximation of $X$.

\subsubsection{Relationship between PCA and SVD}

To prove the
relationship between PCA and SVD, we assume that the dataset $X$ has been {\em
centered}, i.e. where each column has mean 0. The covariance matrix for a
centered dataset X is $C = \frac{1}{n} X'X$. Principal Component Analysis is a
spectral decomposition of $C$, i.e.  \[ C = E \Lambda E' \]

where $E$ is a $d \times d$ matrix, and $\Lambda$ is a diagonal matrix. The $E$
is an orthonormal matrix (column vectors are norm 1, and orthogonal to each
other), where columns are called {\em eigenvectors}. The diagonal of $\Lambda$,
${\bf \lambda}=\mathtt{diag}(\Lambda)$ contains the corresponding {\em
eigenvalues}. The matrix $E$ acts as a basis for reconstructing the covariance
matrix E. The fraction of variance explained by eigenvector $i$ is
$\frac{\lambda_i}{\sum_{j=1}^d \lambda_j}$.

The relationship between SVD and PCA becomes clear if we consider that C can be
expressed in terms of the SVD matrices, as follows

\begin{eqnarray*} C = \frac{1}{n} X'X & = & \frac{1}{n} VS'U'USV' \\ & = & V
\frac{1}{n} S'SV'\ (\mathrm{as\ } U'U=I) \\ & = & V \frac{1}{n} S^2 V'\
(\mathrm{as\ } S=S') \end{eqnarray*}

so the eigenvector matrix $E$ is the same as the right singular vector matrix
$V$ (with some caveats if X is not full rank).

\newpage
\newgeometry{landscape, top=1.5cm, bottom=1.7cm, left=.5cm, right=.5cm}

\subsection{List of all the phenotypes scores used from HCP and PNC}

The tables below contain the variable names used in our study, the description of what it refers to, and which category the questionnaire comes from. 

\paragraph{Used phenotypes from the HCP Dataset}
\begin{longtable}{ l l l }
\toprule
\textbf{Variable} & \textbf{Description} & \textbf{Category}\\
\midrule
AngAffect\_Unadj*\footnote{{*: Those were used by \cite{Ooi-2022Comparison, Kong-2019Spatial, He-2020Deep}. All papers used the same variables.}} & Anger-Affect Survey: Unadjusted & Emotion (NIH Toolbox)\\
AngAggr\_Unadj* & Anger-Physical Aggression Survey: Unadjusted & Emotion (NIH Toolbox)\\
AngHostil\_Unadj* & Anger-Hostility Survey: Unadjusted & Emotion (NIH Toolbox)\\
ASR\_Aggr\_T & Aggressive Behavior Gender and Age Adjusted T-score & Psychiatric/Life Function\\
ASR\_Anxd\_Pct & Anxious/Depressed Gender and Age Adjusted T-score & Psychiatric/Life Function\\
ASR\_Attn\_T & Attention Problems Gender and Age Adjusted T-score & Psychiatric/Life Function\\
ASR\_Extn\_T & Externalizing Gender and Age Adjusted T-score & Psychiatric/Life Function\\
ASR\_Intn\_T & Internalizing Gender and Age Adjusted T-score & Psychiatric/Life Function\\
ASR\_Intr\_T & Intrusive Gender and Age Adjusted T-score & Psychiatric/Life Function\\
ASR\_Rule\_T & Rule Breaking Behavior Gender and Age Adjusted T-score & Psychiatric/Life Function\\
ASR\_Soma\_T & Somatic Complaints Gender and Age Adjusted T-score & Psychiatric/Life Function\\
ASR\_TAO\_Sum & Sum of Thought, Attention, and Other Problems Raw Score & Psychiatric/Life Function\\
ASR\_Thot\_T & Thought Problems Gender and Age Adjusted T-score & Psychiatric/Life Function\\
ASR\_Totp\_T & Total Problems Gender and Age Adjusted T-score & Psychiatric/Life Function\\
ASR\_Witd\_T & Withdrawn Gender and Age Adjusted T-score & Psychiatric/Life Function\\
CardSort\_Unadj* & Dimensional Change Card Sort Test: Unadjusted & Cognition (NIH Toolbox)\\
CogCrystalComp\_AgeAdj & Cognition Crystallized Composite: Age Adjusted & Cognition (NIH Toolbox)\\
CogEarlyComp\_AgeAdj & Cognition Early Childhood Composite: Age Adjusted & Cognition (NIH Toolbox)\\
CogFluidComp\_AgeAdj & Cognition Fluid Composite: Age Adjusted & Cognition (NIH Toolbox)\\
CogTotalComp\_AgeAdj & Cognition Total Composite Score: Age Adjusted & Cognition (NIH Toolbox)\\
DDisc\_AUC\_200 & Area Under the Curve for Delay Discounting of \$200 & Cognition\\
DDisc\_AUC\_40K* & Area Under the Curve for Delay Discounting of \$40,000 & Cognition\\
Dexterity\_Unadj* & 9-hole Pegboard Dexterity Test: Unadjusted & Motor (NIH Toolbox)\\
Emotion\_Task\_Face\_Acc* & Accuracy \% during FACE blocks in EMOTION task & In-Scanner Task\\
EmotSupp\_Unadj* & Emotional Support Survey: Unadjusted & Emotion (NIH Toolbox)\\
Endurance\_Unadj* & 2-minute Walk Endurance Test: Unadjusted & Motor (NIH Toolbox)\\
ER40ANG* & Number of Correct Anger Identifications & Emotion (Penn ER Test)\\
ER40FEAR* & Number of Correct Fear Identifications & Emotion (Penn ER Test)\\
ER40NOE* & Number of Correct Neutral Identifications & Emotion (Penn ER Test)\\
ER40SAD* & Number of Correct Sad Identifications & Emotion (Penn ER Test)\\
ER40\_CR* & Number of Correct Responses & Emotion (Penn ER Test)\\
ER40\_CRT & Correct Responses Median Response Time (ms) & Emotion (Penn ER Test)\\
FearAffect\_Unadj* & Fear-Affect Survey: Unadjusted & Emotion (NIH Toolbox)\\
FearSomat\_Unadj* & Fear-Somatic Arousal Survey: Unadjusted & Emotion (NIH Toolbox)\\
Flanker\_Unadj* & Flanker Test: Unadjusted Scale Score & Cognition (NIH Toolbox)\\
Friendship\_Unadj* & Friendship Survey: Unadjusted & Emotion (NIH Toolbox)\\
GaitSpeed\_Comp* & 4-Meter Walk Gait Speed Test: Computed Score & Motor (NIH Toolbox)\\
InstruSupp\_Unadj* & Instrumental Support Survey: Unadjusted & Emotion (NIH Toolbox)\\
IWRD\_TOT* & Word Memory: Total Number of Correct Responses & Cognition (Penn Tests)\\
Language\_Task\_Math\_Avg & Language Task MATH difficulty level* & Language\\
Language\_Task\_Story\_Avg & Language Task STORY difficulty level* & Language\\
LifeSatisf\_Unadj* & General Life Satisfaction Survey: Unadjusted & Emotion (NIH Toolbox)\\
ListSort\_Unadj* & List Sorting Working Memory Test: Unadjusted & Cognition (NIH Toolbox)\\
Loneliness\_Unadj* & Loneliness Survey: Unadjusted & Emotion (NIH Toolbox)\\
Mars\_Final* & Mars Final Contrast Sensitivity Score & Sensory\\
MeanPurp\_Unadj* & Meaning and Purpose Survey\: Unadjusted & Emotion (NIH Toolbox)\\
MMSE\_Score* & Mini Mental Status Exam Total Score & Alertness\\
NEOFAC\_A* & Agreeableness & Personality (NEO-FFI)\\
NEOFAC\_C* & Conscientiousness & Personality (NEO-FFI)\\
NEOFAC\_E* & Extraversion & Personality (NEO-FFI)\\
NEOFAC\_N* & Neuroticism & Personality (NEO-FFI)\\
NEOFAC\_O* & Openness to Experience & Personality (NEO-FFI)\\
Odor\_Unadj* & Odor Identification Age 3+ Unadjusted & Sensory (NIH Toolbox)\\
PainInterf\_Tscore* & Pain Interference Survey Age 18+: T-score & Sensory (NIH Toolbox)\\
PercHostil\_Unadj* & Perceived Hostility Survey: Unadjusted & Emotion (NIH Toolbox)\\
PercReject\_Unadj* & Perceived Rejection Survey: Unadjusted & Emotion (NIH Toolbox)\\
PercStress\_Unadj* & Perceived Stress Survey: Unadjusted & Emotion (NIH Toolbox)\\
PicSeq\_AgeAdj & Picture Sequence Memory Test: Age-Adjusted & Cognition (NIH Toolbox)\\
PicVocab\_AgeAdj & Picture Vocabulary Test: Age-Adjusted & Cognition (NIH Toolbox)\\
PicVocab\_Unadj* & Picture Vocabulary Test: Unadjusted & Cognition (NIH Toolbox)\\
PMAT24\_A\_CR* & Progressive Matrices\: Number of Correct Responses & Cognition (Penn Tests)\\
PosAffect\_Unadj* & Positive Affect Survey: Unadjusted & Emotion (NIH Toolbox)\\
ProcSpeed\_AgeAdj & Pattern Comparison Processing Speed: Age-Adjusted & Cognition (NIH Toolbox)\\
PSQI\_Score* & Pittsburgh Sleep Questionnaire Total Score & Alertness\\
ReadEng\_AgeAdj & Oral Reading Recognition Test: Age Adjusted & Cognition (NIH Toolbox)\\
ReadEng\_Unadj* & Oral Reading Recognition Test: Unadjusted & Cognition (NIH Toolbox)\\
Relational\_Task\_Acc* & Average accuracy \% during RELATIONAL task & In-Scanner Task\\
Sadness\_Unadj* & Sadness Survey: Unadjusted & Emotion (NIH Toolbox)\\
SCPT\_SEN* & Short Continuous Performance Test: Sensitivity & Cognition (Penn Tests)\\
SCPT\_SPEC* & Short Continuous Performance Test: Specificity & Cognition (Penn Tests)\\
SelfEff\_Unadj* & Self-Efficacy Survey: Unadjusted & Emotion (NIH Toolbox)\\
Social\_Task\_Perc\_Random* & Overall \% of stimuli that the subject rated as random & In-Scanner Task\\
Social\_Task\_Perc\_TOM* & Overall \% of stimuli that received a social rating & In-Scanner Task\\
Strength\_Unadj* & Grip Strength Test: Unadjusted & Motor (NIH Toolbox)\\
Taste\_Unadj* & Regional Taste Intensity Age 12+ Unadjusted & Sensory (NIH Toolbox)\\
VSPLOT\_TC* & Variable Short Penn Line Orientation Test: Total Correct & Cognition (Penn Tests)\\
WM\_Task\_Acc* & Accuracy across all conditions in working memory task & In-Scanner Task\\
\bottomrule
\label{tab:phenotypes-hcp}

\end{longtable}

\paragraph{Used phenotypes from the PNC Dataset}
\begin{longtable}{ l l l }
\toprule
\textbf{Variable} & \textbf{Description} & \textbf{Category} \\
\midrule
LNB\_MCR & Number of Correct Responses to for 1-Back and  & Letter N-Back (LNB) \\
& 2-Back Trials & \\
LNB\_MRTC & Mean of the Median Response Time for Correct  & Letter N-Back (LNB)  \\
& Responses for 1-Back (TP)  and for 2-Back (TP) Trials & \\
MP\_MP2RTCR & Median Response Time for Correct Mouse Click Responses & Motor Praxis (MP)  \\
PADT\_A & Total Correct Responses for All Test Trials by genus & Penn Age Differentiation  (PADT)\\
PADT\_PC & Percent Correct Responses for All Test Trials by genus & Penn Age Differentiation  (PADT) \\
PADT\_SAME\_PC & Percent Correct Responses to Test Trials with   & Penn Age Differentiation  (PADT) \\
& No Age Difference, by genus& \\
PADT\_T & Median Response Time for All Test Trials, by genus & Penn Age Differentiation  (PADT)\\
PCET\_ACC2 & Calculated Accuracy Measure & Penn Conditional Exclusion  (PCET)\\
PCET\_CAT & Number of Categories Achieved & Penn Conditional Exclusion  (PCET)\\
PCPT\_T\_FP & Total of Incorrect Responses to Number Trials (TP) & Penn Continuous Performance  (PCPT)\\
&  and Letter Trials (FP)& \\
PCPT\_T\_FPRT & Median Response Time for Incorrect Responses to & Penn Continuous Performance  (PCPT)\\
&  Number Trials (TP) and Letter Trials (TP) & \\
PCPT\_T\_TP & Total of Correct Responses to Number Trials (TP)  & Penn Continuous Performance  (PCPT)\\
& and Letter Trials (TP) & \\
PCPT\_T\_TPRT & Median Response Time for Correct Responses to Number  & Penn Continuous Performance  (PCPT)\\
& Trials (TP) and Letter Trials (TP) & \\
PEDT\_A & Total Correct Responses for All Test Trials, by genus & Penn Emotion Differentiation  (PEDT)\\
PEDT\_ANG\_PC & Percent of Correct Responses for Anger Trials, by genus & Penn Emotion Differentiation  (PEDT)\\
PEDT\_FEAR\_PC & Percent of Correct Responses for Fear Trials, by genus & Penn Emotion Differentiation  (PEDT)\\
PEDT\_HAP\_PC & Percent of Correct Responses for Happy Trials, by genus & Penn Emotion Differentiation  (PEDT)\\
PEDT\_PC & Percent of Correct Responses for All Test Trials, by genus & Penn Emotion Differentiation  (PEDT)\\
PEDT\_SAD\_PC & Percent of Correct Responses for Sad Trials, by genus & Penn Emotion Differentiation  (PEDT)\\
PEDT\_SAME\_PC & Percent Correct Responses to Test Trials with Neutral & Penn Emotion Differentiation  (PEDT)\\
&  Difference, by genus & \\
PEDT\_T & Median Response Time for All Test Trials, by genus & Penn Emotion Differentiation (PEDT)\\
PEIT\_CR & Total Correct Responses for All Test Trials, by genus & Penn Emotion Identification  (PEIT)\\
PEIT\_CRT & Median Response Time for Total Correct Test Trial  & Penn Emotion Identification (PEIT)\\
&Responses, by genus & \\
PFMT\_IFAC\_RTC & Median Response Time for Total Correct Test & Penn Face Memory (PFMT)\\
&  Trial Responses & \\
PFMT\_IFAC\_TOT & Total Correct Responses for All Test Trials & Penn Face Memory (PFMT)\\
PLOT\_PC & Percent Correct Responses for All Test Trials, by genus & Penn Line Orientation (PLOT)\\
PLOT\_TC & Total Correct Responses for All Test Trials, by genus & Penn Line Orientation (PLOT)\\
PLOT\_TCRT & Median Response Time for Correct Trials, by genus & Penn Line Orientation (PLOT)\\
PMAT\_CR & Total Correct Responses for All Test Trials, by genus & Primary Mental Abilities (PMAT)\\
PVRT\_CR & Total Correct Responses for All Test Trials, by genus & Penn Verbal Reasoning (PVRT)\\
PVRT\_RTCR & Median Response Time for Correct Verbal Reasoning& Penn Verbal Reasoning (PVRT)\\
& Responses, by genus& \\
PVRT\_RTER & Median Response Time for Incorrect Verbal Reasoning & Penn Verbal Reasoning (PVRT)\\
& Responses, by genus& \\
PWMT\_KIWRD\_RTC & Median Response Time for Total Correct Test Trial  & Penn Word Memory (PWMT)\\
& Responses & \\
PWMT\_KIWRD\_TOT & Total Correct Responses for All Test Trials & Penn Word Memory (PWMT)\\
VOLT\_SVT & Total Correct Responses for All Test Trial & Visual Object Learning (VOLT)\\
VOLT\_SVTCRT & Median Response Time for Total Correct Responses  & Visual Object Learning (VOLT)\\
VOLT\_SVTIRT & Median Response Time for Incorrect Responses  & Visual Object Learning (VOLT)\\
WRAT\_CR\_RAW & Total Raw Score & Wide Range Achievement (WRAT)\\
WRAT\_CR\_STD & Total Standard Score & Wide Range Achievement (WRAT)\\
\bottomrule
\end{longtable}
\label{tab:phenotypes-pnc}
\restoregeometry

\subsection{$r^2$ Results}

\begin{figure}[h!]
    \centering
    \includegraphics[width=\linewidth]{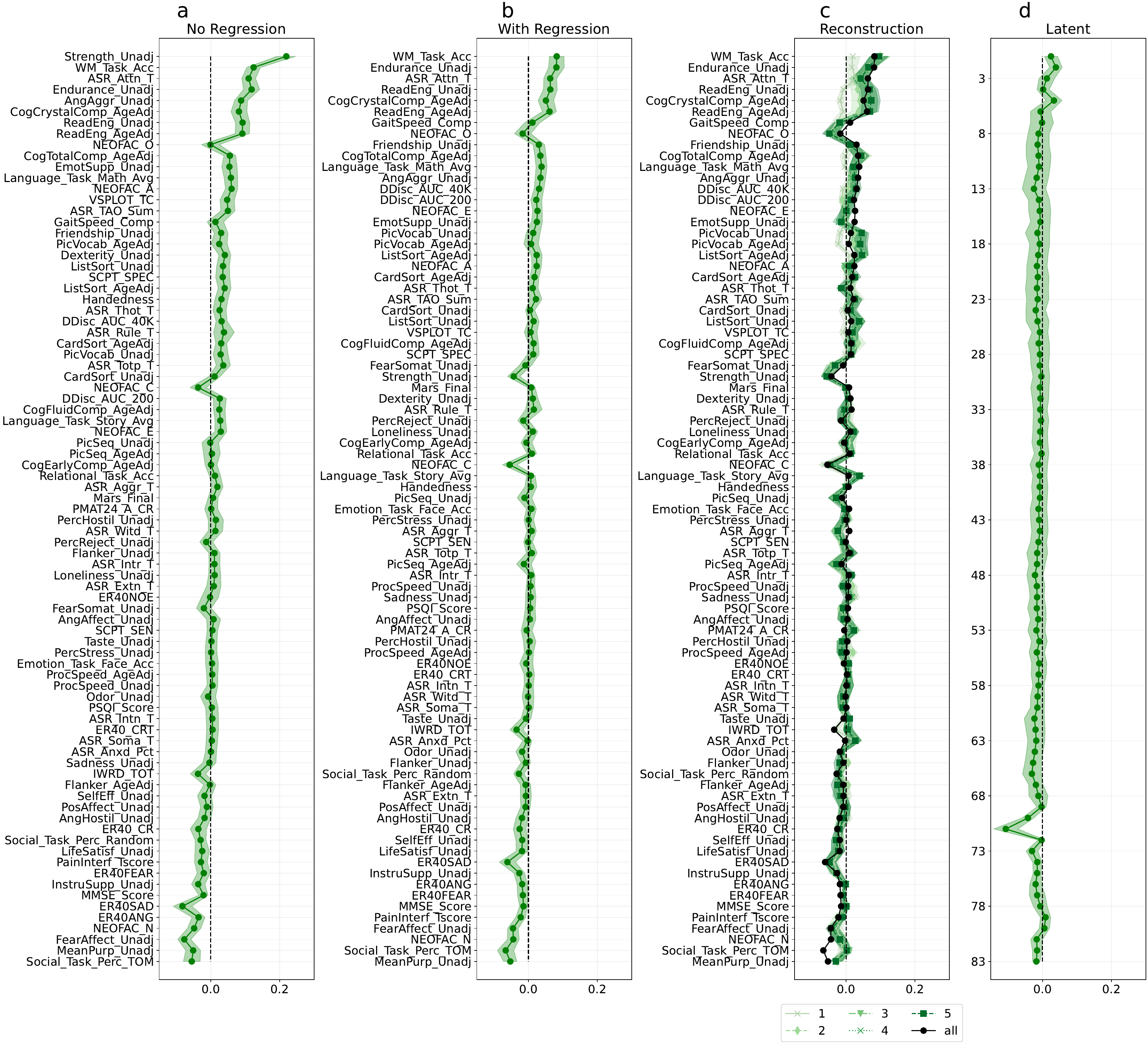}
\caption{Predictability of phenotype variables in the validation set, quantified by
the coefficient of determination ($r^2$, rather than by correlation between predicted and true values as in Figure~\ref{fig:all_hcp}) between predictions and true values, for HCP before regressing the covariates (a), after regression (b), reconstruction (c), and of the latent phenotypes (d). The shaded regions represent the standard deviation across
resamplings.}
\label{fig:all_hcp_r2}
\end{figure}

\begin{figure}
    \centering
    \includegraphics[width=\linewidth]{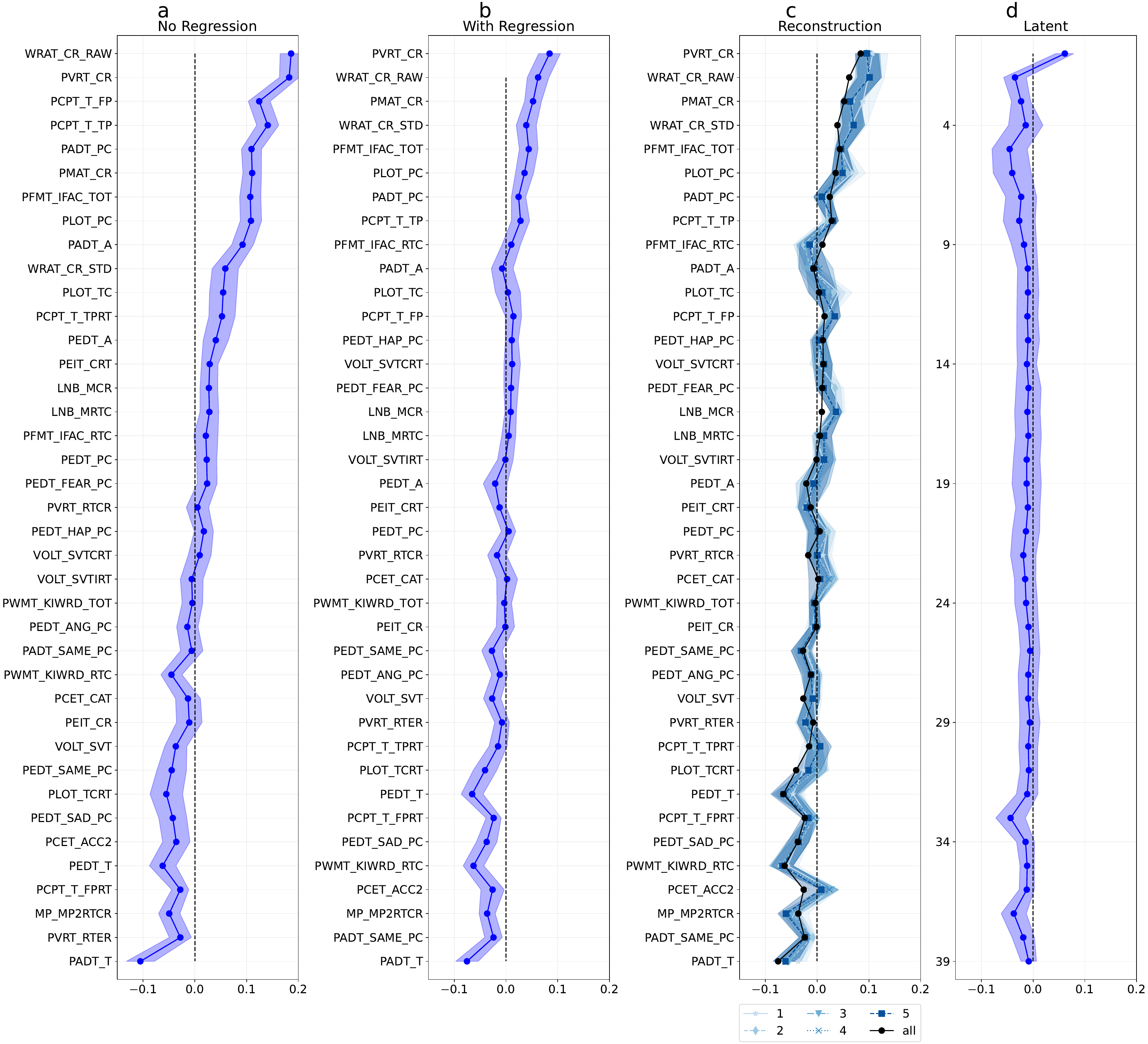}
\caption{Predictability of phenotype variables in the validation set, quantified by
the coefficient of determination ($r^2$ by correlation between predicted and true values as in Figure~\ref{fig:all_pnc})  between predictions and true values, for PNC before regressing the covariates (a), after regression (b), reconstruction (c), and of the latent phenotypes (d). The shaded regions represent the standard deviation across
resamplings.}
\label{fig:all_pnc_r2}
\end{figure}

\subsection{Beta-coefficients for the linear regression}
\begin{figure}
\includegraphics[width=\linewidth]{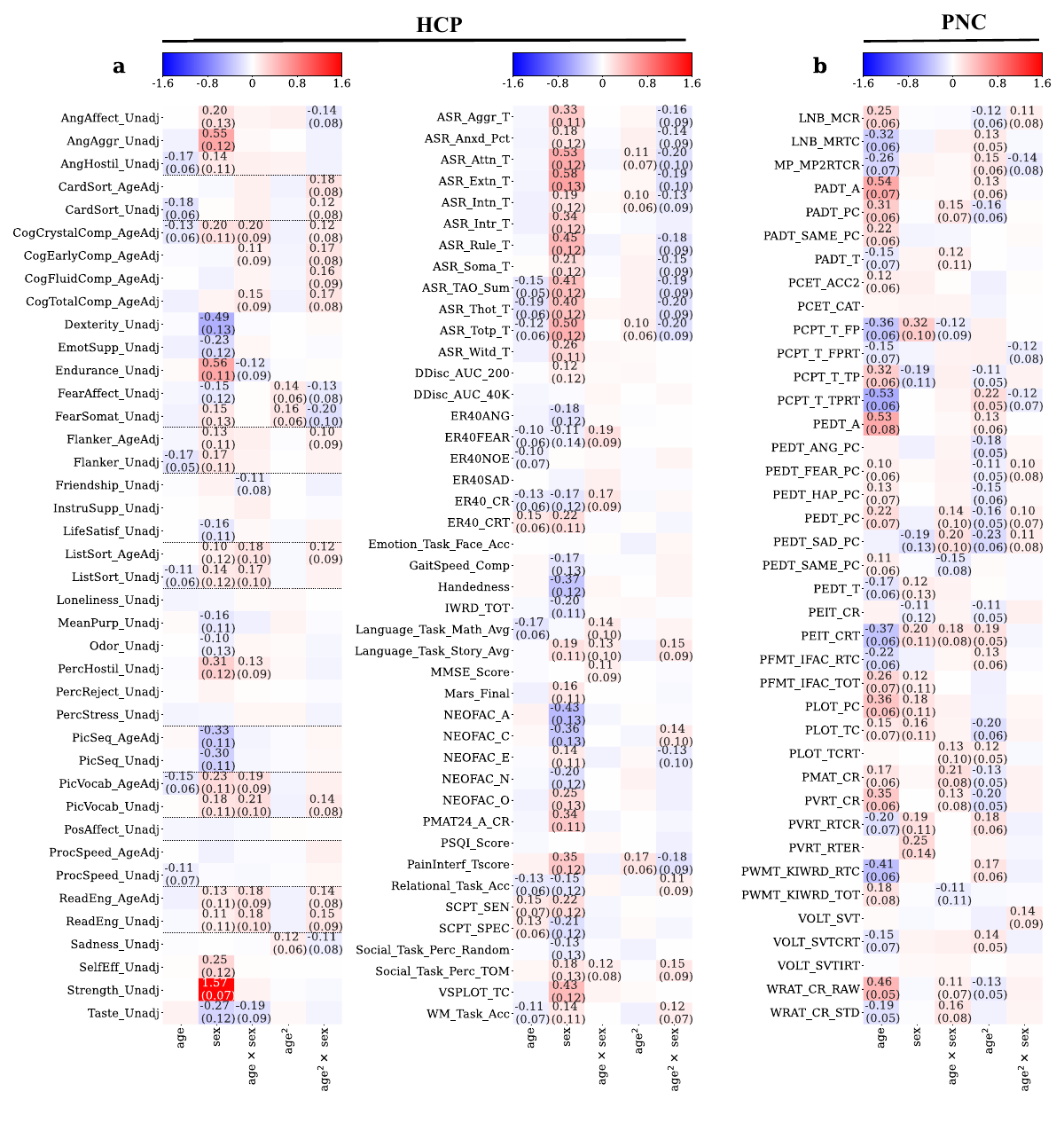}
\caption{Beta-weights of the linear regression used to regress out age, sex,
and their quadratic interactions from HCP (\textbf{a}) PNC (\textbf{b}).
(\textbf{a}) Dotted lines include pairs of phenotypes that were both Unadjusted
and Age-adjusted (\emph{AgeAdj}). These results are the same as those reported in Figure~\ref{fig:beta-coefficients}
but values, where the mean is above $\pm$ 0.1, are annotated in the plot. Each annotated entry corresponds to the mean (std) over the 100 repetitions. } 
\label{fig:beta-coefficients-annotated}
\end{figure}

\newpage
\subsection{How many latent phenotypes do we need to have an accurate prediction?}

\begin{figure}
\centering
\includegraphics[width=.8\linewidth]{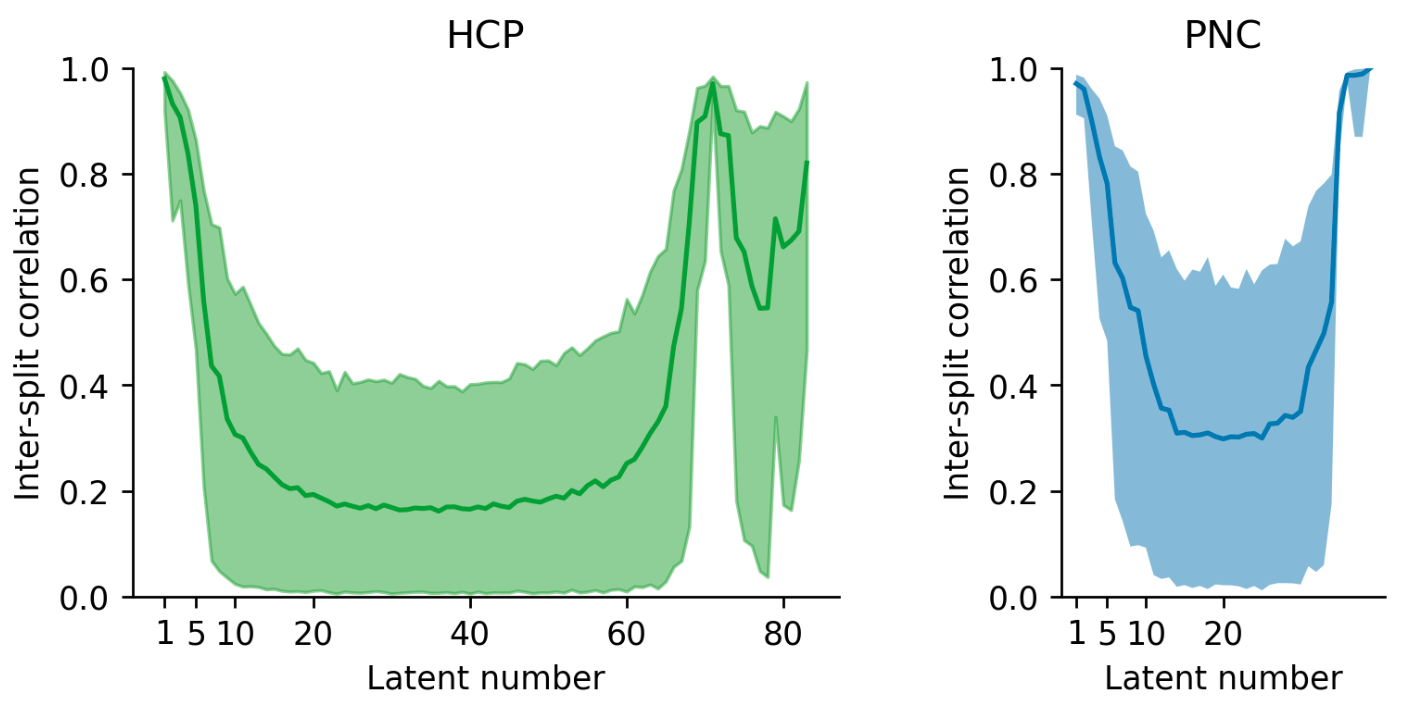}
\caption{Marriage results using the inter-split correlation. In contrast to the plot in the main text, where we report the $r^2$, here we report the inter-split correlation.}
\label{fig:smudge_plot_all}
\end{figure}

\newpage
\section{Pre-registration of replication on HBN dataset}
\input{hbn_preprint}

\end{document}

%% file: fmriprep_boiler.tex
Results included in this manuscript come from preprocessing performed
using \emph{fMRIPrep} 21.0.2 (\cite{fmriprep1}; \cite{fmriprep2};
RRID:SCR\_016216), which is based on \emph{Nipype} 1.6.1
(\citet{nipype1}; \citet{nipype2}; RRID:SCR\_002502).

\begin{description}
\item[Preprocessing of B0 inhomogeneity mappings]
A total of 1 fieldmaps were found available within the input BIDS
structure. A \emph{B0} nonuniformity map (or
\emph{fieldmap}) was estimated from the phase-drift map(s) measure with
two consecutive GRE (gradient-recalled echo) acquisitions. The
corresponding phase-map(s) were phase-unwrapped with \texttt{prelude}
(FSL 6.0.5.1:57b01774).
A total of 1 T1-weighted (T1w) images were found within the input BIDS
dataset. The T1-weighted (T1w) image was corrected for intensity
non-uniformity (INU) with \texttt{N4BiasFieldCorrection} \citep{n4},
distributed with ANTs 2.3.3 \citep[RRID:SCR\_004757]{ants}, and used as
T1w-reference throughout the workflow. The T1w-reference was then
skull-stripped with a \emph{Nipype} implementation of the
\texttt{antsBrainExtraction.sh} workflow (from ANTs), using OASIS30ANTs
as target template. Brain tissue segmentation of cerebrospinal fluid
(CSF), white-matter (WM) and gray-matter (GM) was performed on the
brain-extracted T1w using \texttt{fast} \cite{fsl_fast}. Brain surfaces were reconstructed using
\texttt{recon-all} \citep[FreeSurfer 6.0.1,
RRID:SCR\_001847,][]{fs_reconall}, and the brain mask estimated
previously was refined with a custom variation of the method to
reconcile ANTs-derived and FreeSurfer-derived segmentations of the
cortical gray-matter of Mindboggle
\citep[RRID:SCR\_002438,][]{mindboggle}. Volume-based spatial
normalization to two standard spaces (MNI152NLin2009cAsym,
MNI152NLin6Asym) was performed through nonlinear registration with
\texttt{antsRegistration} (ANTs 2.3.3), using brain-extracted versions
of both T1w reference and the T1w template. The following templates were
selected for spatial normalization: \emph{ICBM 152 Nonlinear
Asymmetrical template version 2009c} {[}\citet{mni152nlin2009casym},
RRID:SCR\_008796; TemplateFlow ID: MNI152NLin2009cAsym{]}, \emph{FSL's
MNI ICBM 152 non-linear 6th Generation Asymmetric Average Brain
Stereotaxic Registration Model} {[}\citet{mni152nlin6asym},
RRID:SCR\_002823; TemplateFlow ID: MNI152NLin6Asym{]}.
\item[Functional data preprocessing]
For each of the 3 BOLD runs found per subject (across all tasks and
sessions), the following preprocessing was performed. First, a reference
volume and its skull-stripped version were generated using a custom
methodology of \emph{fMRIPrep}. Head-motion parameters with respect to
the BOLD reference (transformation matrices, and six corresponding
rotation and translation parameters) are estimated before any
spatiotemporal filtering using \texttt{mcflirt} \citep[FSL
6.0.5.1:57b01774,][]{mcflirt}. BOLD runs were slice-time corrected to
1.47s (0.5 of slice acquisition range 0s-2.94s) using \texttt{3dTshift}
from AFNI \citep[RRID:SCR\_005927]{afni}. The BOLD time-series
(including slice-timing correction when applied) were resampled onto
their original, native space by applying the transforms to correct for
head-motion. These resampled BOLD time-series will be referred to as
\emph{preprocessed BOLD in original space}, or just \emph{preprocessed
BOLD}. The BOLD reference was then co-registered to the T1w reference
using \texttt{bbregister} (FreeSurfer) which implements boundary-based
registration \citep{bbr}. Co-registration was configured with six
degrees of freedom. Several confounding time-series were calculated
based on the \emph{preprocessed BOLD}: framewise displacement (FD),
DVARS and three region-wise global signals. FD was computed using two
formulations following Power (absolute sum of relative motions,
\citet{power_fd_dvars}) and Jenkinson (relative root mean square
displacement between affines, \citet{mcflirt}). FD and DVARS are
calculated for each functional run, both using their implementations in
\emph{Nipype} \citep[following the definitions by][]{power_fd_dvars}.
The three global signals are extracted within the CSF, the WM, and the
whole-brain masks. Additionally, a set of physiological regressors were
extracted to allow for component-based noise correction
\citep[\emph{CompCor},][]{compcor}. Principal components are estimated
after high-pass filtering the \emph{preprocessed BOLD} time-series
(using a discrete cosine filter with 128s cut-off) for the two
\emph{CompCor} variants: temporal (tCompCor) and anatomical (aCompCor).
tCompCor components are then calculated from the top 2\% variable voxels
within the brain mask. For aCompCor, three probabilistic masks (CSF, WM
and combined CSF+WM) are generated in anatomical space. The
implementation differs from that of Behzadi et al.~in that instead of
eroding the masks by 2 pixels on BOLD space, the aCompCor masks are
subtracted a mask of pixels that likely contain a volume fraction of GM.
This mask is obtained by dilating a GM mask extracted from the
FreeSurfer's \emph{aseg} segmentation, and it ensures components are not
extracted from voxels containing a minimal fraction of GM. Finally,
these masks are resampled into BOLD space and binarized by thresholding
at 0.99 (as in the original implementation). Components are also
calculated separately within the WM and CSF masks. For each CompCor
decomposition, the \emph{k} components with the largest singular values
are retained, such that the retained components' time series are
sufficient to explain 50 percent of variance across the nuisance mask
(CSF, WM, combined, or temporal). The remaining components are dropped
from consideration. The head-motion estimates calculated in the
correction step were also placed within the corresponding confounds
file. The confound time series derived from head motion estimates and
global signals were expanded with the inclusion of temporal derivatives
and quadratic terms for each \citep{confounds_satterthwaite_2013}.
Frames that exceeded a threshold of 0.5 mm FD or 1.5 standardised DVARS
were annotated as motion outliers. The BOLD time-series were resampled
into standard space, generating a \emph{preprocessed BOLD run in
MNI152NLin2009cAsym space}. First, a reference volume and its
skull-stripped version were generated using a custom methodology of
\emph{fMRIPrep}. The BOLD time-series were resampled onto the following
surfaces (FreeSurfer reconstruction nomenclature): \emph{fsaverage6},
\emph{fsaverage}. Automatic removal of motion artifacts using
independent component analysis \citep[ICA-AROMA,][]{aroma} was performed
on the \emph{preprocessed BOLD on MNI space} time-series after removal
of non-steady state volumes and spatial smoothing with an isotropic,
Gaussian kernel of 6mm FWHM (full-width half-maximum). Corresponding
``non-aggresively'' denoised runs were produced after such smoothing.
Additionally, the ``aggressive'' noise-regressors were collected and
placed in the corresponding confounds file. \emph{Grayordinates} files
\citep{hcppipelines} containing 91k samples were also generated using
the highest-resolution \texttt{fsaverage} as intermediate standardized
surface space. All resamplings can be performed with \emph{a single
interpolation step} by composing all the pertinent transformations
(i.e.~head-motion transform matrices, susceptibility distortion
correction when available, and co-registrations to anatomical and output
spaces). Gridded (volumetric) resamplings were performed using
\texttt{antsApplyTransforms} (ANTs), configured with Lanczos
interpolation to minimize the smoothing effects of other kernels
\citep{lanczos}. Non-gridded (surface) resamplings were performed using
\texttt{mri\_vol2surf} (FreeSurfer).

\end{description}

Many internal operations of \emph{fMRIPrep} use \emph{Nilearn} 0.8.1
\citep[RRID:SCR\_001362]{nilearn}, mostly within the functional
processing workflow. For more details of the pipeline, see
\href{https://fmriprep.readthedocs.io/en/latest/workflows.html}{the
section corresponding to workflows in \emph{fMRIPrep}'s documentation}.

%% file: hbn_preprint.tex
\subsection{Reliability and predictability of phenotype information from functional connectivity on the Healthy Brain Network (HBN) dataset}
This section follows the guidelines from OSF for a secondary data analysis template, which was inspired by \cite{Van-den-Akker-2021Preregistration}. Where necessary to improve comprehension, the question from the template form is copied into this for clarity in \emph{italic}.
 
\subsection{Description}
Neuroimaging research aims to unravel the intricate relationship between patterns of functional connectivity across the entire brain and various behavioral phenotypes. The pre-registration described in this appendix seeks to build, extend, and evaluate the generalizability of our analyses described in the associated main document 
(see sections \ref{sec:methods} through \ref{sec:discuss}), 
extend our analysis, and evaluate the generalizability of our results to a new dataset, specifically the Healthy Brain Network (HBN). In other words, the work described in the main part of this document was exploratory, and with the analyses that we will carry out with the HBN dataset, we aim to confirm our findings on an independent dataset. Similar to the exploratory work described above, our main objective is to train predictive models for phenotypes and compare them to the prediction of latent phenotypes (obtained from a singular value decomposition (SVD) of the phenotypes). In the associated preprint we evaluated: (i) the predictability of phenotypes and latent phenotypes, (ii) the reliability of the latent phenotypes, and (iii) the impact of controlling for age/sex confounds in the prediction. With this new study, we want to evaluate if the previous findings generalize to an entirely new dataset acquired from different scanners and with different phenotypes.

\subsection{Contributors}
Jessica Dafflon, Dustin Moraczewski, Eric Earl, Dylan M. Nielson, Gabriel Loewinger, Patrick McClure, Adam G. Thomas, and Francisco Pereira
(See \ref{sec:title} for details)

\subsection{License}
CC0 1.0 Universal

\subsection{Subjects}
Medicine and Health Sciences, Psychiatry and Psychology
Computer Science, Life Sciences, Machine Learning

\subsection{Study Information}
\subsubsection{Research Questions}
\emph{
List each research question included in this study. When specifying your research questions, it is good practice to use only two new concepts per research question. For example, split up your questions into a simple format: `Does X lead to Y?' and `Is the relationship between X and Y moderated by Z?'. By splitting up the research questions here, you can more easily describe the statistical test for each research question later.}

\begin{enumerate}
    \item How does controlling for age/sex confounds impact the prediction of phenotypes from functional connectivity in HBN?
    \item How reliable are the latent phenotypes estimated with SVD across different sub-samples of the HBN participants?
    \item Are latent phenotypes obtained by applying linear transformation (SVD)  to phenotypes more predictable from functional connectivity than the original phenotypes on the HBN dataset?
    \item Do models predicting phenotype variables reconstructed from the latent SVD variables with functional connectivity perform as well as models predicting the original phenotype variables with functional connectivity in the HBN data?
\end{enumerate}

\subsubsection{Hypotheses}
\emph{For each of the research questions listed in the previous section, provide one or more specific and testable hypothesis. Please make clear whether the hypotheses are directional (e.g., A $>$ B) or non-directional (e.g., A $\neq$ B). If directional, state the direction. You may also provide a rationale for each hypothesis.}
\begin{enumerate}
    \item \textbf{How does controlling for age/sex confounds impact the phenotypes prediction?} \\
    - Hypothesis: In accordance with the literature \citep{Chyzhyk-2022How} and the results we obtained on our exploratory analysis (Figure~\ref{fig:beta-coefficients}), failure to remove age and sex confounds will lead to a higher correlation coefficient for variables that are strongly correlated with age/sex. Accordingly, we expect that predictive performance will be lower after regressing age/sex out of the phenotypes. Though our expectation is directional, we will test this with a two-sided paired t-test on the predictive performance with and without regressing out age and sex. \\
    - Rationale: Information from sex and age can drive the predictability of specific phenotypes. If not regressed out, those confounds can inflate the predictability of some phenotypes. This is particularly visible in the variable `Strength Unadjusted' from our exploratory analysis of the HCP dataset. If sex is not regressed out, it is the most predictable variable (Figure~\ref{fig:all_hcp}a), but most of its predictability is given by the sex information. 
    \item \textbf{How reliable are the latent phenotypes estimated with SVD across different samples?} \\
    - Hypothesis: Only the first five latent phenotypes (in order of explained variance) will be reliable across different samples, with subsequent latent phenotypes showing decreased reliability. For a complete description of how our reliability metric was computed, see section \ref{sec:gale-algo}. We specifically define a reliable component as having an empirical lower 95\% confidence interval on the $r^2$ greater than 0.2. The empirical confidence interval will be calculated from 1,000 random splits of the data. \\
    - Rationale: Our previous analyses of the HCP and PNC datasets have demonstrated that while the first few latent phenotypes are reliably identified across different samples, the reliability diminishes for subsequent latent phenotypes. We anticipate a similar pattern in the HBN dataset, where the initial latent phenotypes should be reliably identified across different dataset splits. 
    \item \textbf{Can we obtain more predictable phenotypes if we use a linear transformation to latent phenotypes (SVD) on HBN?} \\
    - Hypothesis: Applying a linear dimensionality reduction (SVD) to obtain latent phenotypes on the HBN dataset will not decrease the predictability compared to directly using phenotypes. We will evaluate their predictability by comparing the correlation of the prediction for the untransformed phenotypes and those transformed using SVD with a two-sided paired t-test. Again our expectation is directional but we are using a two-sided case because it will still be of interest if either the latent or the original phenotypes have significantly greater performance. \\
    - Rationale: Given our exploratory analysis with HCP and PNC where we did not observe an increase in predictability by using latent phenotypes, in comparison with the original phenotype variables, we do not expect to see that the former will be more predictable. 
    \item \textbf{Are phenotype variables reconstructed from the SVD latent variables as informative for training a model as the original phenotype variables?} \\
    - Hypothesis: The reconstruction of phenotypes using a subset of latent phenotypes, particularly the first few components, will exhibit similar or enhanced predictability compared to using all latent phenotypes. To evaluate if there is any statistical difference in the performance of the prediction algorithms on using all of the components or a subset thereof, we will use the autorank library \citep{herbold2020autorank, demvsar2006statistical}. See sections \ref{sec:critical-diff} and \ref{sec:prediction-recons} for a more detailed description. \\
    - Rationale: Linear transformations such as SVD aim to capture the underlying structure within complex datasets by creating uncorrelated variables (latent phenotypes) that explain as much variance as possible. Conversely, the original data can be reconstructed from the most informative latent phenotypes, retaining essential information while discarding redundant or noisy signals. If used to denoise the phenotype variables in the training set of prediction models, the denoising can potentially lead to improved predictive performance. Therefore, we expect that prediction performance of models trained on reconstructed phenotypes will not decrease and may even surpass that of models trained using the original phenotype variables. 
    
\end{enumerate}

\subsection{Data Description}
\subsubsection{Datasets used}
\label{sec:pre-print-dataset}
\emph{Name and briefly describe the dataset(s), and if applicable, the subsets of the data you plan to use. Useful information to include here is the type of data (e.g., cross-sectional or longitudinal), the general content of the questions, and some details about the respondents. In the case of longitudinal data, information about the survey’s waves is useful as well. Mention the most relevant information so that readers do not have to search for the information themselves.} \\
In the first part of this manuscript, we conducted an exploratory analysis using the HCP \citep{Van-Essen:2012aa} and PNC dataset \citep{Satterthwaite-2016The}, for our confirmatory analysis, described in this pre-registration, we will use the Healthy Brain Network (HBN) dataset \citep{Alexander-2017An}. While the HCP dataset includes young adults aged between 22 and 35, the PNC dataset includes children aged 8 to 21. Both PNC and HBN are neurodevelopmental cohorts acquired in the Philadelphia and New York metropolitan areas, respectively. The HBN dataset contains neuroimaging and phenotypic data gathered across a diverse population of children and adolescents (ages 5-21). The dataset includes, among other modalities, functional MRI, diffusion MRI and electroencephalography collected longitudinally. It also includes extensive behavioral assessments, cognitive tests and psychiatric diagnostics. For this analysis, we will use the resting state MRI images and cognitive phenotypes, and only used cross-sectional information. The HBN data was acquired in 3 sites, using different scanner manufacturers, a number of time points and length of scans; an overview of the MRI protocol can be found via the \href{https://fcon_1000.projects.nitrc.org/indi/cmi_healthy_brain_network/MRI%20Protocol.html}{NITRC website} \citep{Kennedy-2016The}.

\subsubsection{Data availability}
\emph{Specify the degree to which the datasets are open or publicly available.} \\
Images and phenotypic description are publicly available and can be downloaded \href{https://fcon_1000.projects.nitrc.org/indi/cmi_healthy_brain_network/sharing_neuro.html#Data%20License}{here}. To access the phenotypical information, however,  a Data Use Agreement (DUA) needs to be filled out. 
\subsubsection{Data access}
\emph{If there are any restrictions to accessing the dataset, please describe this here.}
A Data Use Agreement (DUA) is required to obtain the phenotypic data. Details of the DUA are available via the \href{https://fcon_1000.projects.nitrc.org/indi/cmi_healthy_brain_network/Pheno_Access.html}{NITRC website} \citep{Kennedy-2016The}.

\subsubsection{Data identifiers}
\emph{Please provide a URL, DOI, or other persistent, unique identifier of the dataset.} \\
We will use Release 10.0 of the HBN dataset, which was released on April 13, 2022, and is the most recent release as of this writing. The \href{https://fcon_1000.projects.nitrc.org/indi/cmi_healthy_brain_network/Release%20information.html}{full release history} is available via the NITRC website as is the 
\href{http://www.nitrc.org/account/login.php?return_to=http://fcon_1000.projects.nitrc.org/indi/cmi_healthy_brain_network/downloads/downloads_MRI_R10.html}{Neuroimaging} and \href{https://fcon_1000.projects.nitrc.org/indi/cmi_healthy_brain_network/File/_pheno/HBN_R10_Pheno.csv}{Phenotypic} data. The linked phenotypic file provides a list of subjects in the HBN dataset and a flag if the full phenotype is available. To obtain the full phenotypic data, a Data Use Agreement (DUA) is required. The DOI of the dataset descriptor is 10.1038/sdata.2017.181 \citep{Alexander-2017An}

\subsubsection{Access date}
\emph{Specify the download or data access date. If the data were accessed multiple times by different team members, specify the download date for that data that will be used in the statistical analysis.}
We download Release 10.0 of the HBN dataset in July of 2022. Release 10.0 is the most recent release as of this writing (April 2024). 

\subsubsection{Data collection procedures}
\emph{If the data collection procedure is well documented, provide a link to that information. If the data collection procedure is not well documented, describe, to the best of your ability, how data were collected. Describe the representativeness of the sample and any possible biases stemming from the data collection.} \\
The data collection for the HBN dataset is described in the \href{https://fcon_1000.projects.nitrc.org/indi/cmi_healthy_brain_network/Citation.html}{designated dataset descriptor}, \citep{Alexander-2017An}. 

\subsubsection{Codebook}
\emph{Some studies offer codebooks to describe their data. If such a codebook is publicly available, link, cite, or upload the document. If not, provide other available documentation. Also provide guidance on what parts of the codebook or other documentation are most relevant.} \\
In addition to the original paper describing the dataset, the dataset is further described on its \href{https://fcon_1000.projects.nitrc.org/indi/cmi_healthy_brain_network/About.html}{entry} on the NITRC website. A data dictionary for the phenotypes can also be downloaded from \href{https://coins.trendscenter.org/}{COINS}. 

\subsection{Variables}
\subsubsection{Measured variables}
\emph{Describe both outcome measures as well as predictors and covariates and label them accordingly. If you are using a scale or an index, state the construct the scale/index represents, which items the scale/index will consist of, and how these items will be aggregated. When the aggregation is based on exploratory factor analysis (EFA) or confirmatory factor analysis (CFA), also specify the relevant details (EFA: rotation, how the number of factors will be determined, how best fit will be selected, CFA: how loadings will be specified, how fit will be assessed, which residuals variance terms will be correlated). If you are using any categorical variables, state how you will code them in the statistical analyses.} \\
We will use a subset of the Healthy Brain Network (HBN) dataset, comprising 895 subjects. The dataset consists of a training set with 672 individuals (386 males and 286 females) with an average age of $12.40 \pm 3.60$  years, and a separate test set containing 105 subjects (66 males and 39 females) with an average age of $12.32 \pm 3.60$ years, and a hold-out set containing 114 subjects (67 males and 47 females) with an average age of $12.52 \pm 3.76$ years. 
Subjects were assigned to the training, test, and hold-out sets in a stratified sampling procedure. Subjects were sorted by age and sex at birth, and randomly within each sex. They were then assigned in a proportional round-robin fashion to training, test, and hold-out datasets. Although the HBN dataset contains 11 diagnostic labels, we opted not to use this information for splitting the dataset, as some conditions are represented by very few subjects. Additionally, incorporating extra criteria for comorbidity would have been necessary, complicating the dataset division process. 
Also, note that according to the pre-processing notation we adopted for all datasets, we encoded female as 1 and male as 2; however, when regressing out the covariates, we encoded male as 1 and female as 0. A list with the subject IDs for the train, test, and hold-out dataset will be provided on the gitrepo together with the code for the analyses (\href{https://github.com/JessyD/brain-phenotypes/commit/8d9c075d7165c70d7bbd89718f18b1c91a358621}{https://github.com/JessyD/brain-phenotypes}) and on the \href{https://osf.io/7hr9w/}{OSF description of this project}.

\paragraph{Imaging data}
We have already preprocessed the HBN imaging data with the same preprocessing described in our preprint for the PNC dataset. The preprocessing of the HBN dataset was conducted using \href{https://fmriprep.org/en/stable/}{fmriprep} version 21.0.2. Motion censoring was performed at the individual time point level, with points exceeding 0.2 mm frame-wise displacement or a derivative root mean square (DVARS) above 75 marked for censoring. Intervals of less than five points between censor points were also flagged for censoring. The six estimated head-motion parameters, their derivatives, and average signals within anatomically-derived white matter and cerebrospinal fluid masks obtained from fmriprep, were regressed from the functional magnetic resonance imaging (fMRI) signal prior to functional connectivity (FC) computation. We generated functional connectivity matrices by calculating Pearson correlations among the average time series of each brain region pair. The regions were delineated using the Schaefer 400 parcellation, which consists of 17 networks \citep{Schaefer-2018Local-global}. If any of the runs had more than 50\% of the runs flagged as time points to be censored, we do not use that run for computing the FC. Some subjects from the HBN dataset contain more than one resting state runs, if more than one run is
available for a subject, we will use an average of the computed functional connectivity. For a complete description of how the functional connectivity matrix will be computed, please refer to section~\ref{sec:fc-matrix}.

\begin{figure}[h]
  \centering
  \includegraphics[width=\textwidth]{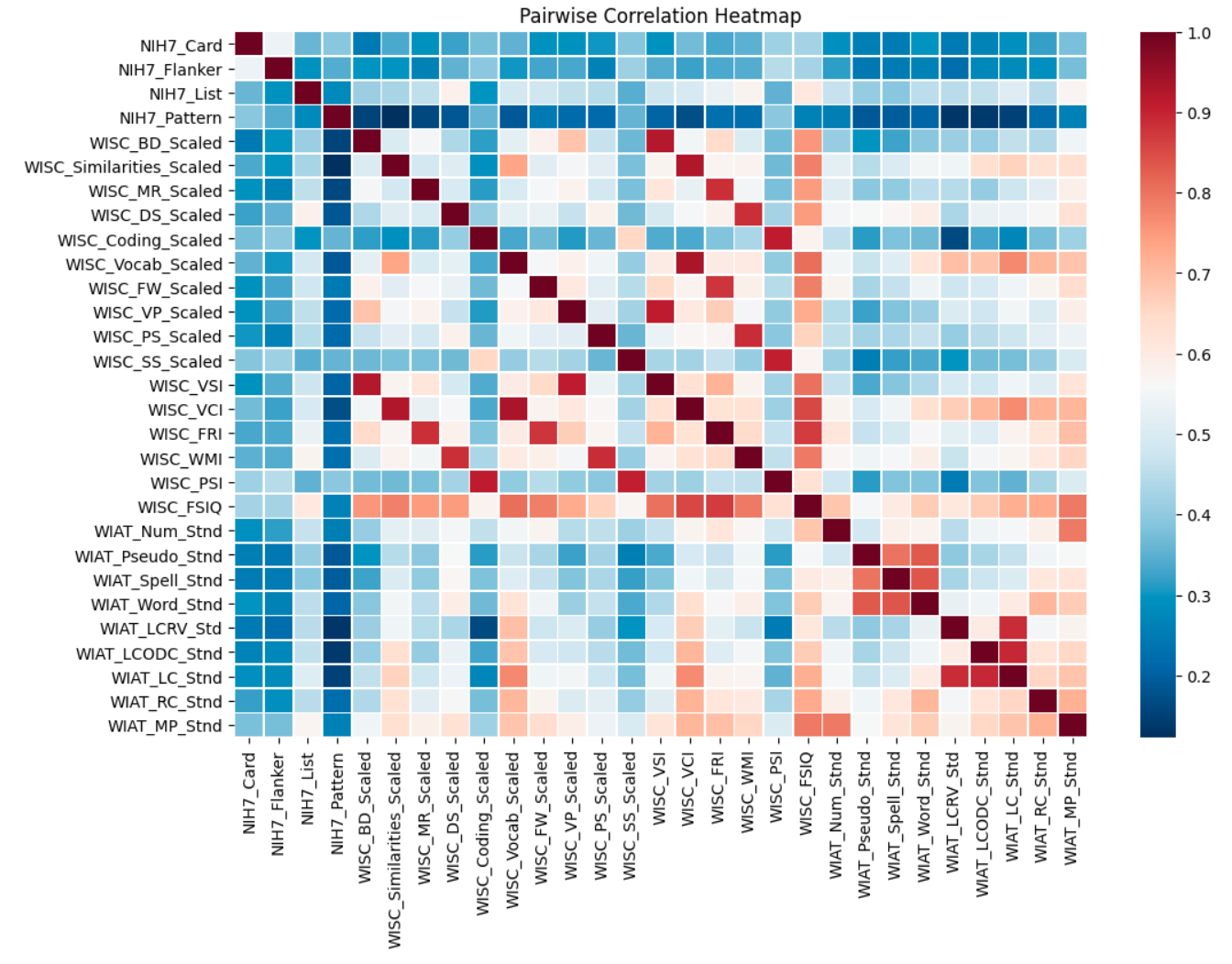}
  \caption{Pairwise correlation between all the HBN phenotypes that we will include in our analysis}
  \label{fig:corr-pheno-hbn}
\end{figure}

\paragraph{Phenotypes}
While the HBN dataset encompasses both behavioral and clinical phenotypes, our analysis focused exclusively on behavioral measures. We specifically selected cognitive phenotypes that cover constructs similar to those found in the HCP and PNC datasets. For each specific instrument, we selected the summary metric it provided, rather than individual answers or subscales, to ensure consistency with the other datasets in our analysis. Ultimately, while the original phenotypes cover behavioral, personality, and mental health domains, we curated a set of 29 behavioral scores (see section \ref{tab:hbn-pheno} for a full list). As a final check to identify problems or redundancy, we also calculated the correlation between the different phenotypes selected (Figure~\ref{fig:corr-pheno-hbn}). As described below, missing phenotypes or phenotypes that are above and below 3 standard deviations will be treated as missing data and imputed. All phenotype values will be z-scored to be on a similar scale and avoid scores that are higher to dominate the model prediction. For a complete description, please refer to section~\ref{sec:behavioral-hcp} and section \ref{sec:behavioral-pnc}, where we describe how the behavioral phenotypes are preprocessed for the HCP and PNC datasets.

\begin{figure}[h]
  \centering
  \includegraphics[width=\textwidth]{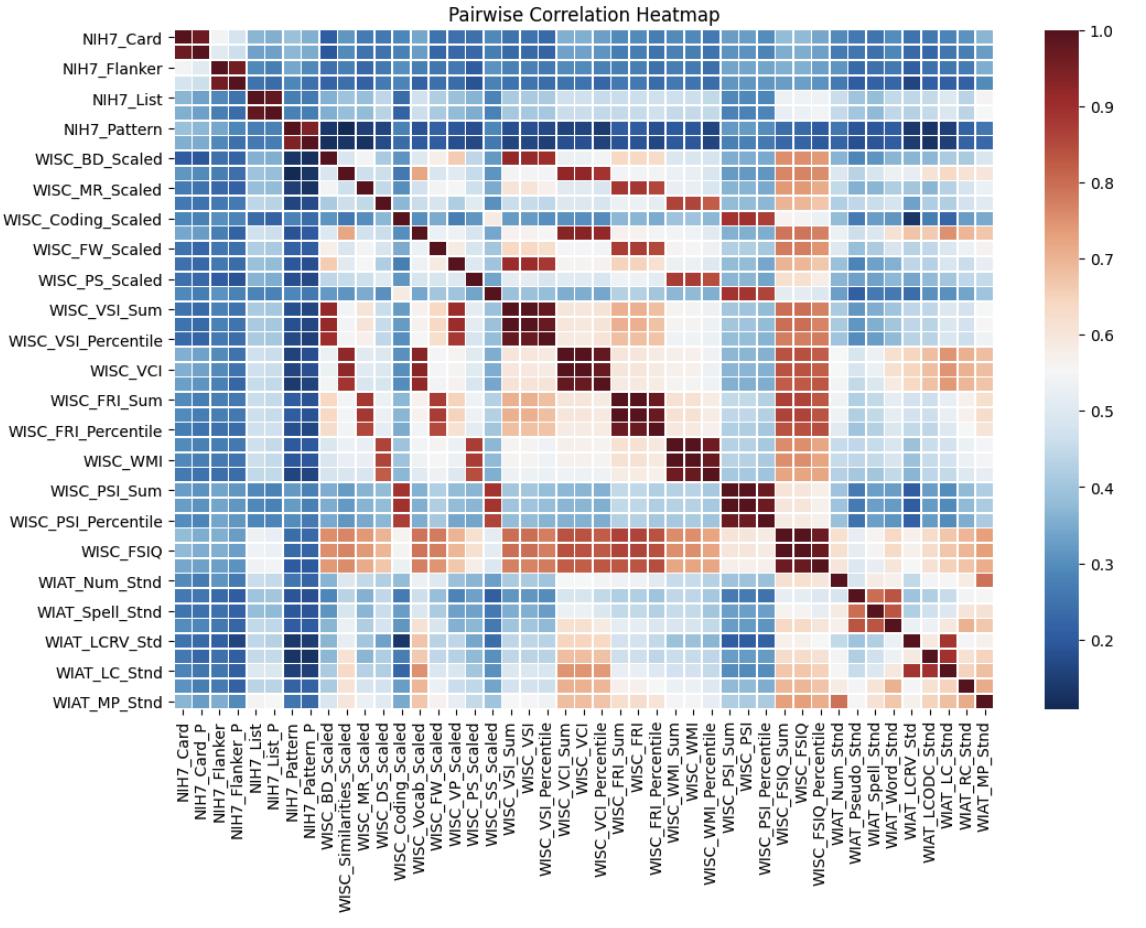}
  \caption{Pairwise correlation between all the HBN phenotypes. This plot includes some of the variables that we defined as redundant as they include the same information, just slightly modified by any type of normalisation (percentile, scale, sum)}
  \label{fig:corr-pheno-hbn-red}
\end{figure}

\begin{table}[ht]
\caption{This table lists the phenotype variables from the Healthy Brain Network (HBN) dataset that will be utilized for analysis. The variables are ordered by the instrument they belong to, marked in bold.}
\label{tab:hbn-pheno}
\begin{tabularx}{\textwidth}{lX}
\toprule
\textbf{Test} & \textbf{Description} \\
\hline
\textbf{NIH Scores} & \\
NIH7\_Card & Dimensional Change Card Sort Age 3+ Age-adjusted Scale Score \\
NIH7\_Flanker & Flanker Inhibitory Control and Attention Age 3+ Age-adjusted Score \\
NIH7\_List & List Sorting Working Memory Age 7+ Age-adjusted Score \\
NIH7\_Pattern & Pattern Comparison Process Speed 7+ Age-adjusted Score \\
\hline
\textbf{WISC} & \\
\hline
WISC\_BD\_Scaled & Block Design Scaled Score \\
WISC\_Similarities\_Scaled & Similarities Scaled Score \\
WISC\_MR\_Scaled & Matrix Reasoning Scaled Score \\
WISC\_DS\_Scaled & Digit Span Scaled Score \\
WISC\_Coding\_Scaled & Coding Scaled Score \\
WISC\_Vocab\_Scaled & Vocabulary Scaled Score \\
WISC\_FW\_Scaled & Figure Weights Scaled Score \\
WISC\_VP\_Scaled & Visual Puzzles Scaled Score \\
WISC\_PS\_Scaled & Picture Span Scaled Score \\
WISC\_SS\_Scaled & Symbol Search Scaled Score \\
WISC\_VSI & VSI Composite Score \\
WISC\_VC & VCI Composite Score \\
WISC\_FRI & FRI Composite Score \\
WISC\_WM & WMI Composite Score \\
WISC\_PSI & PSI Composite Score \\
WISC\_FSIQ & FSIQ Composite Score \\
\hline
\textbf{WIAT} & \\
\hline
WIAT\_Num\_Stnd & Numerical Operations Standard Score \\
WIAT\_Pseudo\_Stnd & Pseudo-word Decoding Standard Score \\
WIAT\_Spell\_Stnd & Spelling Standard Score \\
WIAT\_Word\_Stnd & Word Reading Standard Score \\
WIAT\_LCRV\_Std & Listening Comprehension Receptive Vocabulary Standard Score \\
WIAT\_LCODC\_Stnd & Listening Comprehension Oral Discourse Comprehension Standard Score \\
WIAT\_LC\_Stnd & Listening Comprehension Standard Score \\
WIAT\_RC\_Stnd & Reading Comprehension Standard Score \\
WIAT\_MP\_Stnd & Math Problem Solving Standard Score \\
\hline
\end{tabularx}
\end{table}

\subsubsection{Unit of analysis}
\emph{Which units of analysis (respondents, cases, etc.) will be included or excluded in your study? Taking these inclusion and exclusion criteria into account, indicate the expected sample size of the data you’ll be using for your statistical analyses. If you have a research question about a certain group you may need to exclude participants based on one or more characteristics. Be very specific when describing these characteristics so that readers will be able to redo your moves easily.} \\
Subjects were chosen if their imaging preprocessing pipeline was completed without error and if any of the runs had more than 50\% of the time points flagged as points to be censured (due to the recessive motion), we do not use that run for computing the FC. The numbers described in section~\ref{sec:pre-print-dataset}, describe the dataset that will be used for this confirmatory analysis. Missing phenotypes were not used as criteria for exclusion, as we imputed the missing phenotype values. 

\subsubsection{Missing data}
\emph{What do you know about missing data in the dataset (i.e., overall missingness rate, information about differential dropout)? How will you deal with incomplete or missing data? Provide descriptive information, if available, on the amount of missing data for each variable you will use in the statistical analyses. Based on this information, provide a new expected sample size.} \\
Similar to the analysis described in our preprint for the HCP and PNC dataset (section~\ref{sec:behavioral-hcp} and \ref{sec:behavioral-pnc}, each phenotype measure underwent z-scoring across the phenotypic measures. We identified missing values and removed outliers (i.e., variables that were above/below 3 standard deviations for that phenotype), and imputed both using the IterativeImputer function from the scikit-learn library \citep{Pedregosa-2011Scikit-learn:}. The main idea behind the IterativeImputer is to model each missing value in one variable as a function of all the other variables available for a subject that is present. The algorithm works iteratively in a round-robin matter, where during each iteration, one phenotype is chosen as (y), and the remaining features are used as input to the imputation algorithm. Subjects that had no phenotypical information were excluded; however, if one phenotype was present, all the remaining values were imputed.

\subsubsection{Sampling weights}
\emph{Are there sampling weights available with this dataset? If so, are you using them or are you using your own sampling weights? Sampling weights can be useful in secondary data analysis because the sample may not be entirely representative of the population you are interested in.} \\
There are no sampling weights. Models are trained and evaluated on samples from the same population.

\subsection{Knowledge of Data}

\subsubsection{Prior Publication/Dissemination}
\emph{List the publications, working papers, and conference presentations you have worked on that are based on the dataset you will use. For each work, list the variables you analyzed, but limit yourself to variables that are relevant to the proposed analysis. If the dataset is longitudinal, also state which wave of the dataset you analyzed. Specify the previous works for each co-author separately.}
The work described in the main manuscript, in which all co-authors were involved, analysed the predictability of phenotypes from the Human Connectome Project (HCP) and the Philadelphia Neurodevelopmental Cohort (PNC). Building on these findings, we are testing the generalizability of our results by incorporating a third dataset. Therefore, the analyses described in this preregistration will be the same as those described in our preprint (the main part of this document).

\subsubsection{Prior knowledge}
Disclose any prior knowledge you may have about the dataset that is relevant for the proposed analysis. If you do not have any prior knowledge of it, please state so. Your prior knowledge could stem from working with the data first-hand, from reading previously published research, or from codebooks. Provide prior knowledge for every author separately.
\paragraph{All co-authors}
are familiar with the original publication provided by the HBN consortia where the recruitment criteria and data description are detailed \citep{Alexander-2017An}. We have already preprocessed the neuroimaging data as described above, selected the phenotypes, and formatted the phenotypic data to be compatible with our analytical pipeline.
\paragraph{Jessica Dafflon}
has worked with the structural MRI from the HBN. 

\paragraph{Dylan M. Nielson}
 has worked with the HBN data before and published a pre-print Validation of CBCL depression scores of adolescents in three independent datasets \citep{Zelenina-2023Validation}.

\paragraph{Dustin Moraczewski}
has interacted with the dataset through familiarity with the relevant data papers and other literature, as well as the dataset contents, and has engaged in downloading, preprocessing, quality checking, and preparing the data for collaborators but has not implemented any further analyses.

\paragraph{Eric Earl} 
has interacted with the dataset only to reshape the tabular phentoype data for easier ingestion into the project's analysis \citep{Earl-2023Big}.

\paragraph{Gabriel Loewinger} has never worked with the HBN dataset before.

\paragraph{Patrick McClure} has never worked with the HBN dataset before.

\paragraph{Adam G. Thomas} has familiarity with the dataset via analysis presented in \cite{lam2022interpretable} which has been presented at lab meetings.

\paragraph{Francisco Pereira}
 was involved in a separate study \citep{lam2022interpretable}  where they used the HBN data to test a new method for creating interpretable latent variables. However, in this study, the authors have used only the clinical phenotypes from the HBN. Neither the behavior nor the imaging data which we will use were analyzed.

\subsection{Analyses}
\subsubsection{Statistical models}
\emph{
For each hypothesis, describe the statistical model you will use to test the hypothesis. Include the type of model (e.g., ANOVA, multiple regression, SEM) and the specification of the model. Specify any interactions and post-hoc analyses and remember that any test not included here must be labeled as an exploratory test in the final paper.} \\
\begin{enumerate}
    \item \textbf{How does controlling for age/sex confounds impact the phenotypes prediction?} \\
    To answer this question, we will fit a ridge regression to predict phenotypes from resting state functional connectivity before and after regressing out age and sex, along with the non-linear term $\textit{age}^2$ and the interaction terms $\textit{age} \times \textit{sex}$ and $\textit{age}^2 \times \textit{sex}$. See section~\ref{sec:confounds} from the main manuscript for a detailed description of how we control for those covariates. The correlation between predicted and true phenotypes before and after the covariate regression will be compared along with the beta-coefficients obtained for each phenotype measure. We will use a paired t-test to ensure that the differences observed are not noise with an alpha of 0.05.
        \item \textbf{How reliable are the latent phenotypes estimated with SVD across different samples?} \\
 To assess the consistency of the Singular Value Decomposition (SVD) derived latent phenotypes, we will analyze how many of these latent phenotypes would replicate when the SVD is applied to two independent samples from the same population (for a detailed description, please refer to section~\ref{sec:svd-reliability}). To do this, we will divide the phenotype dataset into two halves 1000 times and independently perform the SVD on each split. Due to the mathematical properties of the SVD, it is not guaranteed that we will get the same component order. We will use a reference SVD obtained from using the entire data and employ the Gale-Shapely stable marriage algorithm \citep{gale-shapely} to match each component to its counterpart in the splits (refer to the preprint section~\ref{sec:gale-algo} for a more detailed description). Our cutoff for defining a reliable component will be an empirical lower 95\% confidence interval on the $r^2$ greater than 0.2.
    \item \textbf{Can we obtain more predictable phenotypes if we use a linear transformation to latent phenotypes (SVD) on HBN? } \\
    We will also train a ridge regression to predict the latent phenotypes (SVD latent variables derived from the phenotypes) from resting state functional connectivity. The main idea of this experiment is to see if, by compressing the phenotypes into a latent repression, we will lose any predictive information. We will evaluate the change in performance by comparing the prediction performance for the untransformed variables and those after the SVD. We motivated the selection of SVD in section \ref{sec:svd-pheno} and go into detail about the SVD definition in section \ref{sec:svd-maths}. We will use a paired t-test to ensure that the differences in error for the predicted and true values between the latent and phenotype populations observed are not noise with an alpha of 0.05.
    \item \textbf{Is the reconstruction of latent phenotypes more predictable than using all latent phenotypes?}
    Similar to what we described in the main section of this manuscript (section \ref{sec:prediction-recons}), to evaluate if ridge regression models trained using reconstructed phenotype variables (using a range of 1 to maximum of reliable latent phenotypes, which we expect to be 5, latent components and all components) performed as well or better than one trained using the original phenotypes, we will use the Autorank library \citep{herbold2020autorank, demvsar2006statistical} to check if there is a significant performance between the reconstructed results. The choice of training 6 models is motivated by our exploratory analyses, where we saw that only the first five latent phenotypes in each dataset
were reliable, and while most of the variance could be explained by the first two latent phenotypes,
explaining 95\% of the variance would require most of the latent phenotypes.

\end{enumerate}

\subsubsection{Effect size}
\emph{If applicable, specify a predicted effect size or a minimum effect size of interest for all the effects tested in your statistical analyses.} \\

\begin{enumerate}
    \item \textbf{How does controlling for age/sex confounds impact the phenotypes prediction?} \\
    The effect size (Cohen's d) we observed in HCP was 0.62 and in PNC it was 1.03. We will be conservative and use the smaller of the two effect sizes to estimate our statistical power in HBN.
    \item \textbf{How reliable are the latent phenotypes estimated with SVD across different samples?} \\
    Our evaluation of latent phenotype reliability is based on the lower 95\% confidence interval of the inter-split $r^2$. We set a threshold of 0.2. 
    \item \textbf{Can we obtain more predictable phenotypes if we use a linear transformation to latent phenotypes (SVD) on HBN?} \\
    The sample size for our test set in HBN will be 105 participants. With 105 participants we will have 80\% power to detect an effect with a Cohen's d of 0.275.
    
    \item \textbf{Are phenotype variables reconstructed from the SVD latent variables as informative for training a model as the original phenotype variables?} \\
    In the HCP data we observed an effect size (Cohen's f) of 0.13. In the PNC data we observed an effect size of 0.10. The PNC data more closely resembles the HBN data in the number of phenotypes and relative homogeneity of those phenotypes, so we will use the lower and more conservative effect size from PNC to estimate our power in the HBN data.
    
\end{enumerate}

\subsubsection{Statistical power}
\emph{Present the statistical power available to detect the predicted effect size or the smallest effect size of interest. Use the sample size after updating for missing data and outliers.} \\
\begin{enumerate}
    \item \textbf{How does controlling for age/sex confounds impact the phenotypes prediction?} \\
    We are testing this hypothesis with a two-sided paired t-test across phenotypes. HBN has 29 phenotypes that we are considering. With an effect size of 0.62 and alpha of 0.05, this gives us a power of 94.6\%.
    \item \textbf{How reliable are the latent phenotypes estimated with SVD across different samples?} \\
    Our test for each component is that the lower 95\% confidence interval of the inter-split $r^2$ is greater than 0.2. In the HBN data we will have 681 participants in which to test component reliability, so each split will have 390 participants. We ran simulations drawing 1000 samples with n = 390 per split from a 100000 simulated population in which we set the true correlation between splits. From these simulations we found that if the true inter-split $r^2$ is 0.28, we have a 99\% chance of detecting it as being above 0.2.  
    \item \textbf{Can we obtain more predictable phenotypes if we use a linear transformation to latent phenotypes (SVD) on HBN?} \\
    As stated above:
    The sample size for our test set in HBN will be 105 participants. With 105 participants, we will have 80\% power to detect an effect with a Cohen's d of 0.275.
    \item \textbf{Are phenotype variables reconstructed from the SVD latent variables as informative for training a model as the original phenotype variables?} \\
    We will use the Autorank library to test the overall hypothesis that there are differences in predictive performance depending on the number of factors used to reconstruct the phenotypes in the training dataset. The Autorank library automatically selects a test based on the normality and hetroscedasticity of the data, but the HCP and PNC datasets were suitable for testing with a repeated measures ANOVA, so that is what we used for determining power. The PNC data had a Cohen's f of 0.10, nonsphericity correction of 0.5, and the median correlation between repeated measures was 0.9. There are 30 phenotypes in the HBN data and we will be using an alpha of 0.05 and assuming 6 measurements (5 reconstructed plus the original). With these parameters we estimate a power for the HBN dataset of 0.69.
\end{enumerate}

\subsection{Inference criteria}
\emph{What criteria will you use to make inferences? Describe the information you will use (e.g. specify the p-values, effect sizes, confidence intervals, Bayes factors, specific model fit indices),as well as cut-off criteria, where appropriate. Will you be using one-or two-tailed tests for each of your analyses? If you are comparing multiple conditions or testing multiple hypotheses, will you account for this, and if so, how?}
Unless otherwise specified, we will use an alpha of 0.05.

\subsubsection{Assumption Violation/ Model Non-Convergence}
\emph{What will you do should your data violate assumptions, your model not converge, or some other analytic problem arises?} \\
The HBN dataset is similar in its demographics to the PNC dataset, and hence, we do not expect big deviations in predictability. However, if any deviations arise we will investigate the reason for deviations and will report those as exploratory analyses.

\subsubsection{Reliability and Robustness testing}
\emph{Provide a series of decisions or tests about evaluating the strength, reliability, or robustness of your finding. This may include within-study replication attempts, additional covariates, cross-validation, applying weights, selectively applying constraints in an SEM context (e.g., comparing model fit statistics), overfitting adjustment techniques used, or some other simulation/sampling/bootstrapping method.} \\
We assessed the reliability and robustness using three main strategies:
\begin{enumerate}
    \item \textbf{Evaluate our previous findings with a third dataset:} Our major effort is to assess whether previous findings for PNC and HCP would generalize to an independent dataset. This is why we are pre-registering this analysis where we re-implement the same methods on a separate dataset.
    \item \textbf{Assess the reliability of the latent phenotypes:} As we mentioned above (Analysis, statistical models), there is no guarantee that if we repeat the SVD on different splits of the data, we would obtain the same results. Therefore, we repeated the analysis on 1,000 different splits and assessed how reliably we could identify the components on different splits (See section Analysis, statistical methods or our preprint for a more detailed description).
    \item \textbf{Resampling within datasets:} All of our reported results have been conducted using bootstrapping (i.e., we repeated our analysis 100 times with different training datasets resampled from the original one -- see section \ref{sec:prediction-model} for a detailed description). We use these bootstraps to reduce the variance of our predictive performance estimates.

\end{enumerate}

\subsubsection{Exploratory analysis}
\emph{If you plan to explore your dataset to look for unexpected differences or relationships, describe those tests here. If reported, add them to the final paper under a heading that clearly differentiates this exploratory part of your study from the confirmatory part.} \\
\begin{itemize}
    \item How predictable are phenotypes from functional connective in HBN, and how do they compare to the predictability of the other datasets we analyzed? 
    To answer this question we will fit a ridge regression to try to predict phenotypes from resting state functional connectivity. We will check if the obtained predictions are within the same range $(r=[0, 0.4]$) as those obtained for the other two datasets. For a detailed description of how the ridge regression model will be trained and which variables will be used as input, please refer to section~\ref{sec:prediction-model} of the main manuscript, as the same model architecture (retrained for the HBN dataset) and pre-processing used for HCP and PNC will be used for the new dataset
    
\end{itemize}

%% file: main.bbl
\begin{thebibliography}{82}
\providecommand{\natexlab}[1]{#1}
\providecommand{\url}[1]{\texttt{#1}}
\expandafter\ifx\csname urlstyle\endcsname\relax
  \providecommand{\doi}[1]{doi: #1}\else
  \providecommand{\doi}{doi: \begingroup \urlstyle{rm}\Url}\fi

\bibitem[Abraham et~al.(2014)Abraham, Pedregosa, Eickenberg, Gervais, Mueller,
  Kossaifi, Gramfort, Thirion, and Varoquaux]{nilearn}
A.~Abraham, F.~Pedregosa, M.~Eickenberg, P.~Gervais, A.~Mueller, J.~Kossaifi,
  A.~Gramfort, B.~Thirion, and G.~Varoquaux.
\newblock Machine learning for neuroimaging with scikit-learn.
\newblock \emph{Frontiers in Neuroinformatics}, 8, 2014.
\newblock ISSN 1662-5196.
\newblock \doi{10.3389/fninf.2014.00014}.
\newblock URL
  \url{https://www.frontiersin.org/articles/10.3389/fninf.2014.00014/full}.

\bibitem[Alexander et~al.(2017)Alexander, Escalera, Ai, Andreotti, Febre,
  Mangone, Vega-Potler, Langer, Alexander, Kovacs, Litke, O'Hagan, Andersen,
  Bronstein, Bui, Bushey, Butler, Castagna, Camacho, Chan, Citera, Clucas,
  Cohen, Dufek, Eaves, Fradera, Gardner, Grant-Villegas, Green, Gregory, Hart,
  Harris, Horton, Kahn, Kabotyanski, Karmel, Kelly, Kleinman, Koo, Kramer,
  Lennon, Lord, Mantello, Margolis, Merikangas, Milham, Minniti, Neuhaus,
  Levine, Osman, Parra, Pugh, Racanello, Restrepo, Saltzman, Septimus, Tobe,
  Waltz, Williams, Yeo, Castellanos, Klein, Paus, Leventhal, Craddock,
  Koplewicz, and Milham]{Alexander-2017An}
L.~M. Alexander, J.~Escalera, L.~Ai, C.~Andreotti, K.~Febre, A.~Mangone,
  N.~Vega-Potler, N.~Langer, A.~Alexander, M.~Kovacs, S.~Litke, B.~O'Hagan,
  J.~Andersen, B.~Bronstein, A.~Bui, M.~Bushey, H.~Butler, V.~Castagna,
  N.~Camacho, E.~Chan, D.~Citera, J.~Clucas, S.~Cohen, S.~Dufek, M.~Eaves,
  B.~Fradera, J.~Gardner, N.~Grant-Villegas, G.~Green, C.~Gregory, E.~Hart,
  S.~Harris, M.~Horton, D.~Kahn, K.~Kabotyanski, B.~Karmel, S.~P. Kelly,
  K.~Kleinman, B.~Koo, E.~Kramer, E.~Lennon, C.~Lord, G.~Mantello, A.~Margolis,
  K.~R. Merikangas, J.~Milham, G.~Minniti, R.~Neuhaus, A.~Levine, Y.~Osman,
  L.~C. Parra, K.~R. Pugh, A.~Racanello, A.~Restrepo, T.~Saltzman, B.~Septimus,
  R.~Tobe, R.~Waltz, A.~Williams, A.~Yeo, F.~X. Castellanos, A.~Klein, T.~Paus,
  B.~L. Leventhal, R.~C. Craddock, H.~S. Koplewicz, and M.~P. Milham.
\newblock An open resource for transdiagnostic research in pediatric mental
  health and learning disorders.
\newblock \emph{Sci Data}, 4:\penalty0 170181, Dec 2017.
\newblock \doi{10.1038/sdata.2017.181}.

\bibitem[Avants et~al.(2008)Avants, Epstein, Grossman, and Gee]{ants}
B.~Avants, C.~Epstein, M.~Grossman, and J.~Gee.
\newblock Symmetric diffeomorphic image registration with cross-correlation:
  Evaluating automated labeling of elderly and neurodegenerative brain.
\newblock \emph{Medical Image Analysis}, 12\penalty0 (1):\penalty0 26--41,
  2008.
\newblock ISSN 1361-8415.
\newblock \doi{10.1016/j.media.2007.06.004}.
\newblock URL
  \url{http://www.sciencedirect.com/science/article/pii/S1361841507000606}.

\bibitem[Barch et~al.(2013)Barch, Burgess, Harms, Petersen, Schlaggar,
  Corbetta, Glasser, Curtiss, Dixit, Feldt, Nolan, Bryant, Hartley, Footer,
  Bjork, Poldrack, Smith, Johansen-Berg, Snyder, Van~Essen, and {WU-Minn HCP
  Consortium}]{Barch-2013Function}
D.~M. Barch, G.~C. Burgess, M.~P. Harms, S.~E. Petersen, B.~L. Schlaggar,
  M.~Corbetta, M.~F. Glasser, S.~Curtiss, S.~Dixit, C.~Feldt, D.~Nolan,
  E.~Bryant, T.~Hartley, O.~Footer, J.~M. Bjork, R.~Poldrack, S.~Smith,
  H.~Johansen-Berg, A.~Z. Snyder, D.~C. Van~Essen, and {WU-Minn HCP
  Consortium}.
\newblock Function in the human connectome: task-fmri and individual
  differences in behavior.
\newblock \emph{Neuroimage}, 80:\penalty0 169--89, Oct 2013.
\newblock \doi{10.1016/j.neuroimage.2013.05.033}.

\bibitem[Behzadi et~al.(2007)Behzadi, Restom, Liau, and Liu]{compcor}
Y.~Behzadi, K.~Restom, J.~Liau, and T.~T. Liu.
\newblock A component based noise correction method ({CompCor}) for {BOLD} and
  perfusion based fmri.
\newblock \emph{NeuroImage}, 37\penalty0 (1):\penalty0 90--101, 2007.
\newblock ISSN 1053-8119.
\newblock \doi{10.1016/j.neuroimage.2007.04.042}.
\newblock URL
  \url{http://www.sciencedirect.com/science/article/pii/S1053811907003837}.

\bibitem[Bertolero and Bassett(2020)]{Bertolero-2020Deep}
M.~A. Bertolero and D.~S. Bassett.
\newblock Deep neural networks carve the brain at its joints.
\newblock \emph{arXiv preprint arXiv:2002.08891}, 2020.

\bibitem[Botvinik-Nezer et~al.(2020)Botvinik-Nezer, Holzmeister, Camerer,
  Dreber, Huber, Johannesson, Kirchler, Iwanir, Mumford, Adcock,
  et~al.]{Botvinik-Nezer-2020Variability}
R.~Botvinik-Nezer, F.~Holzmeister, C.~F. Camerer, A.~Dreber, J.~Huber,
  M.~Johannesson, M.~Kirchler, R.~Iwanir, J.~A. Mumford, R.~A. Adcock, et~al.
\newblock Variability in the analysis of a single neuroimaging dataset by many
  teams.
\newblock \emph{Nature}, 582\penalty0 (7810):\penalty0 84--88, 2020.

\bibitem[Carp(2012)]{Carp-2012On}
J.~Carp.
\newblock On the plurality of (methodological) worlds: estimating the analytic
  flexibility of fmri experiments.
\newblock \emph{Frontiers in neuroscience}, 6:\penalty0 149, 2012.

\bibitem[Cecchetti and Handjaras(2022)]{Cecchetti-2022Reproducible}
L.~Cecchetti and G.~Handjaras.
\newblock Reproducible brain-wide association studies do not necessarily
  require thousands of individuals.
\newblock \emph{PsyArXiv}, 2022.

\bibitem[Chen et~al.(2022)Chen, Tam, Kebets, Orban, Ooi, Asplund, Marek,
  Dosenbach, Eickhoff, Bzdok, et~al.]{Chen-2022Shared}
J.~Chen, A.~Tam, V.~Kebets, C.~Orban, L.~Q.~R. Ooi, C.~L. Asplund, S.~Marek,
  N.~U. Dosenbach, S.~B. Eickhoff, D.~Bzdok, et~al.
\newblock Shared and unique brain network features predict cognitive,
  personality, and mental health scores in the abcd study.
\newblock \emph{Nature communications}, 13\penalty0 (1):\penalty0 2217, 2022.

\bibitem[Chen et~al.(2023)Chen, Ooi, Tan, Zhang, Li, Asplund, Eickhoff, Bzdok,
  Holmes, and Yeo]{Chen-2023Relationship}
J.~Chen, L.~Q.~R. Ooi, T.~W.~K. Tan, S.~Zhang, J.~Li, C.~L. Asplund, S.~B.
  Eickhoff, D.~Bzdok, A.~J. Holmes, and B.~T. Yeo.
\newblock Relationship between prediction accuracy and feature importance
  reliability: An empirical and theoretical study.
\newblock \emph{NeuroImage}, 274:\penalty0 120115, 2023.

\bibitem[Chopra et~al.(2022)Chopra, Dhamala, Lawhead, Ricard, Orchard, An,
  Chen, Wulan, Kumar, Rubenstein, Moses, Chen, Levi, Holmes, Aquino, Fornito,
  Harpaz-Rotem, Germine, Baker, Yeo, and Holmes]{Chopra-2022Reliable}
S.~Chopra, E.~Dhamala, C.~Lawhead, J.~A. Ricard, E.~R. Orchard, L.~An, P.~Chen,
  N.~Wulan, P.~Kumar, A.~Rubenstein, J.~Moses, L.~Chen, P.~Levi, A.~Holmes,
  K.~Aquino, A.~Fornito, I.~Harpaz-Rotem, L.~T. Germine, J.~T. Baker, B.~T.
  Yeo, and A.~J. Holmes.
\newblock Reliable and generalizable brain-based predictions of cognitive
  functioning across common psychiatric illness.
\newblock \emph{medRxiv}, 2022.
\newblock \doi{10.1101/2022.12.08.22283232}.
\newblock URL
  \url{https://www.medrxiv.org/content/early/2022/12/15/2022.12.08.22283232}.

\bibitem[Chyzhyk et~al.(2022)Chyzhyk, Varoquaux, Milham, and
  Thirion]{Chyzhyk-2022How}
D.~Chyzhyk, G.~Varoquaux, M.~Milham, and B.~Thirion.
\newblock How to remove or control confounds in predictive models, with
  applications to brain biomarkers.
\newblock \emph{GigaScience}, 11:\penalty0 giac014, 2022.

\bibitem[Cox and Hyde(1997)]{afni}
R.~W. Cox and J.~S. Hyde.
\newblock Software tools for analysis and visualization of fmri data.
\newblock \emph{NMR in Biomedicine}, 10\penalty0 (4-5):\penalty0 171--178,
  1997.
\newblock
  \doi{10.1002/(SICI)1099-1492(199706/08)10:4/5<171::AID-NBM453>3.0.CO;2-L}.

\bibitem[Dafflon et~al.(2022)Dafflon, F.~Da~Costa, V{\'a}{\v{s}}a, Monti,
  Bzdok, Hellyer, Turkheimer, Smallwood, Jones, and Leech]{Dafflon-2022A}
J.~Dafflon, P.~F.~Da~Costa, F.~V{\'a}{\v{s}}a, R.~P. Monti, D.~Bzdok, P.~J.
  Hellyer, F.~Turkheimer, J.~Smallwood, E.~Jones, and R.~Leech.
\newblock A guided multiverse study of neuroimaging analyses.
\newblock \emph{Nature Communications}, 13\penalty0 (1):\penalty0 3758, 2022.

\bibitem[Dale et~al.(1999)Dale, Fischl, and Sereno]{fs_reconall}
A.~M. Dale, B.~Fischl, and M.~I. Sereno.
\newblock Cortical surface-based analysis: I. segmentation and surface
  reconstruction.
\newblock \emph{NeuroImage}, 9\penalty0 (2):\penalty0 179--194, 1999.
\newblock ISSN 1053-8119.
\newblock \doi{10.1006/nimg.1998.0395}.
\newblock URL
  \url{http://www.sciencedirect.com/science/article/pii/S1053811998903950}.

\bibitem[Dem{\v{s}}ar(2006)]{demvsar2006statistical}
J.~Dem{\v{s}}ar.
\newblock Statistical comparisons of classifiers over multiple data sets.
\newblock \emph{The Journal of Machine learning research}, 7:\penalty0 1--30,
  2006.

\bibitem[Earl et~al.(2023)Earl, Dafflon, Salamanca-Giron, Faskowitz, Basavaraj,
  Moraczewski, Pereira, and Thomas]{Earl-2023Big}
E.~Earl, J.~Dafflon, R.~Salamanca-Giron, J.~Faskowitz, A.~Basavaraj,
  D.~Moraczewski, F.~Pereira, and A.~Thomas.
\newblock {Big neuroimaging dataset BIDS tabular phenotype tools}, June 2023.
\newblock URL \url{https://doi.org/10.5281/zenodo.8084023}.

\bibitem[Esteban et~al.(2018{\natexlab{a}})Esteban, Blair, Markiewicz,
  Berleant, Moodie, Ma, Isik, Erramuzpe, Kent, DuPre, Sitek, Gomez, Lurie, Ye,
  Poldrack, and Gorgolewski]{fmriprep2}
O.~Esteban, R.~Blair, C.~J. Markiewicz, S.~L. Berleant, C.~Moodie, F.~Ma, A.~I.
  Isik, A.~Erramuzpe, M.~Kent, James D.~andGoncalves, E.~DuPre, K.~R. Sitek,
  D.~E.~P. Gomez, D.~J. Lurie, Z.~Ye, R.~A. Poldrack, and K.~J. Gorgolewski.
\newblock fmriprep.
\newblock \emph{Software}, 2018{\natexlab{a}}.
\newblock \doi{10.5281/zenodo.852659}.

\bibitem[Esteban et~al.(2018{\natexlab{b}})Esteban, Markiewicz, Blair, Moodie,
  Isik, Erramuzpe~Aliaga, Kent, Goncalves, DuPre, Snyder, Oya, Ghosh, Wright,
  Durnez, Poldrack, and Gorgolewski]{fmriprep1}
O.~Esteban, C.~Markiewicz, R.~W. Blair, C.~Moodie, A.~I. Isik,
  A.~Erramuzpe~Aliaga, J.~Kent, M.~Goncalves, E.~DuPre, M.~Snyder, H.~Oya,
  S.~Ghosh, J.~Wright, J.~Durnez, R.~Poldrack, and K.~J. Gorgolewski.
\newblock {fMRIPrep}: a robust preprocessing pipeline for functional {MRI}.
\newblock \emph{Nature Methods}, 2018{\natexlab{b}}.
\newblock \doi{10.1038/s41592-018-0235-4}.

\bibitem[Esteban et~al.(2019)Esteban, Markiewicz, Blair, Moodie, Isik,
  Erramuzpe, Kent, Goncalves, DuPre, Snyder, et~al.]{Esteban-2019fMRIPrep:}
O.~Esteban, C.~J. Markiewicz, R.~W. Blair, C.~A. Moodie, A.~I. Isik,
  A.~Erramuzpe, J.~D. Kent, M.~Goncalves, E.~DuPre, M.~Snyder, et~al.
\newblock fmriprep: a robust preprocessing pipeline for functional mri.
\newblock \emph{Nature methods}, 16\penalty0 (1):\penalty0 111--116, 2019.

\bibitem[Evans et~al.(2012)Evans, Janke, Collins, and Baillet]{mni152nlin6asym}
A.~Evans, A.~Janke, D.~Collins, and S.~Baillet.
\newblock Brain templates and atlases.
\newblock \emph{NeuroImage}, 62\penalty0 (2):\penalty0 911--922, 2012.
\newblock \doi{10.1016/j.neuroimage.2012.01.024}.

\bibitem[Finn and Rosenberg(2021)]{Finn-2021Beyond}
E.~S. Finn and M.~D. Rosenberg.
\newblock Beyond fingerprinting: Choosing predictive connectomes over reliable
  connectomes.
\newblock \emph{NeuroImage}, 239:\penalty0 118254, 2021.

\bibitem[Finn et~al.(2015)Finn, Shen, Scheinost, Rosenberg, Huang, Chun,
  Papademetris, and Constable]{Finn-2015Functional}
E.~S. Finn, X.~Shen, D.~Scheinost, M.~D. Rosenberg, J.~Huang, M.~M. Chun,
  X.~Papademetris, and R.~T. Constable.
\newblock Functional connectome fingerprinting: identifying individuals using
  patterns of brain connectivity.
\newblock \emph{Nature neuroscience}, 18\penalty0 (11):\penalty0 1664--1671,
  2015.

\bibitem[Fonov et~al.(2009)Fonov, Evans, McKinstry, Almli, and
  Collins]{mni152nlin2009casym}
V.~Fonov, A.~Evans, R.~McKinstry, C.~Almli, and D.~Collins.
\newblock Unbiased nonlinear average age-appropriate brain templates from birth
  to adulthood.
\newblock \emph{NeuroImage}, 47, Supplement 1:\penalty0 S102, 2009.
\newblock \doi{10.1016/S1053-8119(09)70884-5}.

\bibitem[Gale and Shapley(1962)]{gale-shapely}
D.~Gale and L.~S. Shapley.
\newblock College admissions and the stability of marriage.
\newblock \emph{The American Mathematical Monthly}, 69\penalty0 (1):\penalty0
  9--15, 1962.

\bibitem[Gell et~al.(2023)Gell, Eickhoff, Omidvarnia, Kueppers, Patil,
  Satterthwaite, Mueller, and Langner]{Gell-2023The}
M.~Gell, S.~B. Eickhoff, A.~Omidvarnia, V.~Kueppers, K.~R. Patil, T.~D.
  Satterthwaite, V.~I. Mueller, and R.~Langner.
\newblock The burden of reliability: How measurement noise limits
  brain-behaviour predictions.
\newblock \emph{BioRxiv}, pages 2023--02, 2023.

\bibitem[Genon et~al.(2022)Genon, Eickhoff, and Kharabian]{Genon-2022Linking}
S.~Genon, S.~B. Eickhoff, and S.~Kharabian.
\newblock Linking interindividual variability in brain structure to behaviour.
\newblock \emph{Nature Reviews Neuroscience}, 23\penalty0 (5):\penalty0
  307--318, 2022.

\bibitem[Glasser et~al.(2013{\natexlab{a}})Glasser, Sotiropoulos, Wilson,
  Coalson, Fischl, Andersson, Xu, Jbabdi, Webster, Polimeni, {Van Essen}, and
  Jenkinson]{Glasser-2013The-minimal}
M.~F. Glasser, S.~N. Sotiropoulos, J.~A. Wilson, T.~S. Coalson, B.~Fischl,
  J.~L. Andersson, J.~Xu, S.~Jbabdi, M.~Webster, J.~R. Polimeni, D.~C. {Van
  Essen}, and M.~Jenkinson.
\newblock The minimal preprocessing pipelines for the human connectome project.
\newblock \emph{NeuroImage}, 80:\penalty0 105--124, 2013{\natexlab{a}}.
\newblock ISSN 1053-8119.
\newblock \doi{https://doi.org/10.1016/j.neuroimage.2013.04.127}.
\newblock URL
  \url{https://www.sciencedirect.com/science/article/pii/S1053811913005053}.
\newblock Mapping the Connectome.

\bibitem[Glasser et~al.(2013{\natexlab{b}})Glasser, Sotiropoulos, Wilson,
  Coalson, Fischl, Andersson, Xu, Jbabdi, Webster, Polimeni, Van~Essen, and
  Jenkinson]{hcppipelines}
M.~F. Glasser, S.~N. Sotiropoulos, J.~A. Wilson, T.~S. Coalson, B.~Fischl,
  J.~L. Andersson, J.~Xu, S.~Jbabdi, M.~Webster, J.~R. Polimeni, D.~C.
  Van~Essen, and M.~Jenkinson.
\newblock The minimal preprocessing pipelines for the human connectome project.
\newblock \emph{NeuroImage}, 80:\penalty0 105--124, 2013{\natexlab{b}}.
\newblock ISSN 1053-8119.
\newblock \doi{10.1016/j.neuroimage.2013.04.127}.
\newblock URL
  \url{http://www.sciencedirect.com/science/article/pii/S1053811913005053}.

\bibitem[Gorgolewski et~al.(2011)Gorgolewski, Burns, Madison, Clark, Halchenko,
  Waskom, and Ghosh]{nipype1}
K.~Gorgolewski, C.~D. Burns, C.~Madison, D.~Clark, Y.~O. Halchenko, M.~L.
  Waskom, and S.~Ghosh.
\newblock Nipype: a flexible, lightweight and extensible neuroimaging data
  processing framework in python.
\newblock \emph{Frontiers in Neuroinformatics}, 5:\penalty0 13, 2011.
\newblock \doi{10.3389/fninf.2011.00013}.

\bibitem[Gorgolewski et~al.(2018)Gorgolewski, Esteban, Markiewicz, Ziegler,
  Ellis, Notter, Jarecka, Johnson, Burns, Manh{\~a}es-Savio, Hamalainen,
  Yvernault, Salo, Jordan, Goncalves, Waskom, Clark, Wong, Loney, Modat, Dewey,
  Madison, Visconti~di Oleggio~Castello, Clark, Dayan, Clark, Keshavan,
  Pinsard, Gramfort, Berleant, Nielson, Bougacha, Varoquaux, Cipollini,
  Markello, Rokem, Moloney, Halchenko, Wassermann, Hanke, Horea, Kaczmarzyk,
  de~Hollander, DuPre, Gillman, Mordom, Buchanan, Tungaraza, Pauli, Iqbal,
  Sikka, Mancini, Schwartz, Malone, Dubois, Frohlich, Welch, Forbes, Kent,
  Watanabe, Cumba, Huntenburg, Kastman, Nichols, Eshaghi, Ginsburg, Schaefer,
  Acland, Giavasis, Kleesiek, Erickson, K{\"u}ttner, Haselgrove, Correa,
  Ghayoor, Liem, Millman, Haehn, Lai, Zhou, Blair, Glatard, Renfro, Liu, Kahn,
  P{\'e}rez-Garc{\'\i}a, Triplett, Lampe, Stadler, Kong, Hallquist,
  Chetverikov, Salvatore, Park, Poldrack, Craddock, Inati, Hinds, Cooper,
  Perkins, Marina, Mattfeld, Noel, Snoek, Matsubara, Cheung, Rothmei, Urchs,
  Durnez, Mertz, Geisler, Floren, Gerhard, Sharp, Molina-Romero, Weinstein,
  Broderick, Saase, Andberg, Harms, Schlamp, Arias, Papadopoulos~Orfanos,
  Tarbert, Tambini, De~La~Vega, Nickson, Brett, Falkiewicz, Podranski,
  Linkersd{\"o}rfer, Flandin, Ort, Shachnev, McNamee, Davison, Varada,
  Schwabacher, Pellman, Perez-Guevara, Khanuja, Pannetier, McDermottroe, and
  Ghosh]{nipype2}
K.~J. Gorgolewski, O.~Esteban, C.~J. Markiewicz, E.~Ziegler, D.~G. Ellis, M.~P.
  Notter, D.~Jarecka, H.~Johnson, C.~Burns, A.~Manh{\~a}es-Savio,
  C.~Hamalainen, B.~Yvernault, T.~Salo, K.~Jordan, M.~Goncalves, M.~Waskom,
  D.~Clark, J.~Wong, F.~Loney, M.~Modat, B.~E. Dewey, C.~Madison,
  M.~Visconti~di Oleggio~Castello, M.~G. Clark, M.~Dayan, D.~Clark,
  A.~Keshavan, B.~Pinsard, A.~Gramfort, S.~Berleant, D.~M. Nielson,
  S.~Bougacha, G.~Varoquaux, B.~Cipollini, R.~Markello, A.~Rokem, B.~Moloney,
  Y.~O. Halchenko, D.~Wassermann, M.~Hanke, C.~Horea, J.~Kaczmarzyk,
  G.~de~Hollander, E.~DuPre, A.~Gillman, D.~Mordom, C.~Buchanan, R.~Tungaraza,
  W.~M. Pauli, S.~Iqbal, S.~Sikka, M.~Mancini, Y.~Schwartz, I.~B. Malone,
  M.~Dubois, C.~Frohlich, D.~Welch, J.~Forbes, J.~Kent, A.~Watanabe, C.~Cumba,
  J.~M. Huntenburg, E.~Kastman, B.~N. Nichols, A.~Eshaghi, D.~Ginsburg,
  A.~Schaefer, B.~Acland, S.~Giavasis, J.~Kleesiek, D.~Erickson,
  R.~K{\"u}ttner, C.~Haselgrove, C.~Correa, A.~Ghayoor, F.~Liem, J.~Millman,
  D.~Haehn, J.~Lai, D.~Zhou, R.~Blair, T.~Glatard, M.~Renfro, S.~Liu, A.~E.
  Kahn, F.~P{\'e}rez-Garc{\'\i}a, W.~Triplett, L.~Lampe, J.~Stadler, X.-Z.
  Kong, M.~Hallquist, A.~Chetverikov, J.~Salvatore, A.~Park, R.~Poldrack, R.~C.
  Craddock, S.~Inati, O.~Hinds, G.~Cooper, L.~N. Perkins, A.~Marina,
  A.~Mattfeld, M.~Noel, L.~Snoek, K.~Matsubara, B.~Cheung, S.~Rothmei,
  S.~Urchs, J.~Durnez, F.~Mertz, D.~Geisler, A.~Floren, S.~Gerhard, P.~Sharp,
  M.~Molina-Romero, A.~Weinstein, W.~Broderick, V.~Saase, S.~K. Andberg,
  R.~Harms, K.~Schlamp, J.~Arias, D.~Papadopoulos~Orfanos, C.~Tarbert,
  A.~Tambini, A.~De~La~Vega, T.~Nickson, M.~Brett, M.~Falkiewicz, K.~Podranski,
  J.~Linkersd{\"o}rfer, G.~Flandin, E.~Ort, D.~Shachnev, D.~McNamee,
  A.~Davison, J.~Varada, I.~Schwabacher, J.~Pellman, M.~Perez-Guevara,
  R.~Khanuja, N.~Pannetier, C.~McDermottroe, and S.~Ghosh.
\newblock Nipype.
\newblock \emph{Software}, 2018.
\newblock \doi{10.5281/zenodo.596855}.

\bibitem[Goyal et~al.(2022)Goyal, Moraczewski, Bandettini, Finn, and
  Thomas]{Goyal-2022The}
N.~Goyal, D.~Moraczewski, P.~A. Bandettini, E.~S. Finn, and A.~G. Thomas.
\newblock The positive--negative mode link between brain connectivity,
  demographics and behaviour: a pre-registered replication of smith et al .
  (2015).
\newblock \emph{Royal Society Open Science}, 9\penalty0 (2), Feb. 2022.
\newblock ISSN 2054-5703.
\newblock \doi{10.1098/rsos.201090}.
\newblock URL \url{http://dx.doi.org/10.1098/rsos.201090}.

\bibitem[Gratton et~al.(2022)Gratton, Nelson, and
  Gordon]{Gratton-2022Brain-behavior}
C.~Gratton, S.~M. Nelson, and E.~M. Gordon.
\newblock Brain-behavior correlations: Two paths toward reliability.
\newblock \emph{Neuron}, 110\penalty0 (9):\penalty0 1446--1449, 2022.

\bibitem[Greene and Constable(2023)]{Greene-Clinical}
A.~S. Greene and R.~T. Constable.
\newblock Clinical promise of brain-phenotype modeling: A review.
\newblock \emph{JAMA psychiatry}, 2023.

\bibitem[Greve and Fischl(2009)]{bbr}
D.~N. Greve and B.~Fischl.
\newblock Accurate and robust brain image alignment using boundary-based
  registration.
\newblock \emph{NeuroImage}, 48\penalty0 (1):\penalty0 63--72, 2009.
\newblock ISSN 1095-9572.
\newblock \doi{10.1016/j.neuroimage.2009.06.060}.

\bibitem[Gur et~al.(2010)Gur, Richard, Hughett, Calkins, Macy, Bilker,
  Brensinger, and Gur]{Gur-2010A-cognitive}
R.~C. Gur, J.~Richard, P.~Hughett, M.~E. Calkins, L.~Macy, W.~B. Bilker,
  C.~Brensinger, and R.~E. Gur.
\newblock A cognitive neuroscience-based computerized battery for efficient
  measurement of individual differences: standardization and initial construct
  validation.
\newblock \emph{Journal of neuroscience methods}, 187\penalty0 (2):\penalty0
  254--262, 2010.

\bibitem[Haufe et~al.(2014)Haufe, Meinecke, G{\"o}rgen, D{\"a}hne, Haynes,
  Blankertz, and Bie{\ss}mann]{Haufe-2014On}
S.~Haufe, F.~Meinecke, K.~G{\"o}rgen, S.~D{\"a}hne, J.-D. Haynes, B.~Blankertz,
  and F.~Bie{\ss}mann.
\newblock On the interpretation of weight vectors of linear models in
  multivariate neuroimaging.
\newblock \emph{Neuroimage}, 87:\penalty0 96--110, 2014.

\bibitem[He et~al.(2020)He, Kong, Holmes, Nguyen, Sabuncu, Eickhoff, Bzdok,
  Feng, and Yeo]{He-2020Deep}
T.~He, R.~Kong, A.~J. Holmes, M.~Nguyen, M.~R. Sabuncu, S.~B. Eickhoff,
  D.~Bzdok, J.~Feng, and B.~T. Yeo.
\newblock Deep neural networks and kernel regression achieve comparable
  accuracies for functional connectivity prediction of behavior and
  demographics.
\newblock \emph{NeuroImage}, 206:\penalty0 116276, 2020.
\newblock ISSN 1053-8119.
\newblock \doi{https://doi.org/10.1016/j.neuroimage.2019.116276}.
\newblock URL
  \url{https://www.sciencedirect.com/science/article/pii/S1053811919308675}.

\bibitem[He et~al.(2022)He, An, Chen, Chen, Feng, Bzdok, Holmes, Eickhoff, and
  Yeo]{He-2022Meta-matching}
T.~He, L.~An, P.~Chen, J.~Chen, J.~Feng, D.~Bzdok, A.~J. Holmes, S.~B.
  Eickhoff, and B.~T. Yeo.
\newblock Meta-matching as a simple framework to translate phenotypic
  predictive models from big to small data.
\newblock \emph{Nature neuroscience}, 25\penalty0 (6):\penalty0 795--804, 2022.

\bibitem[Herbold(2020)]{herbold2020autorank}
S.~Herbold.
\newblock Autorank: A python package for automated ranking of classifiers.
\newblock \emph{Journal of Open Source Software}, 5\penalty0 (48):\penalty0
  2173, 2020.

\bibitem[Hernan and Robins(2023)]{hernan_causal_2023}
M.~Hernan and J.~Robins.
\newblock \emph{Causal {Inference}: {What} {If}}.
\newblock Chapman \& {Hall}/{CRC} {Monographs} on {Statistics} \& {Applied}
  {Probab}. CRC Press, 2023.
\newblock ISBN 978-1-4200-7616-5.
\newblock URL \url{https://books.google.com/books?id=_KnHIAAACAAJ}.

\bibitem[Holderrieth et~al.(2022)Holderrieth, Smith, and
  Peng]{Holderrieth-2022Transfer}
P.~Holderrieth, S.~Smith, and H.~Peng.
\newblock Transfer learning for neuroimaging via re-use of deep neural network
  features.
\newblock \emph{medRxiv}, pages 2022--12, 2022.

\bibitem[Holmes and Patrick(2018)]{Holmes-2018The}
A.~J. Holmes and L.~M. Patrick.
\newblock The myth of optimality in clinical neuroscience.
\newblock \emph{Trends in cognitive sciences}, 22\penalty0 (3):\penalty0
  241--257, 2018.

\bibitem[Jenkinson et~al.(2002)Jenkinson, Bannister, Brady, and Smith]{mcflirt}
M.~Jenkinson, P.~Bannister, M.~Brady, and S.~Smith.
\newblock Improved optimization for the robust and accurate linear registration
  and motion correction of brain images.
\newblock \emph{NeuroImage}, 17\penalty0 (2):\penalty0 825--841, 2002.
\newblock ISSN 1053-8119.
\newblock \doi{10.1006/nimg.2002.1132}.
\newblock URL
  \url{http://www.sciencedirect.com/science/article/pii/S1053811902911328}.

\bibitem[Jiang et~al.(2020)Jiang, Calhoun, Fan, Zuo, Jung, Qi, Lin, Li, Zhuo,
  Song, et~al.]{Jiang-2020Gender}
R.~Jiang, V.~D. Calhoun, L.~Fan, N.~Zuo, R.~Jung, S.~Qi, D.~Lin, J.~Li,
  C.~Zhuo, M.~Song, et~al.
\newblock Gender differences in connectome-based predictions of individualized
  intelligence quotient and sub-domain scores.
\newblock \emph{Cerebral cortex}, 30\penalty0 (3):\penalty0 888--900, 2020.

\bibitem[Kaiser(1958)]{Kaiser-1958The}
H.~F. Kaiser.
\newblock The varimax criterion for analytic rotation in factor analysis.
\newblock \emph{Psychometrika}, 23\penalty0 (3):\penalty0 187--200, 1958.

\bibitem[Kennedy et~al.(2016)Kennedy, Haselgrove, Riehl, Preuss, and
  Buccigrossi]{Kennedy-2016The}
D.~N. Kennedy, C.~Haselgrove, J.~Riehl, N.~Preuss, and R.~Buccigrossi.
\newblock The nitrc image repository.
\newblock \emph{Neuroimage}, 124\penalty0 (Pt B):\penalty0 1069--1073, Jan
  2016.
\newblock \doi{10.1016/j.neuroimage.2015.05.074}.

\bibitem[Klein et~al.(2017)Klein, Ghosh, Bao, Giard, H{\"a}me, Stavsky, Lee,
  Rossa, Reuter, Neto, and Keshavan]{mindboggle}
A.~Klein, S.~S. Ghosh, F.~S. Bao, J.~Giard, Y.~H{\"a}me, E.~Stavsky, N.~Lee,
  B.~Rossa, M.~Reuter, E.~C. Neto, and A.~Keshavan.
\newblock Mindboggling morphometry of human brains.
\newblock \emph{PLOS Computational Biology}, 13\penalty0 (2):\penalty0
  e1005350, 2017.
\newblock ISSN 1553-7358.
\newblock \doi{10.1371/journal.pcbi.1005350}.
\newblock URL
  \url{http://journals.plos.org/ploscompbiol/article?id=10.1371/journal.pcbi.1005350}.

\bibitem[Kong et~al.(2019)Kong, Li, Orban, Sabuncu, Liu, Schaefer, Sun, Zuo,
  Holmes, Eickhoff, and Yeo]{Kong-2019Spatial}
R.~Kong, J.~Li, C.~Orban, M.~R. Sabuncu, H.~Liu, A.~Schaefer, N.~Sun, X.-N.
  Zuo, A.~J. Holmes, S.~B. Eickhoff, and B.~T.~T. Yeo.
\newblock Spatial topography of individual-specific cortical networks predicts
  human cognition, personality, and emotion.
\newblock \emph{Cereb Cortex}, 29\penalty0 (6):\penalty0 2533--2551, Jun 2019.
\newblock \doi{10.1093/cercor/bhy123}.

\bibitem[Lam et~al.(2022)Lam, Mahony, Raznahan, and
  Pereira]{lam2022interpretable}
K.~C. Lam, B.~W. Mahony, A.~Raznahan, and F.~Pereira.
\newblock Interpretable (meta) factorization of clinical questionnaires to
  identify general dimensions of psychopathology.
\newblock 2022.

\bibitem[Lanczos(1964)]{lanczos}
C.~Lanczos.
\newblock Evaluation of noisy data.
\newblock \emph{Journal of the Society for Industrial and Applied Mathematics
  Series B Numerical Analysis}, 1\penalty0 (1):\penalty0 76--85, 1964.
\newblock ISSN 0887-459X.
\newblock \doi{10.1137/0701007}.
\newblock URL \url{http://epubs.siam.org/doi/10.1137/0701007}.

\bibitem[Marek et~al.(2022)Marek, Tervo-Clemmens, Calabro, Montez, Kay, Hatoum,
  Donohue, Foran, Miller, Hendrickson, et~al.]{Marek-2022Reproducible}
S.~Marek, B.~Tervo-Clemmens, F.~J. Calabro, D.~F. Montez, B.~P. Kay, A.~S.
  Hatoum, M.~R. Donohue, W.~Foran, R.~L. Miller, T.~J. Hendrickson, et~al.
\newblock Reproducible brain-wide association studies require thousands of
  individuals.
\newblock \emph{Nature}, 603\penalty0 (7902):\penalty0 654--660, 2022.

\bibitem[Miller et~al.(2016)Miller, Alfaro-Almagro, Bangerter, Thomas, Yacoub,
  Xu, Bartsch, Jbabdi, Sotiropoulos, Andersson, et~al.]{Miller-2016Multimodal}
K.~L. Miller, F.~Alfaro-Almagro, N.~K. Bangerter, D.~L. Thomas, E.~Yacoub,
  J.~Xu, A.~J. Bartsch, S.~Jbabdi, S.~N. Sotiropoulos, J.~L. Andersson, et~al.
\newblock Multimodal population brain imaging in the uk biobank prospective
  epidemiological study.
\newblock \emph{Nature neuroscience}, 19\penalty0 (11):\penalty0 1523--1536,
  2016.

\bibitem[Nikolaidis et~al.(2022)Nikolaidis, Chen, He, Shinohara, Vogelstein,
  Milham, and Shou]{Nikolaidis-2022Suboptimal}
A.~Nikolaidis, A.~A. Chen, X.~He, R.~Shinohara, J.~Vogelstein, M.~Milham, and
  H.~Shou.
\newblock Suboptimal phenotypic reliability impedes reproducible human
  neuroscience.
\newblock \emph{BioRxiv}, pages 2022--07, 2022.

\bibitem[Ooi et~al.(2022)Ooi, Chen, Zhang, Kong, Tam, Li, Dhamala, Zhou,
  Holmes, and Yeo]{Ooi-2022Comparison}
L.~Q.~R. Ooi, J.~Chen, S.~Zhang, R.~Kong, A.~Tam, J.~Li, E.~Dhamala, J.~H.
  Zhou, A.~J. Holmes, and B.~T.~T. Yeo.
\newblock Comparison of individualized behavioral predictions across
  anatomical, diffusion and functional connectivity mri.
\newblock \emph{Neuroimage}, 263:\penalty0 119636, Nov 2022.
\newblock \doi{10.1016/j.neuroimage.2022.119636}.

\bibitem[Pedregosa et~al.(2011)Pedregosa, Varoquaux, Gramfort, Michel, Thirion,
  Grisel, Blondel, Prettenhofer, Weiss, Dubourg,
  et~al.]{Pedregosa-2011Scikit-learn:}
F.~Pedregosa, G.~Varoquaux, A.~Gramfort, V.~Michel, B.~Thirion, O.~Grisel,
  M.~Blondel, P.~Prettenhofer, R.~Weiss, V.~Dubourg, et~al.
\newblock Scikit-learn: Machine learning in python.
\newblock \emph{the Journal of machine Learning research}, 12:\penalty0
  2825--2830, 2011.

\bibitem[Pervaiz et~al.(2020)Pervaiz, Vidaurre, Woolrich, and
  Smith]{Pervaiz-2020Optimising}
U.~Pervaiz, D.~Vidaurre, M.~W. Woolrich, and S.~M. Smith.
\newblock Optimising network modelling methods for fmri.
\newblock \emph{NeuroImage}, 211:\penalty0 116604, 2020.
\newblock ISSN 1053-8119.
\newblock \doi{https://doi.org/10.1016/j.neuroimage.2020.116604}.
\newblock URL
  \url{https://www.sciencedirect.com/science/article/pii/S1053811920300914}.

\bibitem[Poldrack et~al.(2020)Poldrack, Huckins, and
  Varoquaux]{Poldrack-2020Establishment}
R.~A. Poldrack, G.~Huckins, and G.~Varoquaux.
\newblock Establishment of best practices for evidence for prediction: a
  review.
\newblock \emph{JAMA psychiatry}, 77\penalty0 (5):\penalty0 534--540, 2020.

\bibitem[Power et~al.(2014)Power, Mitra, Laumann, Snyder, Schlaggar, and
  Petersen]{power_fd_dvars}
J.~D. Power, A.~Mitra, T.~O. Laumann, A.~Z. Snyder, B.~L. Schlaggar, and S.~E.
  Petersen.
\newblock Methods to detect, characterize, and remove motion artifact in
  resting state fmri.
\newblock \emph{NeuroImage}, 84\penalty0 (Supplement C):\penalty0 320--341,
  2014.
\newblock ISSN 1053-8119.
\newblock \doi{10.1016/j.neuroimage.2013.08.048}.
\newblock URL
  \url{http://www.sciencedirect.com/science/article/pii/S1053811913009117}.

\bibitem[Pruim et~al.(2015)Pruim, Mennes, van Rooij, Llera, Buitelaar, and
  Beckmann]{aroma}
R.~H.~R. Pruim, M.~Mennes, D.~van Rooij, A.~Llera, J.~K. Buitelaar, and C.~F.
  Beckmann.
\newblock Ica-{AROMA}: A robust {ICA}-based strategy for removing motion
  artifacts from fmri data.
\newblock \emph{NeuroImage}, 112\penalty0 (Supplement C):\penalty0 267--277,
  2015.
\newblock ISSN 1053-8119.
\newblock \doi{10.1016/j.neuroimage.2015.02.064}.
\newblock URL
  \url{http://www.sciencedirect.com/science/article/pii/S1053811915001822}.

\bibitem[Roalf et~al.(2014)Roalf, Gur, Ruparel, Calkins, Satterthwaite, Bilker,
  Hakonarson, Harris, and Gur]{Roalf-2014Within-individual}
D.~R. Roalf, R.~E. Gur, K.~Ruparel, M.~E. Calkins, T.~D. Satterthwaite, W.~B.
  Bilker, H.~Hakonarson, L.~J. Harris, and R.~C. Gur.
\newblock Within-individual variability in neurocognitive performance: age-and
  sex-related differences in children and youths from ages 8 to 21.
\newblock \emph{Neuropsychology}, 28\penalty0 (4):\penalty0 506, 2014.

\bibitem[Rosenberg and Finn(2022)]{Rosenberg-2022How}
M.~D. Rosenberg and E.~S. Finn.
\newblock How to establish robust brain--behavior relationships without
  thousands of individuals.
\newblock \emph{Nature Neuroscience}, 25\penalty0 (7):\penalty0 835--837, 2022.

\bibitem[Satterthwaite et~al.(2013)Satterthwaite, Elliott, Gerraty, Ruparel,
  Loughead, Calkins, Eickhoff, Hakonarson, Gur, Gur, and
  Wolf]{confounds_satterthwaite_2013}
T.~D. Satterthwaite, M.~A. Elliott, R.~T. Gerraty, K.~Ruparel, J.~Loughead,
  M.~E. Calkins, S.~B. Eickhoff, H.~Hakonarson, R.~C. Gur, R.~E. Gur, and D.~H.
  Wolf.
\newblock {An improved framework for confound regression and filtering for
  control of motion artifact in the preprocessing of resting-state functional
  connectivity data}.
\newblock \emph{NeuroImage}, 64\penalty0 (1):\penalty0 240--256, 2013.
\newblock ISSN 10538119.
\newblock \doi{10.1016/j.neuroimage.2012.08.052}.
\newblock URL
  \url{http://linkinghub.elsevier.com/retrieve/pii/S1053811912008609}.

\bibitem[Satterthwaite et~al.(2016)Satterthwaite, Connolly, Ruparel, Calkins,
  Jackson, Elliott, Roalf, Hopson, Prabhakaran, Behr, Qiu, Mentch, Chiavacci,
  Sleiman, Gur, Hakonarson, and Gur]{Satterthwaite-2016The}
T.~D. Satterthwaite, J.~J. Connolly, K.~Ruparel, M.~E. Calkins, C.~Jackson,
  M.~A. Elliott, D.~R. Roalf, R.~Hopson, K.~Prabhakaran, M.~Behr, H.~Qiu, F.~D.
  Mentch, R.~Chiavacci, P.~M.~A. Sleiman, R.~C. Gur, H.~Hakonarson, and R.~E.
  Gur.
\newblock The philadelphia neurodevelopmental cohort: A publicly available
  resource for the study of normal and abnormal brain development in youth.
\newblock \emph{Neuroimage}, 124\penalty0 (Pt B):\penalty0 1115--1119, Jan
  2016.
\newblock \doi{10.1016/j.neuroimage.2015.03.056}.

\bibitem[Schaefer et~al.(2018)Schaefer, Kong, Gordon, Laumann, Zuo, Holmes,
  Eickhoff, and Yeo]{Schaefer-2018Local-global}
A.~Schaefer, R.~Kong, E.~M. Gordon, T.~O. Laumann, X.-N. Zuo, A.~J. Holmes,
  S.~B. Eickhoff, and B.~T. Yeo.
\newblock Local-global parcellation of the human cerebral cortex from intrinsic
  functional connectivity mri.
\newblock \emph{Cerebral cortex}, 28\penalty0 (9):\penalty0 3095--3114, 2018.

\bibitem[Sch{\"o}ttner et~al.(2023)Sch{\"o}ttner, Bolton, Patel, Nah{\'a}lka,
  Vieira, and Hagmann]{Schottner-2023Exploring}
M.~Sch{\"o}ttner, T.~A.~W. Bolton, J.~Patel, A.~T. Nah{\'a}lka, S.~Vieira, and
  P.~Hagmann.
\newblock Exploring the latent structure of behavior using the human connectome
  project's data.
\newblock \emph{Sci Rep}, 13\penalty0 (1):\penalty0 713, Jan 2023.
\newblock \doi{10.1038/s41598-022-27101-1}.

\bibitem[Schulz et~al.(2020)Schulz, Yeo, Vogelstein, Mourao-Miranada, Kather,
  Kording, Richards, and Bzdok]{Schulz-2020Different}
M.-A. Schulz, B.~T.~T. Yeo, J.~T. Vogelstein, J.~Mourao-Miranada, J.~N. Kather,
  K.~Kording, B.~Richards, and D.~Bzdok.
\newblock Different scaling of linear models and deep learning in {UKBiobank}
  brain images versus machine-learning datasets.
\newblock \emph{Nature Communications}, 11\penalty0 (1), aug 2020.
\newblock \doi{10.1038/s41467-020-18037-z}.
\newblock URL \url{https://doi.org/10.1038%2Fs41467-020-18037-z}.

\bibitem[Slotkin et~al.(2012)Slotkin, Kallen, Griffith, Magasi, Salsman,
  Nowinski, et~al.]{Slotkin-2012NIH}
J.~Slotkin, M.~Kallen, J.~Griffith, S.~Magasi, J.~Salsman, C.~Nowinski, et~al.
\newblock Nih toolbox.
\newblock \emph{Technical Manual.[Google Scholar]}, 2012.

\bibitem[Smith et~al.(2015)Smith, Nichols, Vidaurre, Winkler, Behrens, Glasser,
  Ugurbil, Barch, Van~Essen, and Miller]{Smith-2015A}
S.~M. Smith, T.~E. Nichols, D.~Vidaurre, A.~M. Winkler, T.~E. Behrens, M.~F.
  Glasser, K.~Ugurbil, D.~M. Barch, D.~C. Van~Essen, and K.~L. Miller.
\newblock A positive-negative mode of population covariation links brain
  connectivity, demographics and behavior.
\newblock \emph{Nature neuroscience}, 18\penalty0 (11):\penalty0 1565--1567,
  2015.

\bibitem[Spisak et~al.(2023)Spisak, Bingel, and Wager]{Spisak-2023Multivariate}
T.~Spisak, U.~Bingel, and T.~D. Wager.
\newblock Multivariate bwas can be replicable with moderate sample sizes.
\newblock \emph{Nature}, 615\penalty0 (7951):\penalty0 E4--E7, 2023.

\bibitem[Sui et~al.(2020)Sui, Jiang, Bustillo, and
  Calhoun]{Sui-2020Neuroimaging-based}
J.~Sui, R.~Jiang, J.~Bustillo, and V.~Calhoun.
\newblock Neuroimaging-based individualized prediction of cognition and
  behavior for mental disorders and health: methods and promises.
\newblock \emph{Biological psychiatry}, 88\penalty0 (11):\penalty0 818--828,
  2020.

\bibitem[Tian and Zalesky(2021)]{Tian-2021Machine}
Y.~Tian and A.~Zalesky.
\newblock Machine learning prediction of cognition from functional
  connectivity: Are feature weights reliable?
\newblock \emph{NeuroImage}, 245:\penalty0 118648, 2021.
\newblock ISSN 1053-8119.
\newblock \doi{https://doi.org/10.1016/j.neuroimage.2021.118648}.
\newblock URL
  \url{https://www.sciencedirect.com/science/article/pii/S1053811921009216}.

\bibitem[Tong et~al.(2022)Tong, Xie, Carlisle, Fonzo, Oathes, Jiang, and
  Zhang]{tong2022transdiagnostic}
X.~Tong, H.~Xie, N.~Carlisle, G.~A. Fonzo, D.~J. Oathes, J.~Jiang, and
  Y.~Zhang.
\newblock Transdiagnostic connectome signatures from resting-state fmri predict
  individual-level intellectual capacity.
\newblock \emph{Translational psychiatry}, 12\penalty0 (1):\penalty0 367, 2022.

\bibitem[Tustison et~al.(2010)Tustison, Avants, Cook, Zheng, Egan, Yushkevich,
  and Gee]{n4}
N.~J. Tustison, B.~B. Avants, P.~A. Cook, Y.~Zheng, A.~Egan, P.~A. Yushkevich,
  and J.~C. Gee.
\newblock N4itk: Improved n3 bias correction.
\newblock \emph{IEEE Transactions on Medical Imaging}, 29\penalty0
  (6):\penalty0 1310--1320, 2010.
\newblock ISSN 0278-0062.
\newblock \doi{10.1109/TMI.2010.2046908}.

\bibitem[Van~den Akker et~al.(2021)Van~den Akker, Weston, Campbell, Chopik,
  Damian, Davis-Kean, Hall, Kosie, Kruse, Olsen,
  et~al.]{Van-den-Akker-2021Preregistration}
O.~Van~den Akker, S.~Weston, L.~Campbell, B.~Chopik, R.~Damian, P.~Davis-Kean,
  A.~Hall, J.~Kosie, E.~Kruse, J.~Olsen, et~al.
\newblock Preregistration of secondary data analysis: A template and tutorial.
\newblock \emph{Meta-psychology}, 5:\penalty0 2625, 2021.

\bibitem[Van~Essen et~al.(2012)Van~Essen, Ugurbil, Auerbach, Barch, Behrens,
  Bucholz, Chang, Chen, Corbetta, Curtiss, Della~Penna, Feinberg, Glasser,
  Harel, Heath, Larson-Prior, Marcus, Michalareas, Moeller, Oostenveld,
  Petersen, Prior, Schlaggar, Smith, Snyder, Xu, Yacoub, and {WU-Minn HCP
  Consortium}]{Van-Essen:2012aa}
D.~C. Van~Essen, K.~Ugurbil, E.~Auerbach, D.~Barch, T.~E.~J. Behrens,
  R.~Bucholz, A.~Chang, L.~Chen, M.~Corbetta, S.~W. Curtiss, S.~Della~Penna,
  D.~Feinberg, M.~F. Glasser, N.~Harel, A.~C. Heath, L.~Larson-Prior,
  D.~Marcus, G.~Michalareas, S.~Moeller, R.~Oostenveld, S.~E. Petersen,
  F.~Prior, B.~L. Schlaggar, S.~M. Smith, A.~Z. Snyder, J.~Xu, E.~Yacoub, and
  {WU-Minn HCP Consortium}.
\newblock The human connectome project: a data acquisition perspective.
\newblock \emph{Neuroimage}, 62\penalty0 (4):\penalty0 2222--31, Oct 2012.
\newblock \doi{10.1016/j.neuroimage.2012.02.018}.

\bibitem[Wu et~al.(2022)Wu, Li, Eickhoff, Hoffstaedter, Hanke, Yeo, and
  Genon]{Wu-2022Cross-cohort}
J.~Wu, J.~Li, S.~B. Eickhoff, F.~Hoffstaedter, M.~Hanke, B.~T. Yeo, and
  S.~Genon.
\newblock Cross-cohort replicability and generalizability of connectivity-based
  psychometric prediction patterns.
\newblock \emph{Neuroimage}, 262:\penalty0 119569, 2022.

\bibitem[Wulan et~al.(2024)Wulan, An, Zhang, Kong, Chen, Bzdok, Eickhoff,
  Holmes, and Yeo]{Wulan-2024Translating}
N.~Wulan, L.~An, C.~Zhang, R.~Kong, P.~Chen, D.~Bzdok, S.~B. Eickhoff, A.~J.
  Holmes, and B.~T. Yeo.
\newblock Translating phenotypic prediction models from big to small anatomical
  mri data using meta-matching.
\newblock \emph{bioRxiv}, pages 2023--12, 2024.

\bibitem[Zelenina et~al.(2023)Zelenina, Pine, Stringaris, and
  Nielson]{Zelenina-2023Validation}
M.~Zelenina, D.~Pine, A.~Stringaris, and D.~Nielson.
\newblock Validation of cbcl depression scores of adolescents in three
  independent datasets.
\newblock 2023.

\bibitem[Zhang et~al.(2001)Zhang, Brady, and Smith]{fsl_fast}
Y.~Zhang, M.~Brady, and S.~Smith.
\newblock Segmentation of brain {MR} images through a hidden markov random
  field model and the expectation-maximization algorithm.
\newblock \emph{IEEE Transactions on Medical Imaging}, 20\penalty0
  (1):\penalty0 45--57, 2001.
\newblock ISSN 0278-0062.
\newblock \doi{10.1109/42.906424}.

\bibitem[Zuo et~al.(2019)Zuo, Xu, and Milham]{Zuo-2019Harnessing}
X.-N. Zuo, T.~Xu, and M.~P. Milham.
\newblock Harnessing reliability for neuroscience research.
\newblock \emph{Nature human behaviour}, 3\penalty0 (8):\penalty0 768--771,
  2019.

\end{thebibliography}
